\documentclass[a4paper,fleqn,usenatbib]{mnras}

\usepackage{amsmath}
\usepackage{amssymb}
\usepackage{txfonts}

\usepackage[T1]{fontenc}
\usepackage{ae,aecompl}

\usepackage{graphicx,fixltx2e,amsmath,amssymb,amsfonts,latexsym,wasysym}

\usepackage{natbib}
\usepackage{bm}
\usepackage{times}
\usepackage{color}
\usepackage{amssymb}
\usepackage{fancyvrb}

\title[PMS star X-ray emission with radiative core growth]
{The influence of radiative core growth on coronal X-ray emission from pre-main sequence stars}

\author[Scott G. Gregory et al.]{
Scott G. Gregory,$^{1}$\thanks{E-mail: sg64@st-andrews.ac.uk (SGG), fca@umich.edu (FCA), cdavies@astro.ex.ac.uk (CLD)} 
Fred C. Adams$^{2,3}$ 
and Claire L. Davies$^{4,1}$
\\
$^1$SUPA, School of Physics \& Astronomy, University of St Andrews, St Andrews, KY16 9SS, U.K.\\
$^2$Physics Department, University of Michigan, Ann Arbor, MI 48109, U.S.A. \\
$^3$Astronomy Department, University of Michigan, Ann Arbor, MI 48109, U.S.A.\\
$^4$School of Physics, University of Exeter, Exeter, EX4 4QL, U.K. 
}

\date{Accepted XXX. Received YYY; in original form ZZZ}

\pubyear{2015}

\begin{document}
\label{firstpage}
\pagerange{\pageref{firstpage}--\pageref{lastpage}}
\maketitle

\begin{abstract}
Pre-main sequence (PMS) stars of mass $\gtrsim0.35\,{\rm M}_\odot$ transition from hosting fully convective interiors to configurations with a radiative core and outer convective envelope during their gravitational contraction. This stellar structure change influences the external magnetic field topology and, as we demonstrate herein, affects the coronal X-ray emission as a stellar analog of the solar tachocline develops. We have combined archival X-ray, spectroscopic, and photometric data for $\sim$1000 PMS stars from five of the best studied star forming regions; the ONC, NGC~2264, IC~348, NGC~2362, and NGC~6530. Using a modern, PMS calibrated, spectral type-to-effective temperature and intrinsic colour scale, we deredden the photometry using colours appropriate for each spectral type, and determine the stellar mass, age, and internal structure consistently for the entire sample.  We find that PMS stars on Henyey tracks have, on average, lower fractional X-ray luminosities ($L_{\rm X}/L_\ast$) than those on Hayashi tracks, where this effect is driven by changes in $L_{\rm X}$.  X-ray emission decays faster with age for higher mass PMS stars. There is a strong correlation between $L_\ast$ and $L_{\rm X}$ for Hayashi track stars but no correlation for Henyey track stars. There is no correlation between $L_{\rm X}$ and radiative core mass or radius. However, the longer stars have spent with radiative cores, the less X-ray luminous they become. The decay of coronal X-ray emission from young early K to late G-type PMS stars, the progenitors of main sequence A-type stars, is consistent with the dearth of X-ray detections of the latter.

\end{abstract}




\section{Introduction}
\label{intro}

X-ray emission from low-mass ($\lesssim$3$\,{\rm M}_\odot$) pre-main sequence (PMS) stars is dominantly coronal in origin (e.g. \citealt{sta06}).  Compared to the contemporary Sun, they have coronae that are an order of magnitude hotter on average, $\sim$30${\rm MK}$ compared to $\sim$2${\rm MK}$, and are $\sim$10-10$^5$ times more X-ray luminous \citep{pre05b}.  How the coronae of stars attain such high temperatures is the subject of debate.  It may be caused by the dissipation of magnetohydrodynamic waves \citep{nar96}, a result of numerous, unresolved, nanoflares \citep{par88}, or a combination of both \citep{kli14}.

PMS star X-ray light curves consist of a characteristic (often called ``quiescent'' although it is variable) level of X-ray emission (e.g. \citealt{wol05}), superposed on which are larger, more energetic flares. Flares are the result of particle acceleration along magnetic loops following the energy release from reconnection events.  In turn, reconnection arises from the twisting and tangling of coronal magnetic field lines induced by the photospheric footpoint motions driven by stellar surface transport effects (e.g. surface differential rotation, meridional flow and supergranular diffusion).    

Flares in the X-ray light curves of PMS stars follow similar temporal behaviour to those observed on the Sun, with an almost linear rise phase followed by a slower exponential decay, but can be far more energetic \citep{wol05,fav05}.  The most energetic flare detected from a PMS star would be classified as $\sim$X40,000 if it were to take place on the Sun (A. Aarnio, private communication).  The largest flares detected from PMS stars were once thought to be driven by the star-disc interaction, where magnetic field lines connecting stars to their inner disc were twisted due to the differential rotation rate between the footpoint of the loop at the disc and that on the stellar surface \citep{mon00,fav05}.  However, it was subsequently established by examining excess infra-red emission, a result of the reprocessing of the stellar photons from dusty circumstellar discs, that the largest and most violent X-ray flares also occur on PMS stars whose discs had dispersed (\citealt{get08}; \citealt*{aar10}).  

PMS star X-ray luminosity $L_{\rm X}$ varies from star to star and is known to correlate with stellar mass and bolometric luminosity $L_\ast$ (e.g. \citealt*{fla03b}; \citealt{pre05b,tel07}).  $L_{\rm X}$ can also vary over short timescales of $\sim$hours- to-days (due to flares), and over longer $\sim$Myr timescales with \citet{pre05a} reporting a weak decrease in $L_{\rm X}$ with increasing age, at least in stellar mass stratified subsamples. In contrast to main sequence dwarfs, however, the stellar rotation rate does not appear to play a significant role in PMS star X-ray emission. PMS stars do not follow the well-known main sequence rotation-activity relation, whereby $\log(L_{\rm X}/L_\ast)$ increases with rotation rate until it saturates at $\log(L_{\rm X}/L_\ast)\approx-3$ (e.g. \citealt{wri11}). Instead, almost all PMS stars show saturated X-ray emission, but with orders of magnitude more scatter in $\log(L_{\rm X}/L_\ast)$ than what is found for stars in main sequence clusters \citep{pre05b}. This scatter has been shown to reduce to main sequence levels by a cluster age of $\sim$30$\,{\rm Myr}$, and has been speculatively related, although in a currently unknown way, to the stellar internal structure transition from fully to partially convective as PMS stars evolve across the Hertzsprung-Russell (H-R) diagram \citep{ale12}.      

After emerging from their natal dust clouds, PMS stars are located in the upper right of the [$\log(L_\ast/L_\odot)$ vs $\log T_{\rm eff}$] H-R diagram on the birth line \citep{sta83}.  They initially host fully convective interiors.  By the virial theorem, as the star contracts half of the gravitational energy heats the stellar interior while the other half is radiated away.  While the stellar interior is fully convective the contraction is homologous, such that the mass fraction $m(r)/M_\ast$ contained within radius $r$ inside the star is the same as the radius fraction $r/R_\ast$ at all times \citep{bod11}, and all interior mass shells are losing heat. This global loss of heat across all mass shells results in an internal luminosity profile that increases from the centre of the star to the surface \citep{pal93}.  During this time, the stellar radius decreases with age approximately as $R_\ast\propto t^{-1/3}$ \citep{lam05,bat13}. 

\begin{figure}
   \centering
   \includegraphics[width=0.35\textwidth]{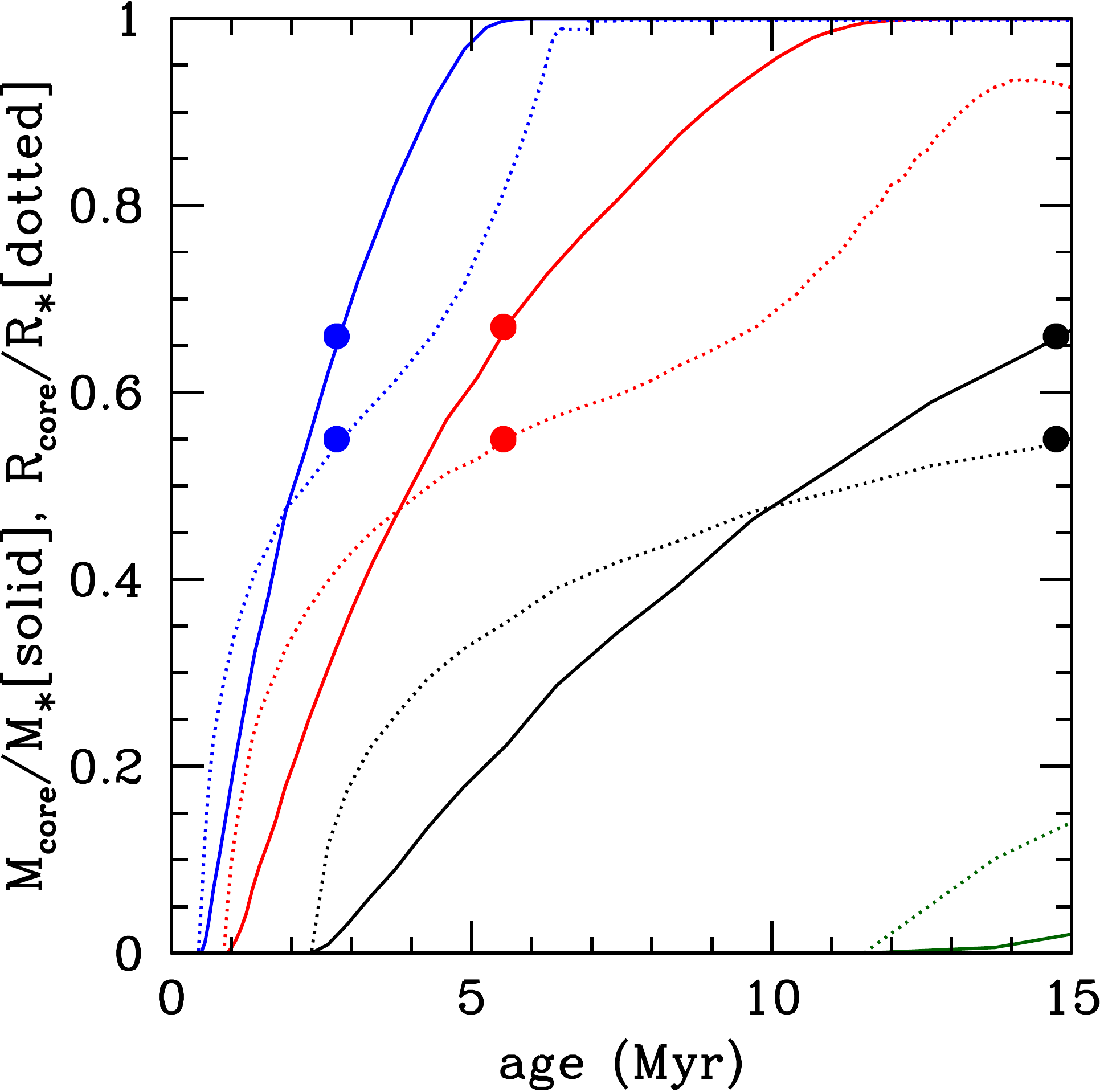}
    \includegraphics[width=0.35\textwidth]{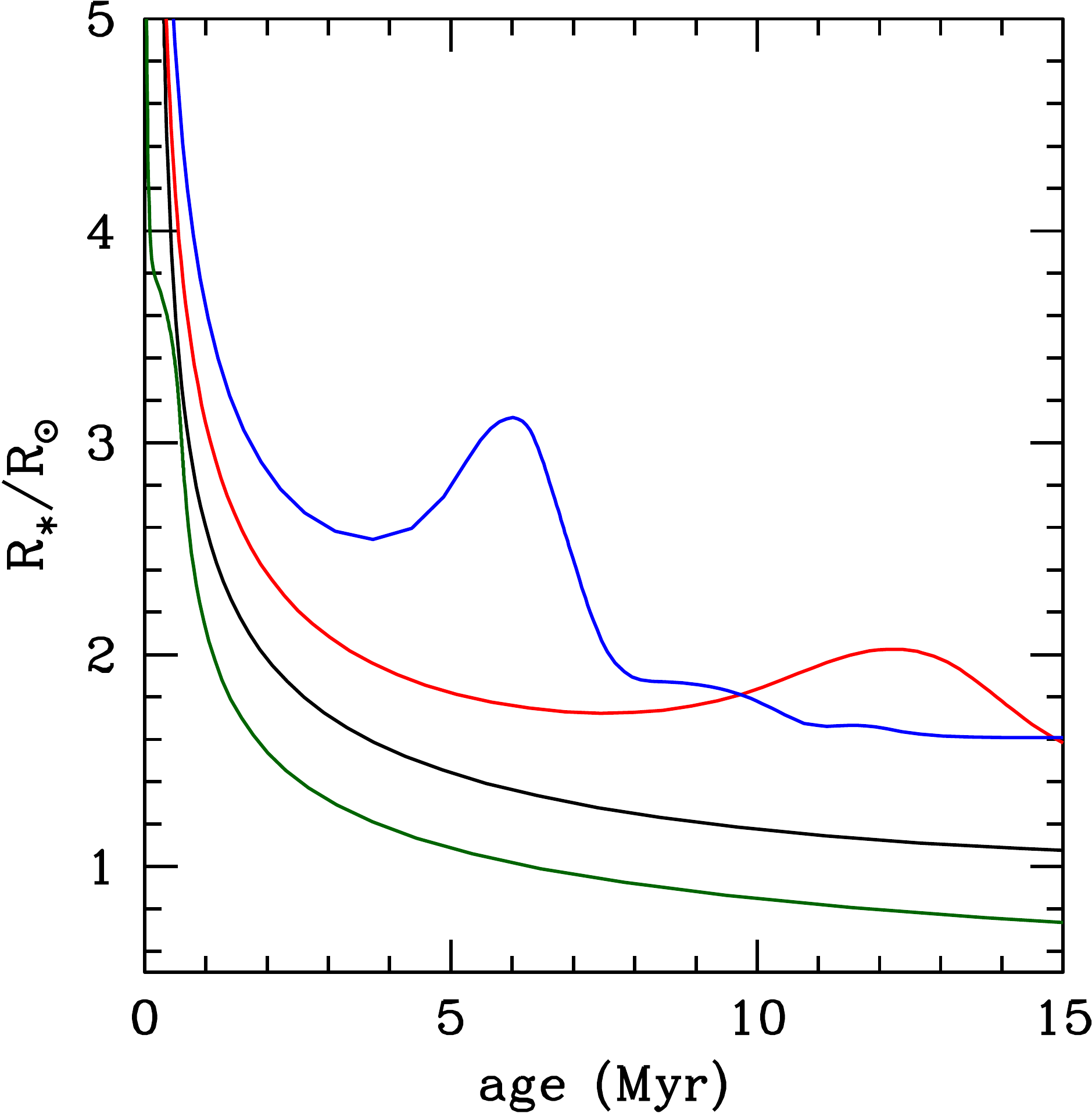}
   \caption{(upper) The fraction of the stellar mass contained within the radiative core (solid lines) and its radius relative to the stellar radius (dotted lines) as a function of age for PMS stars of mass 0.5 (green), 1 (black), 1.5 (red), and $2\,{\rm M}_\odot$ (blue) from the models of \citet{sie00} [solar metallicity with convective overshooting]. The points on the curves mark the approximate age of the beginning of the Henyey track for stars of mass (left-to-right): 2, 1.5, and 1$\,{\rm M}_\odot$), estimated from the second term on the right-hand-side of equation (\ref{tsincehen}). A 0.5$\,{\rm M}_\odot$ star never evolves onto a Henyey track.  Partially convective PMS stars become more centrally condensed once the radiative core has grown to occupy a sufficient proportion of the stellar interior, with the core containing a greater fraction of the stellar mass while occupying a smaller fraction of the stellar radius ($M_{\rm core}/M_\ast>R_{\rm core}/R_\ast$). (lower) The contraction of the stellar radius.  Colours match the upper panel. The slight slowing of the contraction at $<1\,{\rm Myr}$ of a $0.5\,{\rm M}_\odot$ star is caused by deuterium burning and occurs at earlier ages for the higher mass stars.  The expansion of the stellar radius is clear during the Henyey track evolution of the 2 and 1.5$\,{\rm M}_\odot$ stars. A 2 (1.5)$\,{\rm M}_\odot$ star reaches the zero-age main-sequence at approximately 7.2 (13.5)$\,{\rm Myr}$.}
   \label{mcorercore}
\end{figure}

As PMS contraction progresses, the opacity (which is an inverse function of temperature) drops and the central regions of the star become stable against convection. A radiative core begins to grow, at least for stars of mass $\gtrsim$0.35$\,{\rm M}_\odot$ \citep{cha97}, with lower mass stars retaining their fully convective interiors for the entirety of their PMS, and main sequence, evolution.  The lowest mass stars considered in this work (0.1$\,{\rm M_\odot}$, see section \ref{siessstuff}) will remain fully convective for their main sequence lifetime of trillions of years (e.g. \citealt*{lau97}).  When a radiative core develops, the base of the convective zone is heated, and a maximum in the internal luminosity profile is generated (the internal luminosity increases from the centre of the star to a maximum, then drops towards the stellar surface; \citealt{pal93,ibe13}). Mass shells interior to this maximum lose heat while those at larger radii up to the stellar surface gain heat.  This internal luminosity maximum radiatively diffuses towards the stellar surface over time \citep{mae09}, which eventually leads to a rise in the surface luminosity and the star moving from its Hayashi track (where $L_\ast$ is decreasing with age) to its Henyey track (where $L_\ast$ is increasing with age) in the H-R diagram.  During radiative core development, where heat is absorbed by the convective envelope, the stellar contraction becomes non-homologous, with the central regions continuing to contract while the outer layers initially expand \citep{pal93}.  Eventually the PMS star becomes more centrally condensed, with the radiative core containing a greater fraction of the stellar mass while occupying a smaller fraction of the stellar radius (i.e. $m(R_{\rm core})/M_\ast\equiv M_{\rm core}/M_\ast>R_{\rm core}/R_\ast$).    The behaviour described above is illustrated in Figure \ref{mcorercore} for stars of mass 0.5, 1, 1.5, and 2$\,{\rm M}_\odot$, constructed from the models of \citet*{sie00}.       

The stellar structure transition from fully convective to partially convective interiors has been found to influence the large-scale external magnetic field topology of PMS stars (e.g. \citealt{don11,don12}).  Evidence is beginning to emerge for a magnetic evolutionary scenario, whereby stars (at least those of mass $\gtrsim$0.5$\,{\rm M}_\odot$) are born with simple (with dominant low $\ell$ number mode) axisymmetric large-scale magnetic fields that become significantly more complex, multipolar, and dominantly non-axisymmetric once a substantial radiative core has developed \citep{gre12,gre14}.\footnote{This magnetic topology change mirrors what has been reported previously for main-sequence dwarf stars that lie on either side of the fully convective divide \citep{mor08,don08,gre12}.}  Although more data is required to confirm the reported trends, they likely reflect the change in dynamo magnetic field generation process as PMS stars develop stellar analogs of the solar tacholine, a shear layer between the radiative core and convective envelope.  

In this paper we examine the influence of radiative core development on the coronal X-ray emission from PMS stars.  Observationally, the best studied PMS stars reside in star forming regions of average age of a few million years, and are therefore young and predominantly fully convective.  As such, in order to obtain a large sample of partially convective PMS stars we consider data from five of the best studied star forming regions, as detailed in section \ref{section2}.  Rather than taking literature values for the stellar luminosities and effective temperatures, we re-calculate/reassign them based on the observed photometry and estimate spectral types in a consistent manner across our entire sample of $\sim$1000 stars.  During this process we adopt the latest spectral type estimates (e.g. \citealt*{hil13}) and use the latest generation intrinsic colour/effective temperature scale as calibrated for PMS stars \citep{pec13}. Using PMS evolutionary models to estimate their mass, age, and internal structure from their location in the H-R diagram (section \ref{siessstuff}), we throughly examine the difference in the X-ray emission properties of fully and partially convective stars in section \ref{xraycompare}.  In section \ref{sectionevolve} we consider how the X-ray emission changes as partially convective PMS stars evolve across the H-R diagram, before concluding in section \ref{conclusion}.


\section{Stellar parameters: spectral types \& bolometric/X-ray luminosities}\label{section2}
The majority of PMS stars within young star forming regions have fully convective interiors. Therefore, in order to examine the X-ray properties of partially convective PMS stars, it is necessary to consider data from more than one star forming region to increase the sample size. We have collated published data from five of the most extensively studied star forming regions, the Orion Nebula Cluster (ONC), IC~348, NGC~2264, NGC~2362, and NGC~6530, as described in the following subsections.  In order to determine whether or not a given PMS star has a fully or partially convective interior we must position the star in the H-R diagram. This requires an effective temperature, $T_{\rm eff}$, and a stellar luminosity, $L_\ast$.  In turn, $T_{\rm eff}$ can be assigned given a stellar spectral type, while $L_\ast$ can be calculated by dereddening observed photometry, the application of a bolometric correction appropriate for the stellar spectral type, and with an assumed distance modulus (usually taken to be the same for every star in a cluster).   

For each of the five star forming regions we obtained X-ray luminosities, spectral types and observed magnitudes/colours from published catalogs, as detailed below. Rather than taking published values of stellar effective temperatures and luminosities, we assigned and calculated the values ourselves.  This approach has the major advantage that effective temperatures are assigned across all stars in all of the star forming regions in a fully consistent manner, eliminating differences between the published catalogs and allowing them to be more directly compared. The spectral type to effective temperature scale that we adopt in this paper is the recently published PMS scale of \citet{pec13}.  Although they used stars of age $5-30\,{\rm Myr}$ to derive their spectral type to $T_{\rm eff}$ (-intrinsic colour-bolometric correction) scale, they are more appropriate for younger PMS stars than the commonly adopted dwarf star scales. Concerns have long been raised over the use of the latter for PMS stars, which (given their larger radii) have lower surface gravities than dwarf stars (e.g. \citealt{hil97,dah07}).  Furthermore, the \citet{pec13} scale takes into account the greater spot coverage on PMS stars compared to main sequence dwarfs.  

\begin{figure}
   \centering
   \includegraphics[width=0.35\textwidth]{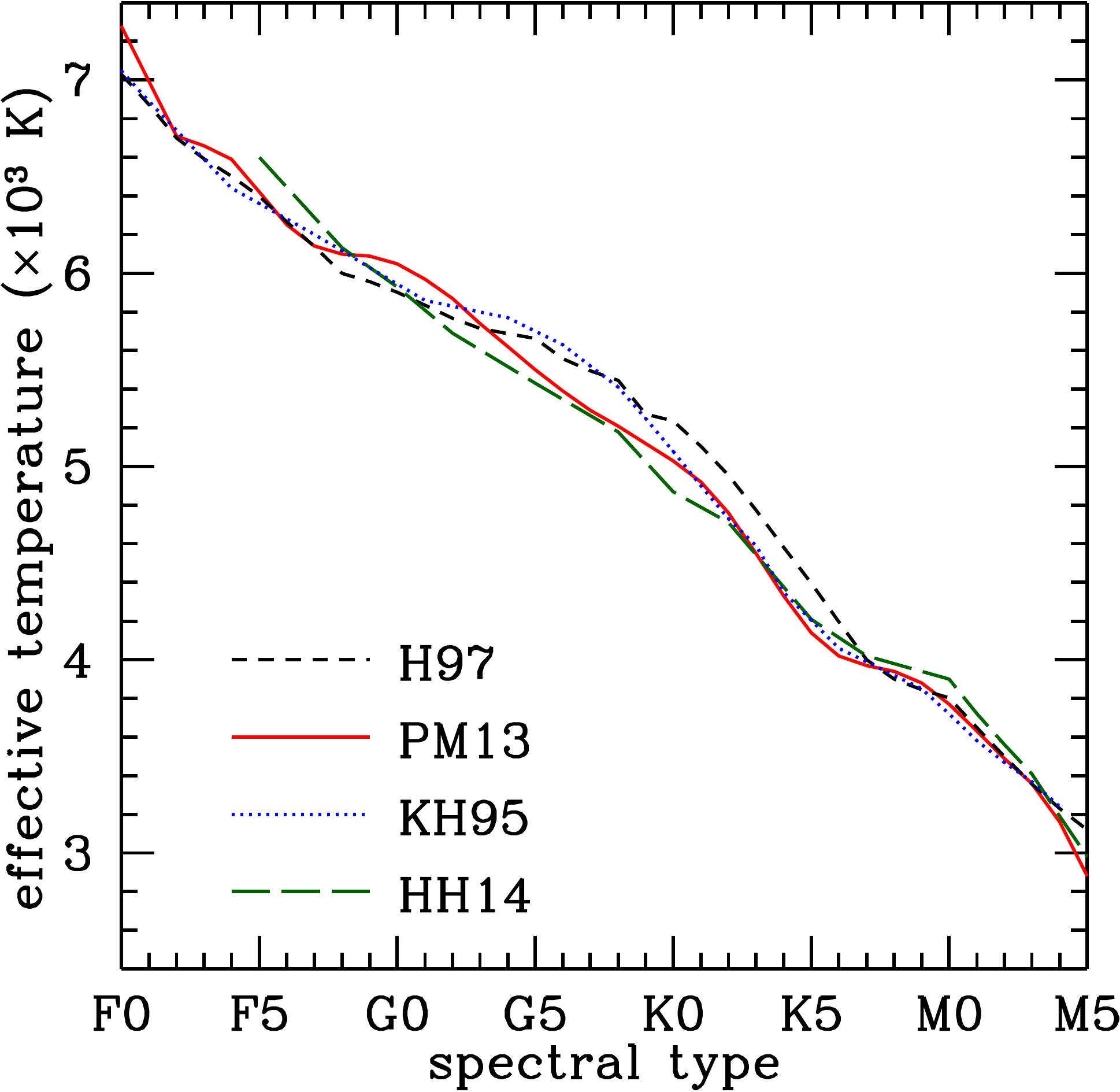}
   \caption{A comparison between the modern PMS spectral type-to-effective temperature conversion scales of \citet{pec13} (solid red line, labelled PM13) and \citet{her14} (long dashed green line, from F5 and later, labelled HH14), and the dwarf star conversion scales used in the ONC survey of \citet{hil97} (short dashed black line, labelled H97, which is a slightly modified version of the dwarf star temperature scale of \citealt{coh79}) and that listed by \citet{ken95} (dotted blue line, labelled KH95).  The effective temperatures of the modern PMS scale range from $\sim150-200\,{\rm K}$ less than the dwarf scale from mid-G to mid-K spectral types when compared to that of \citet{hil97}, whereas the difference is largest for G-types when compared to \citet{ken95}.}
   \label{PM13}
\end{figure}

Figure \ref{PM13} shows a comparison between the modern PMS spectral type-$T_{\rm eff}$ conversion scales of \citet{pec13} and \citet{her14}, and the dwarf star scale used by \citet{hil97} in the acclaimed survey of young stars in the ONC [which, between the considered spectral types, is a slightly modified version of the conversion scale of \citet{coh79}].  Also shown is another commonly used scale, again based on dwarf stars, published by \citet{ken95}. The conversion scale of \cite{pec13} is limited to spectral types F0 to M5, while \citet{her14} extends from F5 to later than M5.  Stars of earlier spectral type are not of interest for this study, as they are too massive to be T Tauri stars.  The M5 cut-off of the \citet{pec13} scale is of little consequence to this study of partially convective PMS stars, as such late type (low mass) stars retain their fully convective interiors for the entirety of their PMS evolution. It is immediately evident from Figure \ref{PM13} that PMS stars are cooler than dwarf stars by $\sim$200$\,{\rm K}$ from roughly mid-G to mid-K [for G-types] when comparing them to the dwarf scales of \citet{hil97} [\citet{ken95}]. This correction will move them to lower masses and to younger ages in the H-R diagram.     

From the photometry and observed colours we have calculated stellar luminosities as described for each of the star forming regions in the following subsections.  This requires bolometric corrections and intrinsic colours for which we also use the PMS conversion scale of \citet{pec13}, thus ensuring that our luminosities have been calculated in a uniform manner across all the star forming regions and using a fully consistent effective temperature/bolometric correction combination. \citet{tor10} discuss the importance of this, which also requires the use of a value for the absolute bolometric magnitude of the Sun (this enters into the calculation of the stellar bolometric luminosity).  For this, we use $M_{\rm bol,\odot}=4.755$, as \citet{pec13}.  This should be used with a modern estimate of the solar effective temperature of $T_{\rm eff,\odot}=5772\,{\rm K}$ and $R_\odot = 6.9566\times10^{8}\,{\rm m}$ - see footnote 2 of \citet{mam12}.      

In summary, for each star forming region we have selected stars with an X-ray luminosity, $L_{\rm X}$ (or an X-ray flux from which $L_{\rm X}$ can be derived given a distance), spectral type between F0 and M5 (inclusive) from which we assign an effective temperature, and photometric magnitudes and colours from which we have calculated bolometric luminosities. Further details for each star forming region are given in the following subsections.  We have removed stars that are known (or suspected) binaries where the two stars are not sufficiently resolved in the X-ray and/or optical catalogs to derive $L_{\rm X}$ and/or $L_\ast$ separately for each.   


\subsection{The ONC}
The Orion Nebula Cluster (ONC) is one of the most extensively studied star forming regions \citep{mue08,ode08}.  
 
\subsubsection{X-ray luminosities}\label{coup}
We take X-ray luminosities from the Massive Young Star-Forming Complex Study in Infrared and X-ray (MYStIX) project \citep{fei13}, specifically the MYStIX Probable Complex Members (MPCM) catalog of \citet{bro13}.  For the ONC, the MYStIX X-ray luminosities are those from the {\it Chandra} Orion Ultradeep Project (COUP; \citealt{get05}), a deep X-ray survey consisting of 13.2 consecutive days of observing, but adjusted to a distance of $414\,{\rm pc}$ \citep{men07}.  (A distance of $450\,{\rm pc}$ was assumed for the COUP without attribution.)  With an exposure time of $832\,{\rm ks}$ this is one of the deepest X-ray surveys of any star forming region ever conducted. 
 
\subsubsection{Spectral types \& photometry}\label{onc_spectral}
We take spectral types from (table 2 of) \citet{hil13}. This is a compendium of literature values, including many re-assignments of spectral types to stars in the ONC catalog of \citet{hil97}, as well as many which are reported for the first time.\footnote{The majority of spectral types of the ONC stars which we consider in this paper (F0-M5, with $L_{\rm X}$ in the MYStIX catalog (see section \ref{coup}) and $V$ and $I_{\rm c}$ photometry) were originally determined by \citet{hil97} and \citet{hil13}.  However, for the stars considered here, the compendium \citet{hil13} study includes spectral type estimates from \citet{gre46}, \citet{par54}, \citet{joh65}, \citet{pen73}, \citet*{pen75}, \citet{coh79}, \citet{wal83}, \citet{dun93}, \citet{edw93}, \citet{luh00}, \citet{luc01}, \citet*{sle04}, \citet*{wol04}, \citet*{dae12}, as well as some attributed by \citet{hil13} to ``H. C. Stempels 2008, private communication'', ``A.E. Samuel 1993 unpublished PhD thesis'', ``K. Stassun 2005, private communication'', ``Prosser \& Stauffer 1995, private communication'', and to ``C. Hamilton 1994 unpublished masters thesis''.}  

For stars where a range of spectral types are given in the \citet{hil13} table we adopt a spectral type that is the median of the earliest and latest reported, with exceptions as follows. i). For stars listed by \citet{hil13} with a spectral type reference of ``Ste=H.~C.~Stempels 2008, private communication" we take those spectral types over other literature values as they have been assigned based on high dispersion spectra.  ii). If a source has different spectral types in \citet{hil97} and in the enhanced ONC census of \cite{hil13}, we take the \citet{hil13} value. iii). When adopting a median spectral type for sources that have been spectral typed in multiple literature sources we neglect outliers.  For example, the star with SIMBAD identifier [H97b] 193 has been spectral typed in 4 studies, \citet{par54}, \citet{joh65}, \citet{coh79}, and \citet{hil97} - as listed by \citet{hil13}.  In three, the star is between K0 and K2, while in one it is assigned K6.  In this case we ignore the K6 outlier and adopt K1.  iv).  Spectral types taken from \citet{luh00} often span a large range of values.  For such stars, if other spectral types are available that fall within the range of those of \citet{luh00} we consider those values only.  Typically, this resulted in the adoption of a spectral type that was close to the median of the \citet{luh00} range. If the other listed spectral types extend to later or earlier than those of \citet{luh00} then we consider the full range of values.  v). Spectral types determined solely from IR spectra often differ from those found from optical spectra, being later, typically (e.g. \citealt{sle04}).  As discussed by \citet{pec13}, the well studied classical T Tauri star TW~Hya, as one example, has been classified as M2.5 from nIR spectra \citep{vac11}, but as K6-K8 from various optical studies (e.g. \citealt{her78}; \citealt*{hof98}; \citealt{tor06}; \citealt{pec13}).  We therefore neglect spectral types determined from IR spectra by \citet{sle04}, as listed in the \citet{hil13} table, unless there are no estimates from optical spectra.  Likewise, we neglect spectral type determinations from the IR study of \citet{wei09} [the latest spectral types from which ``may be too late'', as noted by \citet{hil13}]. vi). We neglect sources where extremely discrepant spectral types have been reported in the literature (e.g. [H97b] 503 which is reported by \citet{hil13} as F2-F7 [\citealt{hil97}] and K0-K7 [\citealt{luh00}], and [H97b] 809 which is assigned F5-K5 by \citet{hil13}, etc).

We then removed stars that are listed as (or are suspected to be) spectroscopic or close binaries, as there may be some confusion in the photometry and/or X-ray luminosities for the individual stars.  These were identified by \citet{mar88}, \citet{mat94}, \citet*{rho01}, \citet{pra02}, \citet{fur08}, \citet*{car08}, \citet{dae12}, \citet{mor12}, \citet{cor13}, and \citet{tob13} (which includes corrected tables from \citealt{tob09}), with some being attributed to ``H C. Stempels 2008, private communication'' by \citet{hil13}.  

We then assigned an effective temperature to each star based on its spectral type using the \citet{pec13} PMS conversion scale, linearly interpolating the $T_{\rm eff}$ value between spectral subtypes where necessary.

\begin{table}
  \caption{Adopted distances (see Appendix \ref{appendix_distances}).}
  \begin{tabular}{ccc}
  \hline
   Region & Distance (pc) & Reference  \\
  \hline
   ONC            &  414 & \citet{men07} \\
   NGC~2264  &  756 & \citet{gil14} \\
   IC~348        &  316  & \citet{her98} \\
   NGC~2362  &  1480 & \citet{moi01} \\
   NGC~6530  &  1250 &  \citet{pri05} \\
  \hline
\end{tabular}
\label{table_distances}
\end{table}

For the ONC stars with spectral types and X-ray luminosities we obtained $(V-I_{\rm c})$ colours and $I_{\rm c}$ magnitudes from \citet{hil97} [who obtained the majority of the photometry directly, although for a minority of the stars we consider here some magnitudes/colours were taken from \citet{her86} and \citet{pro94}].  We then calculated bolometric luminosities in a similar manner to \citet{hil97}, described below, except we have used the recently published PMS intrinsic colours and bolometric corrections of \citet{pec13}. The $I_{\rm c}$ magnitudes were de-reddened by assuming a common extinction law for the cluster,
\begin{equation}
A_{I_{\rm c}} = 0.61A_V=1.56\left[(V-I_c)-(V-I_c)_0\right],
\label{AIc}
\end{equation}
as derived by \citet{hil97} by converting that of \citet{rie85} from the Johnson $I$ band to the Cousins $I_{\rm c}$ band (e.g. \citealt{bes79}). Bolometric luminosities then follow from,
\begin{equation}
\log\left(\frac{L_\ast}{{\rm L}_\odot}\right)=\frac{2}{5}\left[M_{{\rm bol},\odot}-(I_{\rm c}-A_{I_{\rm c}})+\mu - {\rm BC}_{I_{\rm c}}\right],
\label{Lbol}
\end{equation}
where the bolometric correction at $I_{\rm c}$ is calculated from that at $V$ and the intrinsic $(V-I_{\rm c})_0$ colour, ${\rm BC}_{I_{\rm c}} = (V-I_{\rm c})_0 + {\rm BC_V}$, as tabulated by \citet{pec13}.  The solar bolometric magnitude is given above, and the distance modulus, $\mu$, for the cluster is calculated from our assumed cluster distance (see Table \ref{table_distances} and Appendix \ref{appendix_distances}).
	
For some stars in the sample $(V-I_{\rm c})-(V-I_{\rm c})_0<0$ which results in a (meaningless) negative extinction. This is likely as a result of inaccuracies in the adopted intrinsic colour appropriate for the stellar spectral type, an incorrectly assigned spectral type (resulting in an incorrect assignment of the intrinsic colour for that star), or photometric variability of the star.  We follow \citet{hil97} and set $A_{I_{\rm c}}=A_{\rm V}=0$ for such stars. Their calculated luminosities are therefore (likely) lower limits.  72 of the 496 stars in our final ONC sample (see section \ref{siessstuff}), $\sim$15\%, had a negative extinction.  This is an improvement over the original \citet{hil97} study where $\sim$27\% of their sample of stars with extinction estimates had $A_{\rm V}$ set to zero, indicating that the updated census of the ONC presented by \citet{hil13} has improved the spectral typing of stars in the region. The stars with negative extinctions in our sample have a mean spectral type of M3. At such late spectral types, the variation in the intrinsic colour with effective temperature (spectral type) is at its steepest (e.g. \citealt{pec13}), such that a small shift in the spectral type assignment results in a much bigger difference in $(V-I_{\rm c})_0$, and therefore in $A_{\rm V}$, than it would do at earlier spectral types.  Of the 72 stars with negative extinctions, the majority (52) would have a positive extinction with a change in spectral type of one subtype, which is the typical error.  The remaining 20 have a small range in reported spectral types, or a spectral type estimate from a single literature source only, indicating that the resulting negative extinction is not caused by our adoption of a median spectral type.  It is likely that their reported spectral types are inaccurate.


\subsection{NGC 2264}
The NGC~2264 star forming region, part of the Monoceros~OB1 association, has been extensively observed at all wavelengths \citep{dah08}.  


\subsubsection{X-ray luminosities}
\label{ngc2264LX}
NGC~2264 has been observed several times with both {\it Chandra} and {\it XMM-Newton}. Two separate, slightly overlapping, images were obtained with {\it Chandra} (Obs IDs: 2540 \& 2550, PIs: Sciortino \& Stauffer, $\sim97$ and $\sim48\,{\rm ks}$) with \citet{ram04}\footnote{The same {\it Chandra} observation was also analysed by \citet*{sun04} who identified additional X-ray sources in the field-of-view.} and \citet*{fla06} calculating PMS stellar X-ray luminosities. Additional X-ray luminosities have been derived by \citet{dah05} and \citet{dah07} from {\it XMM-Newton} observations (two pointings, both $\lesssim42\,{\rm ksec}$, Obs IDs: 0011420101 \& 0011420201, PI: Simon).  

\begin{figure*}
   \centering
     \includegraphics[width=0.33\textwidth]{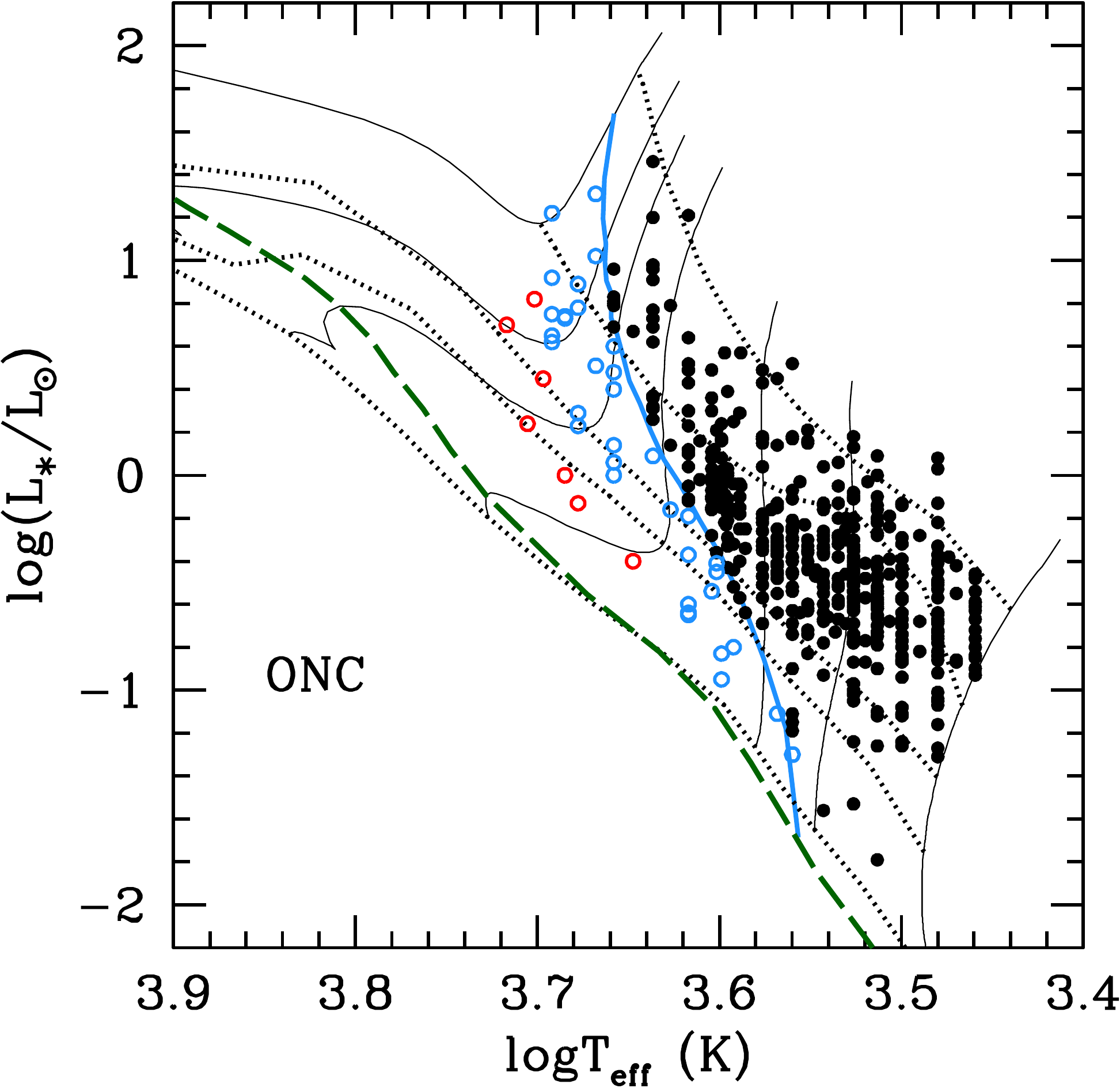}  
      \includegraphics[width=0.33\textwidth]{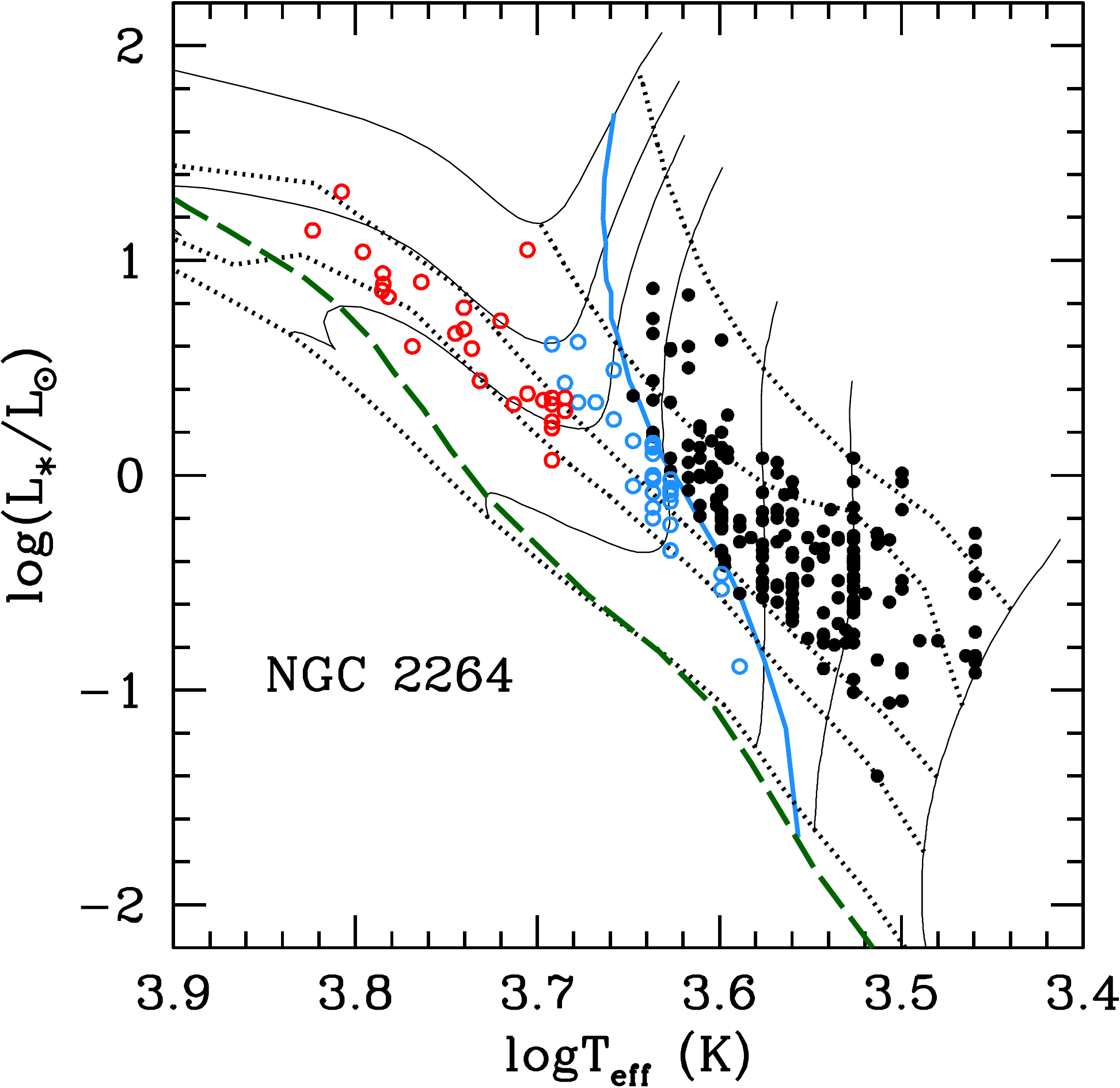} 
       \includegraphics[width=0.33\textwidth]{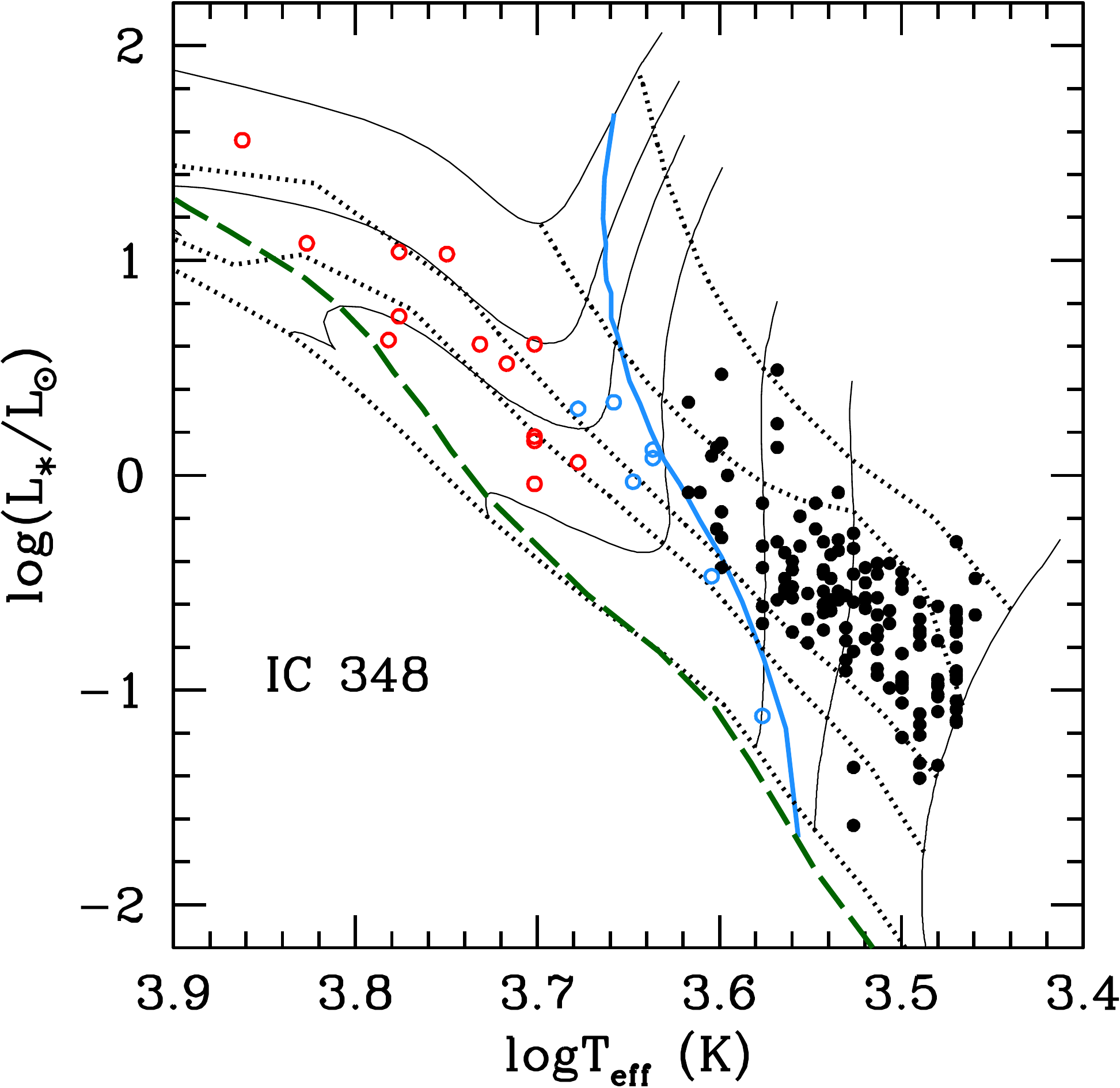}
     \includegraphics[width=0.33\textwidth]{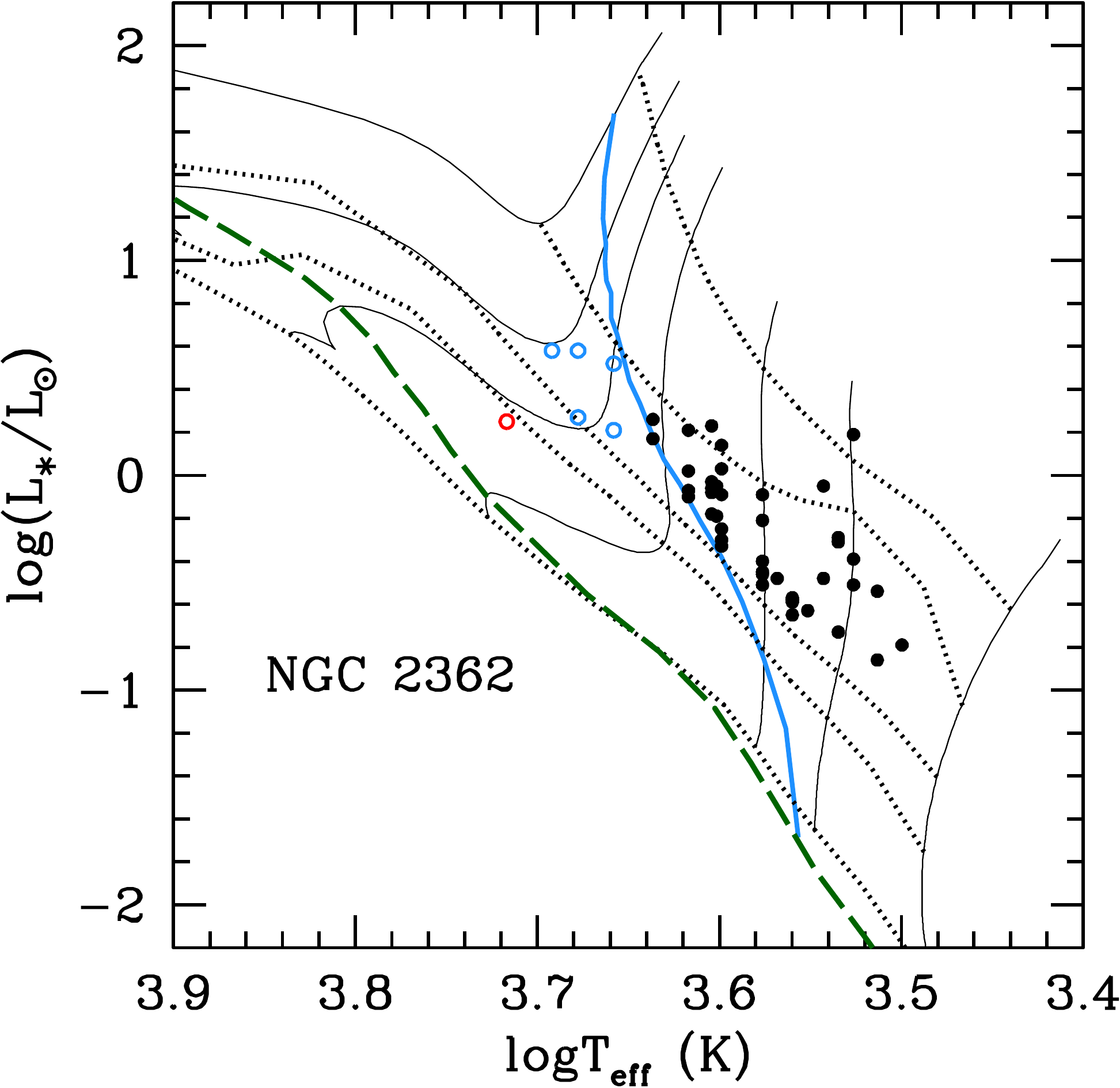}  
      \includegraphics[width=0.33\textwidth]{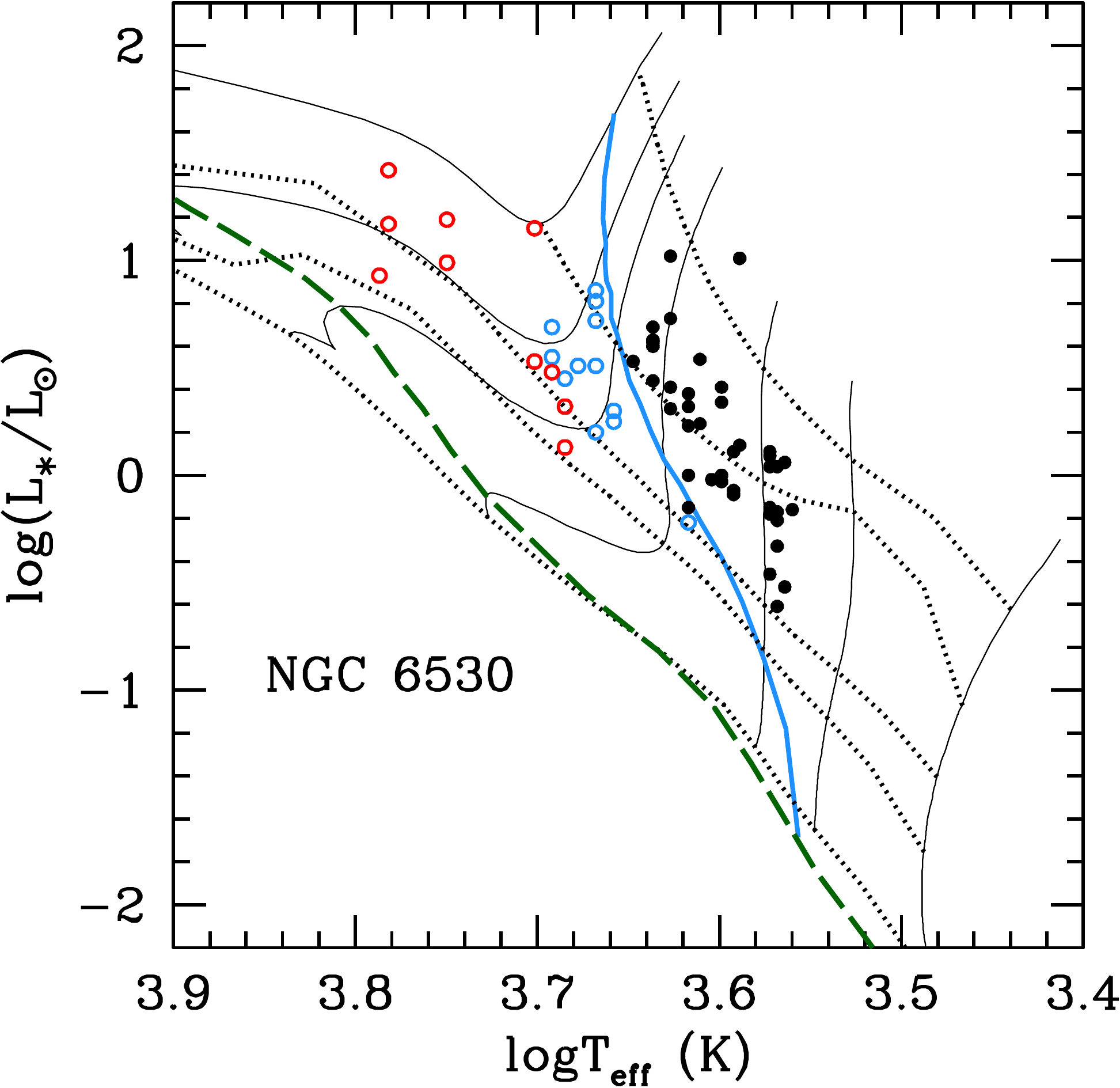} 
   \caption{H-R diagrams for five star forming regions that we consider. Mass tracks (solid black lines; 0.1, 0.3, 0.5, 1, 1.5, 2, 3$\,{\rm M}_\odot$) and isochrones (dotted lines; 0.1, 1, 5, 10 \& 60$\,{\rm Myr}$) are from the models of \citet{sie00}. The dashed green line is the ZAMS.  The solid blue line is the fully convective limit: stars located above and to the right of this line are fully convective (black solid symbols) and those below and to the left are partially convective with a radiative core and outer convective envelope (red and blue open symbols).  Partially convective stars represented by red symbols have evolved onto Henyey tracks (the portion of their mass track where $L_\ast$ is increasing with age) while those represented with blue symbols are still on Hayashi tracks (the portion of their mass track where $L_\ast$ is decreasing with age). 
}
   \label{hrd}
\end{figure*}

Comparison of the papers where the NGC~2264 X-ray data are presented reveals that a variety of differing assumptions and methods have been used by the different authors to calculate $L_{\rm X}$. For example, \citet{ram04} assume a uniform value for the absorbing hydrogen column density $N_{\rm H}$ towards each star, based on ``the most likely value of the observed extinction towards NGC~2264'' and assuming a fixed relationship between $N_{\rm H}$ and $A_{\rm V}$ (e.g. \citealt{vuo03}).  $N_{\rm H}$ estimates from such relationships are highly uncertain, with a large spread apparent from $A_{\rm V}-N_{\rm H}$ diagrams (e.g. \citealt{fei05}). \citet{fla06} note that ``$N_{\rm H}$ is critical for determining $L_{\rm X}$'' and proceed to calculate $L_{\rm X}$ from X-ray spectral fits treating $N_{\rm H}$ as a free parameter in some cases and calculating it from $A_{\rm V}$ in others.  Given the different assumptions, we must be careful when combining the X-ray catalogs.  However, this task has recently been simplified thanks to a re-analysis of the {\it Chandra} images during the MYStIX project \citep{fei13}.  We take $L_{\rm X}$ values for NGC~2264 stars as listed in the MPCM (MYStIX Probable Complex Member) young stars catalog of \citet{bro13}, adjusting them to our adopted distance of $756\,{\rm pc}$ [the \citet{bax09} distance of $913\,{\rm pc}$ was adopted for the MYStIX project \citep{fei13}].  The X-ray luminosities were calculated using the XPHOT method of \citet{get10}. This allows the analysis of faint X-ray sources that cannot be handled by traditional X-ray spectral fitting methods.  Adoption of the MYStIX NGC~2264 X-ray luminosities thus ensures that we are using $L_{\rm X}$ values that have been calculated in a uniform manner for stars throughout the cluster.  We have therefore neglected the {\it XMM-Newton} data of \citet{dah07}. Due to the wider field of view of their observations, the \citet{dah07} catalog contains 43 additional X-ray sources with spectral types and photometry that would have made our final sample, see section \ref{siessstuff}, but which we do not consider here.


\subsubsection{Spectral types \& photometry}
We collated spectral types for NGC~2264 members from the catalogs of \citet{reb02}, \citet{lam04}, \citet{kum04}, \citet{dah05}, \citet*{sun97}, and \citet{mar13}, adopting a spectral type that is the median of the observed range.  As the \citet{mar13} spectral types have been assigned from IR spectra, we neglect their spectral types when they are found to be later than those determined from optical spectra (for the same reason as with the ONC data, see section \ref{onc_spectral}).  The \citet{lam04} catalog is a compendium of values from \citet{reb02} (but an updated version of their catalog with several new spectral types being reported) and spectral types attributed to ``Young et al. (private communication)''.  Some of the \citet{dah05} spectral types were assigned from the author's own observations, while others were taken from the catalogs of \citet{wal56}, \citet{her88}, \citet{reb02}, and \citet{lam04}. 

We again assigned effective temperatures from the conversions of \citet{pec13}, neglecting sources earlier than F0 and later than M5.  We then removed sources which have been identified as, or are suspected, close binary stars, in case of confusion in the assigned photometry and/or X-ray luminosities.  These were SIMBAD ID: Cl$\ast$ NGC 2264 RMS 3323 \citep{pad94}; Cl$\ast$ NGC 2264 RMS 3470 and Cl$\ast$ NGC 2264 RMS 3578 \citep{mat94,kar13}; Cl$\ast$ NGC 2264 LBM 6083 (e.g. \citealt{win14}); Cl$\ast$ NGC 2264 RMS 2465, 3157, 3307, 3370, and 3390, which are identified as ``adaptive optics'' binaries by \citet{dah07}; and stars listed as single and double line spectroscopic binaries by \citet{fur06}.  We have also removed Cl$\ast$ NGC 2264 RMS 3582 which is listed as an eclipsing binary in SIMBAD.  Further investigation reveals that this star is listed in the  International Variable Star Index (VSX) database (as NSV 3128) having been identified as a {\it possible} eclipsing variable by \citet{koc74} (the star is listed as Penna 414 in their table).  The recent study of \citet{kla13}, where several eclipsing and other binaries in NGC~2264 were identified (see below), classified Cl$\ast$ NGC 2264 RMS 3582 as a non-periodic variable (the system is listed as SRa01a\_26062 in their work). Its binary nature is therefore questionable, but we have erred on the side of caution and have removed it from our sample.  Further observations of this object are required.  

Additional close binary systems, and many more exotic variables, have been identified by \citet{kla13}.  From ground based time series photometry they classified periodically variable stars according to the categories of the General Catalogue of Variable stars \citep{sam09}.  They classified many as YSOs (young stellar objects), as one would expect for a star forming region.  Neglecting stars in our final sample that were either not classified by \citet{kla13} or which were classified as YSO or MISC (miscellaneous or non-periodic variables) 43 stars with different types of variable classification remain.  For example, three of the stars in our final sample are classified by \citet{kla13} as GDOR, $\gamma$ Doradus stars.  These are late A to late F-type variable stars that undergo non-radial pulsations (e.g. \citealt*{bal94}; \citealt{kay99}).  However, multiple, independent, spectroscopic studies have classified these stars (Cl$\ast$~NGC~2264~RMS~3440, 3519, \& 3840) as early to mid-M type (e.g. \citealt{reb02,lam04}), and there is little to suggest that they are anything but normal, low mass, PMS stars.  

Thirty three other stars in our final sample have been classified by \citet{kla13} as EA (2/33) or EB (2/33), types of eclipsing binary, or as ELL (27/33 rotating ellipsoidal variables - a type of non-eclipsing close binary system) or ELL/SP (2/33), stars with light curves that could be described as ELL systems, or with variability caused by star spots.  However, a preliminary examination of the photometry and spectra collected as part of the CSI~2264 project (Coordinated Synoptic Investigation of NGC 2264; \citealt{cod14}) reveals no evidence of binarity in $\sim$85\% of these stars, and only tentative evidence in the other cases (A. Cody, private communication). \citet{kla13} mention that their classifications are expected to be $\sim$70\% accurate.  In a region such as NGC~2264, where coordinated multi-wavelength photometric and spectroscopic surveys have been conducted, and where light curves of young stars can show complex variability patterns due to star spots, accretion, and obscuration of stars by warped inner disk material, their classifications based solely on the morphology of light curves may be even less accurate. We have therefore neglected the classifications of \citet{kla13} as they clearly require re-determination.
    
For stars with spectral types, we obtained Cousins $I_{\rm c}$ and $(V-I_{\rm c})$ photometry from \citet{reb02}, \citet{lam04}, \citet{dah05}, and \citet{dah07}, in that order.  We then derived extinctions $A_{I_{\rm c}}$ from $A_{\rm V}$ with the latter being derived with a common extinction law for the cluster, equation (\ref{AIc}), using the intrinsic $(V-I_{\rm c})_0$ colours of \citet{pec13} appropriate for each spectral type.  Bolometric luminosities then followed from equation (\ref{Lbol}).\footnote{During their X-ray study of a portion of NGC~2264 \citet{fla06} derived bolometric luminosities assuming a uniform extinction law for the cluster, but used observed and intrinsic $(R_{\rm c}-I_{\rm c})$ colours and bolometric corrections from the dwarf star effective temperature-spectral type conversion scale of \citet{ken95}, with the intermediate gravity scale of \citet{luh99} for M stars.  As \citet{pec13} do not provide intrinsic colours for PMS stars involving the Cousins $R_{\rm c}$ band we cannot use the observed $(R_{\rm c}-I_{\rm c})$ photometry to derive luminosities, and have used $(V-I_{\rm c})$ instead (as did \citealt{reb06}).}    


\subsection{IC 348}
The open cluster IC~348, a region of ongoing star formation, is part of the Perseus dark cloud complex and thought to be associated with both NGC~1333 and the Perseus~OB2 association \citep{her08}. 


\subsubsection{X-ray luminosities}
 We have taken X-ray fluxes of PMS stars from the IC~348 catalog of \citet{ste12}. It was created by merging four archival {\it Chandra} observations, namely Obs IDs 606 (PI: Preibisch, $\sim$52$\,{\rm ksec}$), 8584 (PI: Calvet, $\sim$50$\,{\rm ksec}$), and 8933 \& 8944 (PI: Wolk, $\sim$40$\,{\rm ksec}$ for each).  The data were originally published by \citealt{pre01} (see also \citealt{pre02}) for 606, and by \citet*{for11} for 8933 \& 8944. To the best that we can determine, the data obtained for Obs ID 8584 were not published prior to the \citet{ste12} work. We converted the X-ray fluxes to X-ray luminosities with our assumed distance to IC~348 of $316\,{\rm pc}$ (\citealt{ste12} assumed a slightly lower value of $310\,{\rm pc}$ in their work). 


\subsubsection{Spectral types \& photometry}
We obtained spectral types of IC~348 PMS stars from \citet{mue07}, most of which were as cataloged by \citet{luh03a} who collated literature estimates from \citet*{har54}, \citet{luh98}, \citet*{str74}, \citet{her98}, and \citet{luh99}, as well as assigning their own for several stars.  Their catalog also includes spectral types from \citet{luh05}.  Once again, we removed O, B, and A-type stars, and stars later than M5, and assigned effective temperatures based on the PMS conversion scale of \citet{pec13}.  

We again removed known or suspected spectroscopic binaries from our sample, as listed by \citet*{duc99}, \citet{nor06}, and \citet{dah08b}.   

\citet{luh03a} argues that the $J$-band provides the best estimates of bolometric luminosities for stars in IC~348 (except for a few binaries that are better resolved in the optical, see below).  For the majority of stars with spectral types we obtained $J$ and $H$-band magnitudes from the 2MASS All-Sky Catalog of Point Sources \citep{cut03,skr06}. 
To de-redden the $J$-band magnitudes we calculated the extinction $A_{\rm J}$ from
the extinction law of \citet{rie85},
\begin{equation}
A_{\rm J} = 2.64\left[ (J-H)-(J-H)_0\right],
\end{equation}
where the intrinsic colour $(J-H)_0$ appropriate for the spectral type is taken from \citet{pec13}. We then calculated luminosities using an equation analogous to equation (\ref{Lbol}), applying spectral type dependent bolometric corrections at the 2MASS $J$-band as listed in \citet{pec13}.  

\citet{luh03a} identified four binary systems (8 stars) which are better resolved at optical wavelengths than in the infrared.  We follow \citet{luh03a} and derive luminosities using the $I_{\rm c}$-band magnitudes rather than $J$ for these stars.  We obtained $V$ magnitudes and $(V-I_{\rm c})$ colours from \citet{her98}, which was available for four of the eight stars.  We de-reddened the $I_{\rm c}$ magnitudes using equation (\ref{AIc}). The bolometric luminosity then follows from equation (\ref{Lbol}) with ${\rm BC}_{I_{\rm c}}$ calculated from the tabulated \citet{pec13} ${\rm BC_V}$ and intrinsic $(V-I_{\rm c})_0$ colours appropriate for the spectral type.


\subsection{NGC 2362}
The open cluster NGC~2362, occasionally referred to as the $\tau$ Canis Majoris Cluster in reference to its most massive member the O9 star $\tau$~CMa, is the oldest star forming region in our sample (e.g. \citealt{dah08c}).


\subsubsection{X-ray luminosities}
We have taken X-ray luminosities from the MYStIX project MPCM young stars catalog of \citet{bro13}. The  MYStIX project was a reanalysis of previously published {\it Chandra} data (Obs ID: 4469, PI: Murray), NGC~2362 having been observed for $\sim100\,{\rm ks}$ in December 2003 \citep{del06,dam06,dah07b}.    


\subsubsection{Spectral types \& photometry}
There are few detailed spectroscopic studies of NGC~2362, with spectral types only available for a handful of cluster members. We take spectral types from \citet{dah07b}, which includes minor updates and adds a  few new spectral types to the catalog of \citet{dah05b}.  As for the other clusters, we neglect stars earlier than F0. $V$-band magnitudes and $(V-I_{\rm c})$ colours are also taken from \citet{dah07b}. We then calculated stellar luminosities by dereddening the observed $I_{\rm c}$-band magnitudes, with the application of appropriate bolometric corrections for each spectral type [see equations (\ref{AIc}) and (\ref{Lbol})].


\subsection{NGC 6530}
NGC~6530 lies towards the Galactic centre, in the Sagittarius--Carina arm, and is associated with the Lagoon Nebula (M8; \citealt{tot08}). 


\subsubsection{X-ray luminosities}
We again take X-ray luminosities from the MPCM MYStIX catalog \citep{bro13}, with a small adjustment to account for our adoption of a distance to NGC~6530 of $1250\,{\rm pc}$ [see Appendix \ref{appendix_distances} and Table \ref{table_distances}; $1300\,{\rm pc}$ was assumed for the MYStIX project \citep{fei13}].  The MYStIX project NGC~6530 data includes a re-analysis of the previously published {\it Chandra} data of the cluster (\citealt{dam04}; Obs ID: 977, PI: Murray, $\sim60\,{\rm ks}$).\footnote{A shorter {\it XMM-Newton} exposure of NGC~6530 was previously published by \citet{rau02}.}  


\subsubsection{Spectral types \& photometry}
Spectral types of cluster members were collated from \citet{pri12}, \citet*{ari07}, \citet{van97}, and \citet{wal61}.\footnote{Spectral types were also considered from \citet{wal57} and \citet{kum04} but none of these stars made our final sample.} We removed sources of spectral type earlier than F0, as well as three stars of luminosity class II.  We again adopted median spectral types of the literature values (see the discussion for the ONC).

We then removed known or suspected spectroscopic binaries from our sample, as listed by \citet{hen12} and \citet{pri07}, in case of confusion in the X-ray luminosities, spectral typing, or photometry.  Non-members, as determined by \citet{pri07,pri12} were also removed.     

We obtained $V$ and $(V-I_{\rm c})$ photometry from (in order of preference) \citet{hen12}, \citet{pri05}, and \citet*{sun00}, and calculated stellar luminosities using equations (\ref{AIc}) and (\ref{Lbol}) with intrinsic colours and bolometric corrections from \citet{pec13}.


\section{Stellar parameters: mass, age \& internal structure}\label{siessstuff}
With effective temperatures and bolometric luminosities of the stars in the five star forming regions that we consider to hand, we calculated the stellar masses and ages using the models of \citet{sie00} - $Z=0.02$ with convective overshooting. In such a way, our stellar mass, $M_\ast$, and age, $t$, estimates are consistent across the entire sample.  The \citet{sie00} mass tracks have somewhat limited stellar mass resolution, so we constructed a series of interpolated mass tracks and produced a code that automatically calculates $M_\ast$ and $t$ given $T_{\rm eff}$ and $L_\ast$ as an input. Our code also estimates whether a star is fully convective or partially convective with a radiative core, and if the latter, the radiative core mass and radius relative to the total stellar mass and radius.  During this process we have not considered how errors in the assigned spectral type, or in the calculated luminosities, translate into errors in $M_\ast$ and $t$.  However, recently \citet*{dav14} did so, and estimated such errors for stars in the ONC using a stellar sample assembled in a similar manner to our own.  Using their study as a guide there is a typical error of 0.1, 0.2, 0.4, and 0.5$\,{\rm M}_\odot$ for stars of mass 0.1-0.6, 0.6-1, 1-2, and 2-3$\,{\rm M}_\odot$ respectively.  For ages $\le2.5\,{\rm Myr}$, errors of $\sim1.1\,{\rm Myr}$ are typical, rising to $\sim$3, $\sim$6.4, and $\sim$11\,{\rm Myr} for stars of age of 2.5-5, 5-10, and 10-15$\,{\rm Myr}$.  The age error increases for older stars because PMS contraction proceeds on the Kelvin-Helmholtz timescale, which scales inversely with stellar radius.  Stellar contraction thus slows with age (see Figure \ref{mcorercore}), decreasing the separation between isochrones in the $\log T_{\rm eff}-\log L_\ast$ plane as stars evolve along their mass tracks.

We adopt the \citet{sie00} models in preference to others as they cover a sufficient range of stellar mass and, crucially for the purposes of this paper, provide stellar internal structure information.  Other, more modern, models often have a limited range in stellar mass.  For example, the publicly available mass tracks and isochrones of \citet{bar15}, which are superb for low-mass and fully convective PMS stars, are limited to $\le$1.4$\,{\rm M}_\odot$.  The \citet{sie00} models are also one of the most commonly used by the PMS star community, and the authors have provided a convenient online interpolator for generating isochrones of desired age.  Furthermore, the \citet{sie00} models were used by several large X-ray surveys (e.g. \citealt{pre05b,tel07}), which allows for a more direct comparison with our results; differences then being attributable to the updated spectral typing and our use of PMS calibrated intrinsic colours and bolometric corrections.  In Appendix \ref{appendix_comparison} we demonstrate, using the models of \citet{jun07} and \citet*{tog11}, that the main results of our paper, described in the following sections, are unaffected by the choice of PMS evolutionary model. 

Before estimating $M_\ast$ and $t$ we removed 37 stars that were to the right of the $0.1\,{\rm M}_\odot$ mass track in the H-R diagram (12 from the ONC, 3 from NGC~2264, and 22 from IC~348).  This is the lowest available mass track in the \citet{sie00} models, and therefore we could not accurately assign $M_\ast$ and $t$ values for such stars.  In practise, this makes little difference to the work in this paper as we are primarily interested in partially convective PMS stars, and stars of mass $\lesssim$0.35$\,{\rm M}_\odot$ remain fully convective for the entirety of their PMS (and subsequent main sequence) evolution (e.g. \citealt{cha97}).  We also removed 3 stars which were more massive than $3\,{\rm M}_\odot$, the upper mass limit of our study (2 from the ONC and 1 from NGC~2264). Although these 3 stars are currently of spectral type G2, G5, and K2, they will evolve into Herbig AeBe stars and, eventually, B-type main sequence stars \citep{sie00}. We neglected 3 stars which appeared to be younger than $0.01\,{\rm Myr}$ (2 from the ONC and 1 from IC~348) and 10 stars 
that were located below the ZAMS (5 from the ONC, 2 from NGC~2264, and 3 from IC~348). It is likely that such stars have inaccurate luminosity estimates and are positioned incorrectly in the H-R diagram.  \citet{ste12} comment that stars which fall significantly below the ZAMS may be PMS stars with edge-on discs observed in scattered light.  
Stars as young as 0.01$\,{\rm Myr}$ may not yet have evolved to the birthline and should not be optically visible \citep{sta83}.  Indeed, estimates suggest that it takes a star about 0.1 to 0.3$\,{\rm Myr}$ to complete the embedded phase of star formation (e.g. \citealt{off11}; \citealt{dun12}).  In practise, stars at such a young age are fully convective, and therefore have no bearing on our study of partially convective PMS stars. 

\begin{table*}
  \caption{A comparison between the mean, standard deviation, median, and median absolute deviation (M.A.D.) of the logarithmic fractional X-ray luminosities, $\log(L_{\rm X}/L_\ast)$, of fully and partially convective PMS stars. The two numbers in the $N_{\rm stars}$ column are the number of fully and partially convective stars considered in each region.  The 34 stars from IC~348 with $L_{\rm X}$ upper limits are not considered here.}
  \begin{tabular}{cccccccccc}
  \hline
    &  & fully convective  & & & & radiative core &&&\\
   region & $N_{\rm stars}$ & mean & std. dev. & median & M.A.D. & mean & std. dev. & median & M.A.D.\\
  \hline
   all               & 803; 147 &  -3.44  & 0.59 & -3.36 & 0.45 & -3.57 & 0.68 & -3.48 & 0.48 \\
   ONC           & 423; 43  & -3.56  & 0.66 & -3.52  & 0.53 & -3.30  & 0.61  & -3.30  & 0.47 \\
   NGC~2264 & 190; 57  &  -3.20  &  0.39 & -3.18  & 0.30 &  -3.58  & 0.54  & -3.44  & 0.38 \\
   IC~348       & 106; 19   &  -3.55  &  0.53 & -3.55  & 0.40 &  -4.00  & 1.01  & -3.61  & 0.69 \\
   NGC~2362 & 43; 6      &  -3.09  &  0.38 & -3.19  & 0.26 &  -3.31  & 0.27  & -3.35  & 0.23 \\
   NGC~6530 & 41; 22    &  -3.37  &  0.40 & -3.34  & 0.32 &  -3.76  & 0.67  & -3.65  & 0.51 \\
  \hline
\end{tabular}
\label{tablestats}
\end{table*}

\begin{table*}
  \caption{As Table \ref{tablestats} but comparing stars which have evolved onto Henyey tracks to those still on their Hayashi tracks in the H-R diagram.  As a radiative core develops before a star moves onto its Henyey track, the Hayashi sample includes all of the fully convective stars within a given region as well as many of the partially convective stars. There is only a single star (in our sample) on a Henyey track in NGC~2362.}
  \begin{tabular}{cccccccccc}
  \hline
   & & Hayashi track  & & & & Henyey track &&\\
   region & $N_{\rm stars}$ & mean & std. dev. & median & M.A.D. & mean & std. dev. & median & M.A.D.\\
  \hline
   all                & 894; 56  &  -3.43  & 0.58 & -3.36 & 0.44 & -3.95 & 0.78 & -3.76 & 0.60 \\
   ONC            & 459; 7    & -3.53  & 0.66 & -3.48  & 0.52 & -3.89  & 0.27  & -3.98  & 0.18 \\
   NGC~2264  & 221; 26  & -3.22 & 0.38 & -3.22  & 0.30 & -3.81 & 0.66  & -3.68 & 0.51 \\
   IC~348         & 113; 12  & -3.54  & 0.54 & -3.54  & 0.40 & -4.30  & 1.10  & -3.75  & 0.85 \\
   NGC~2362   & 48; 1     & -3.12  & 0.38 & -3.22  & 0.26 & -2.94  & - & -2.94  & - \\
   NGC~6530   & 53; 10   & -3.40 & 0.41 & -3.43  & 0.33 & -4.06  & 0.79  & -4.25  & 0.63 \\
  \hline
\end{tabular}
\label{tablestats2}
\end{table*}

Our final sample consists of a total of 984 PMS stars including 466 from the ONC, 247 from NGC~2264, 159 from IC~348 (125 with X-ray luminosities and 34 with $L_{\rm X}$ upper limits), 49 from NGC~2362, and 63 from NGC~6530.  H-R diagrams for the final sample of stars are shown in Figure \ref{hrd}, where the solid blue line separates stars which are partially convective from those that are fully convective. 

It is clear from visual inspection of Figure \ref{hrd} that NGC~6530 and the ONC are the youngest clusters, with stars of median age 1.3 and 1.7$\,{\rm Myr}$ respectively, while NGC~2264, IC~348, and NGC~2362 are older with median ages of 2.2, 2.3, and 2.6$\,{\rm Myr}$ - a similar ordering compared to other studies (e.g. \citealt{may08}).  We also note that for clusters where the spectroscopic surveys are close to complete, such as the ONC, IC~348, and NGC~2264, the fraction of partially convective stars (those to the left of the blue line in Figure \ref{hrd}) is a proxy for the age of the star forming region. For older clusters a larger fraction of stars will have become partially convective compared to younger regions.  


\section{Comparing the X-ray properties of stars on convective and radiative tracks in the H-R diagram}\label{xraycompare}

In this section we compare the X-ray properties, $\log L_{\rm X}$ and $\log (L_{\rm X}/L_\ast)$, of partially and fully convective PMS stars, and of those on Hayashi and Henyey tracks in the H-R diagram. 

We consider four groups of PMS stars: i) those which have fully convective interiors; ii) those which have partially convective interiors consisting of a radiative core with an outer convective envelope; iii) those on Hayashi tracks in the H-R diagram; and iv) those on Henyey tracks in the H-R diagram.  It is clear from PMS stellar evolution models (e.g. \citealt{sie00}), from Figure \ref{hrd}, and the discussion in section \ref{intro} that radiative core development can occur several Myr {\it before} stars transition onto their Henyey tracks. When the radiative core first begins to grow, the luminosity of the star initially continues to decrease before the transition into the Henyey phase where the stellar luminosity rises and the star moves upwards and left in the $\log L_\ast-\log T_{\rm eff}$ plane. Our group iii) (Hayashi track stars) therefore contains a mixture of both fully convective and partially convective stars, whereas our group iv) (Henyey track stars) contains only stars which have evolved to the point where the physics of their internal structure is driving them towards greater luminosities.  Henyey track PMS stars have radiative cores of size $\gtrsim$0.5$\,R_\ast$ and mass $\gtrsim$0.7$\,M_\ast$.      

In terms of the H-R diagrams shown in Figure \ref{hrd}, the filled black points are the fully convective sample, and the open red/blue points the radiative core sample.  The black and blue points together form the Hayashi sample (all still have $L_\ast$ decreasing with age), and the red points form the Henyey sample (which all have $L_\ast$ increasing with age).

Within our sample of 984 PMS stars, the majority, 836, are fully convective (of which 402 are of mass $\le$0.35$\,{\rm M}_\odot$ and will remain fully convective) while 148 have developed radiative cores.  For the ONC, 423 and 43 stars are fully convective and partially convective respectively, with 190 and 57 for NGC~2664, 139 and 20 for IC~348 (106 and 19 with X-ray luminosities; 33 and 1 with $L_{\rm X}$ upper limits), 43 and 6 for NGC~2362, and 41 and 22 for NGC~6530.  

For our entire sample of 984 stars, 927 are on Hayashi tracks in the H-R diagram while 57 have evolved onto Henyey tracks.  For the ONC, 459 and 7 stars are on Hayashi and Henyey tracks  respectively, with 221 and 26 for NGC~2664, 146 and 13 for IC~348 (113 and 12 with X-ray luminosities; 33 and 1 with $L_{\rm X}$ upper limits), 48 and 1 for NGC~2362, and 53 and 10 for NGC~6530.

\begin{figure*}
   \centering
       \includegraphics[width=0.35\textwidth]{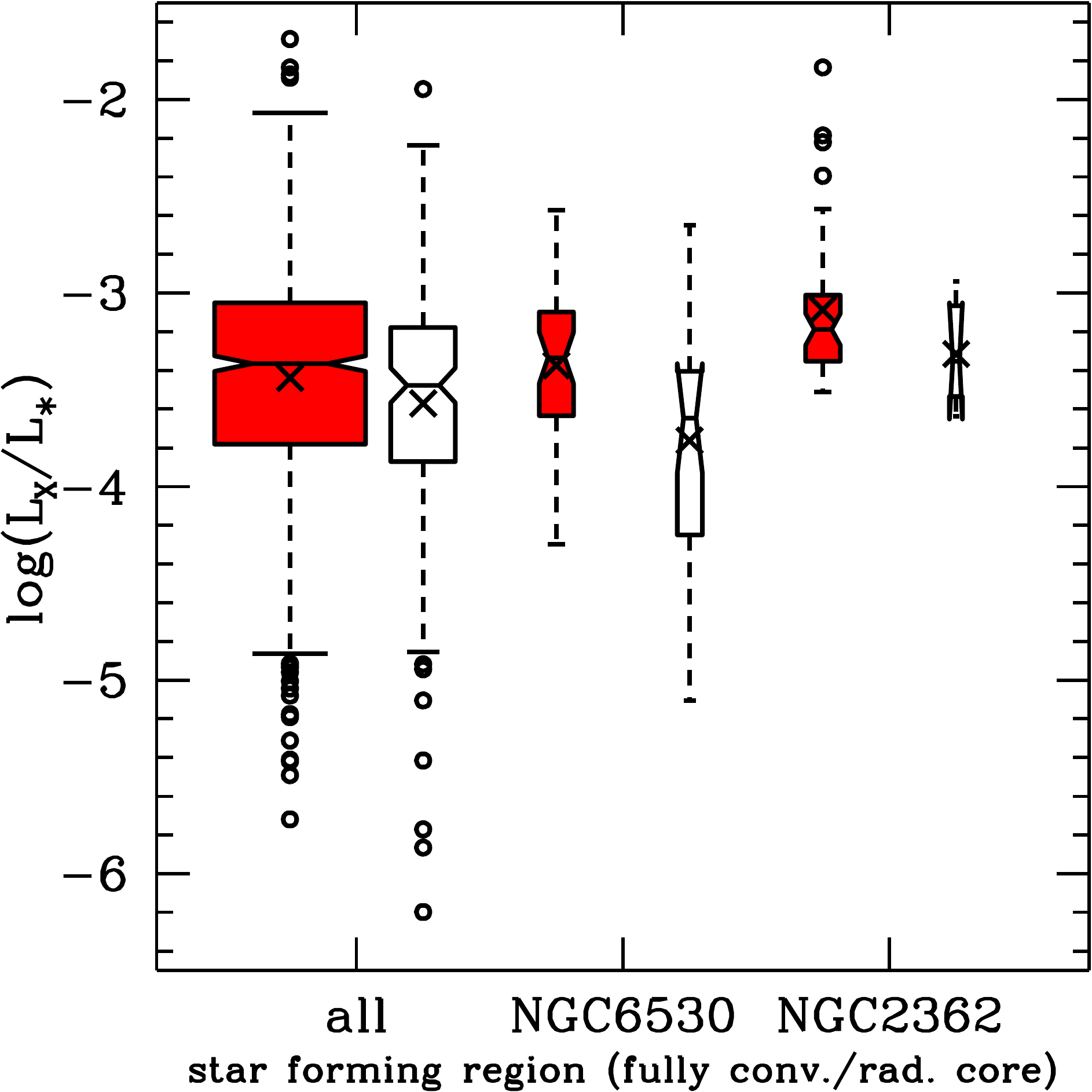}
       \includegraphics[width=0.35\textwidth]{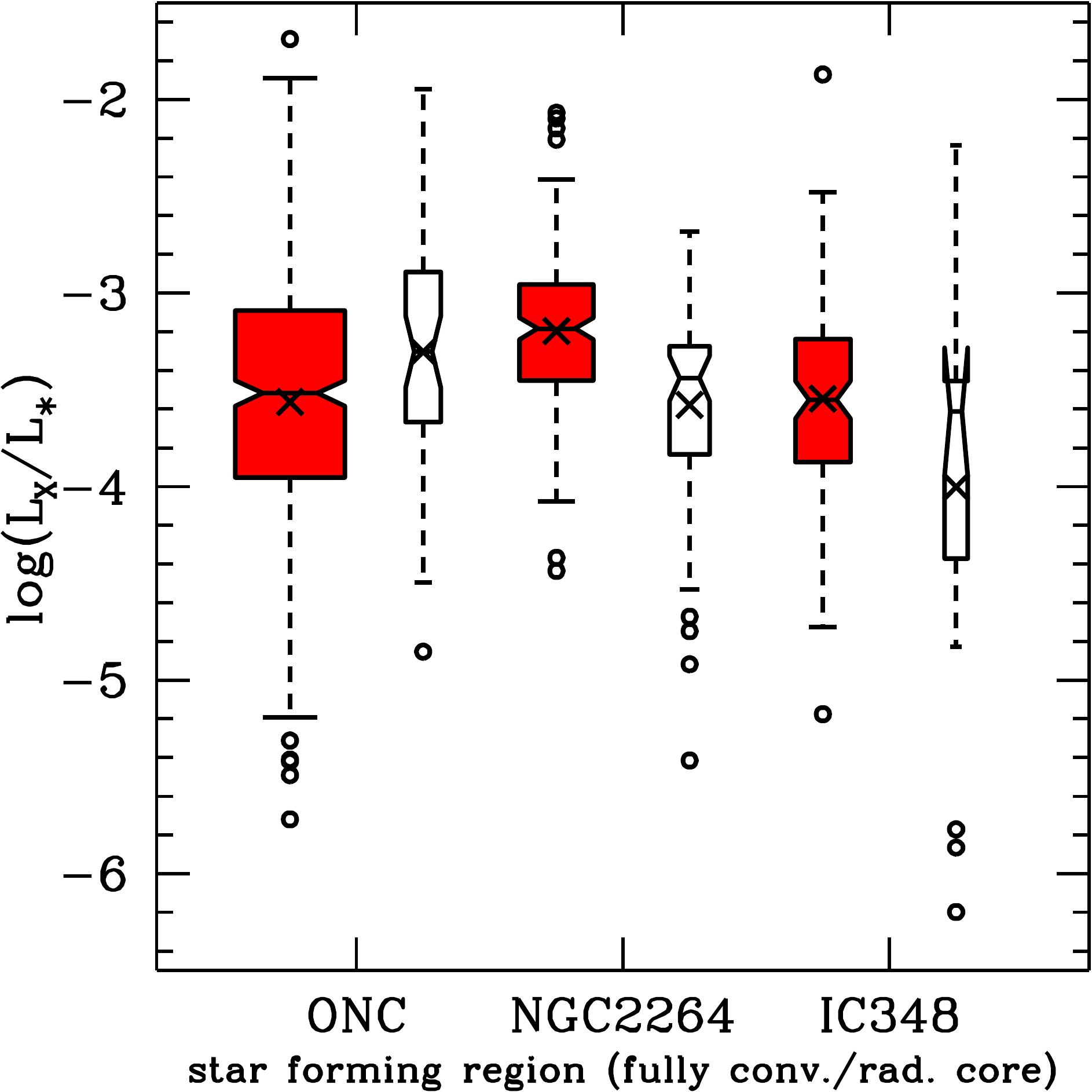} 
       \includegraphics[width=0.35\textwidth]{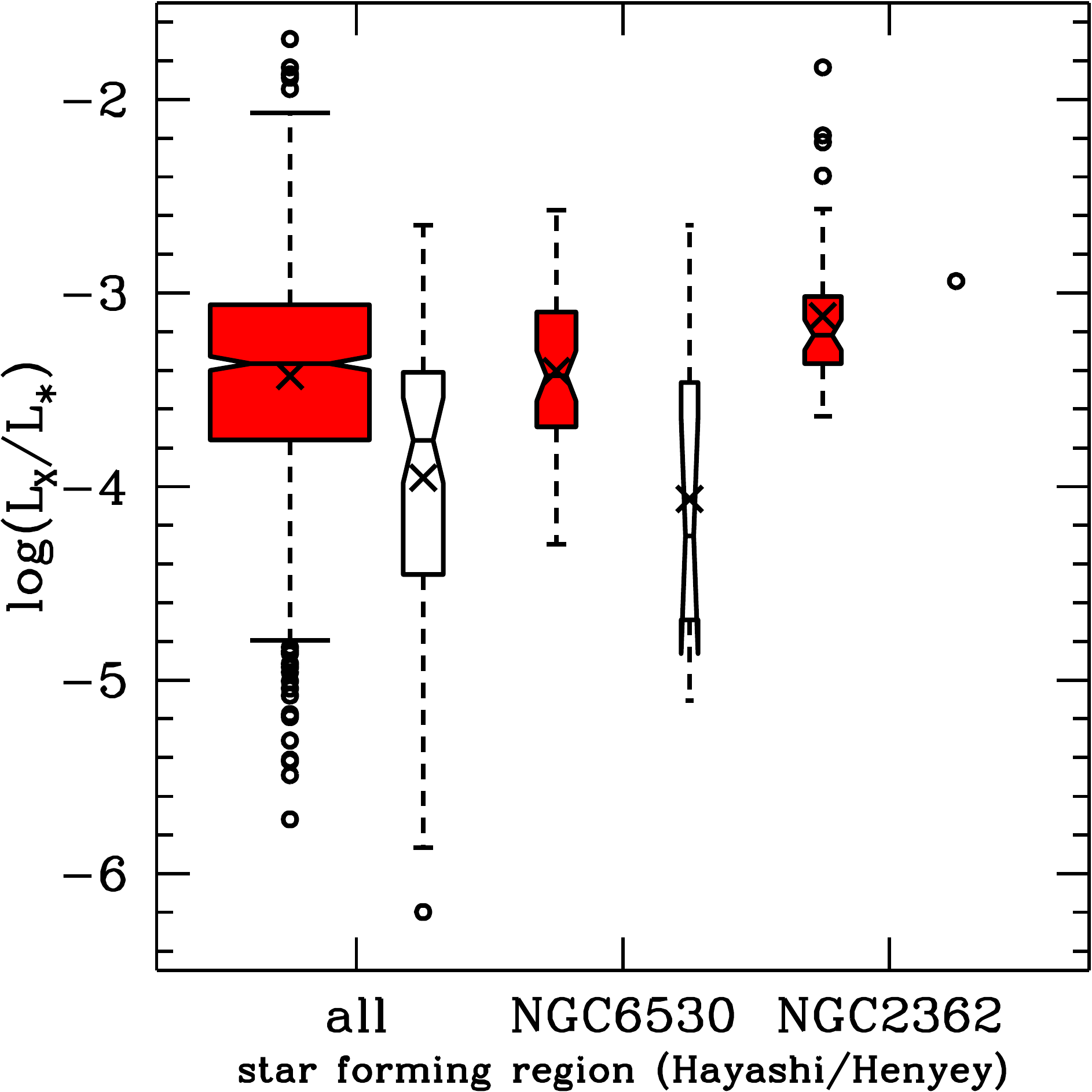}  
      \includegraphics[width=0.35\textwidth]{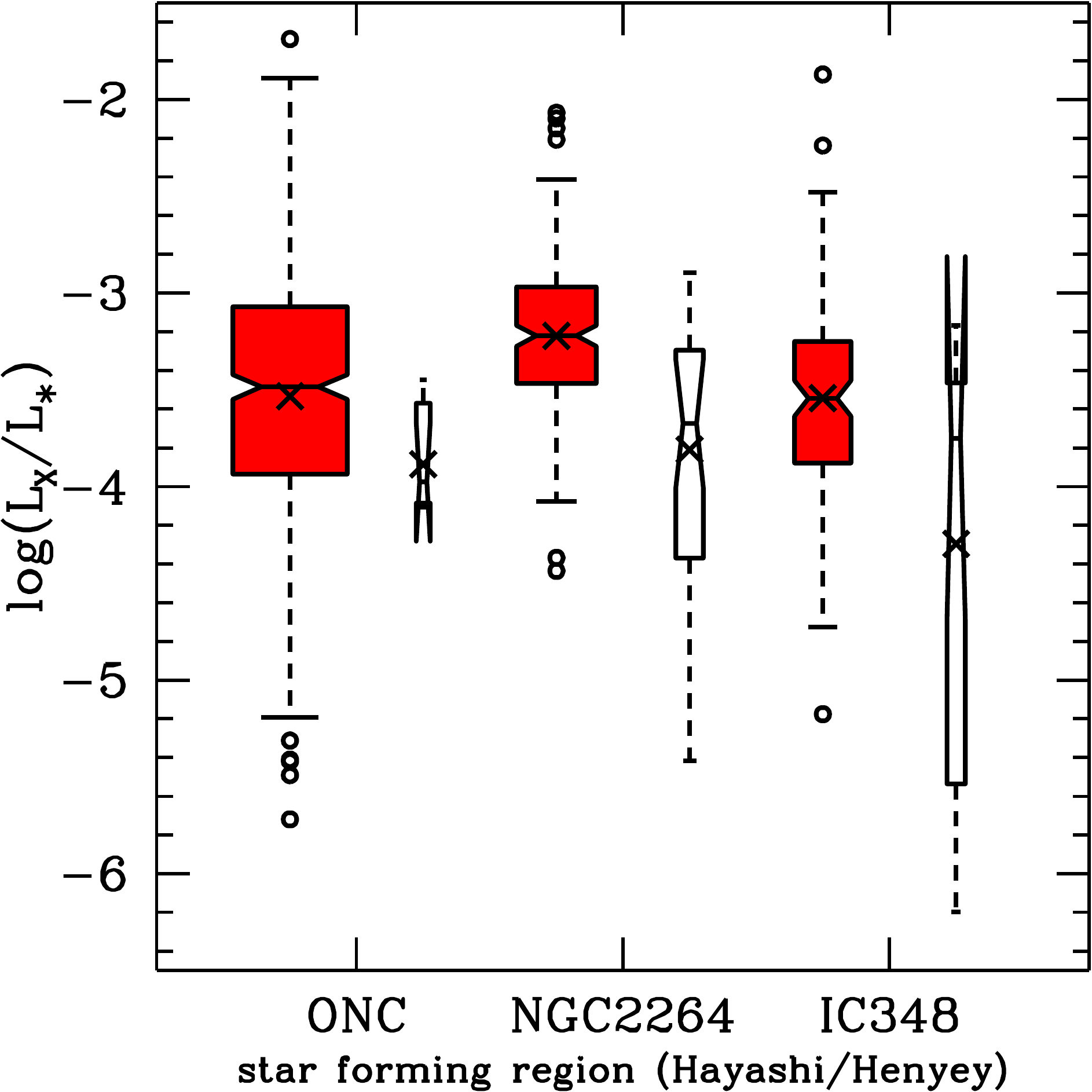}  
   \caption{Variable width notched box plots, with outliers, for the distribution of $\log(L_{\rm X}/L_\ast)$ for stars in all five star forming combined (labelled ``all''), and independently.  The box plot widths, notches, and whisker lengths are defined in the sixth paragraph of section \ref{xraycompare}. In the upper row we are comparing fully convective (red box plots) to partially convective (white box plots) stars, and in the lower row, stars which are on Hayashi tracks (red box plots) to those on Henyey tracks (white box plots) in the H-R diagram.  The black cross denotes the mean value of a sample. For NGC~2362 there is only a single Henyey track star in our sample, rendering a comparison between the Hayashi and Henyey samples meaningless for that region. However, we include it in the figure for completeness.}
   \label{regions_hist_logLXLbol}
\end{figure*}

Tables \ref{tablestats} and \ref{tablestats2} list the mean, standard deviation, median, and median absolute deviation of $\log(L_{\rm X}/L_\ast)$ for the four stellar samples, both for all five star forming regions combined and individually [see Appendix \ref{appendix_logLX} for a comparison of $\log L_{\rm X}$ and $\log(L_\ast/L_\odot)$].  The number of stars in each category is listed as $N_{\rm stars}$. The stars from IC~348 with $L_{\rm X}$ upper limits are not considered in Tables \ref{tablestats} and \ref{tablestats2}, nor in Figure \ref{regions_hist_logLXLbol}, but are considered in the $\log L_\ast-\log L_{\rm X}$ regression fits discussed below.  The sample distributions are plotted graphically as box plots in Figure \ref{regions_hist_logLXLbol} for the regions combined, and individually.  The box plot widths are proportional to the square root of the sample size, which highlights that the bulk of the PMS stars have fully convective interiors.  The whiskers extend from the quartile to the quartile $\pm1.5$ times the interquartile range\footnote{The interquartile range is the difference between the upper and lower quartiles.} ($+$ for the upper whisker, $-$ for the lower whisker), or to the maximum or minimum of the sample if this is smaller than the whisker length would otherwise be, with outliers plotted for the former case.  The box plots are notched with the notch height extending from the median of the sample to $\pm1.57\times(IQR)/\sqrt{N_{\rm stars}}$, where $N_{\rm stars}$ is the sample size and $IQR$ the interquartile range (e.g. \citealt{cha83}).  The notches provide a simple way of determining whether a difference in the medians between samples is significant (roughly when the notches of the box plots being compared do not overlap e.g. \citealt*{mcg78}).       
   
\begin{figure*}
   \centering
 \includegraphics[width=0.35\textwidth]{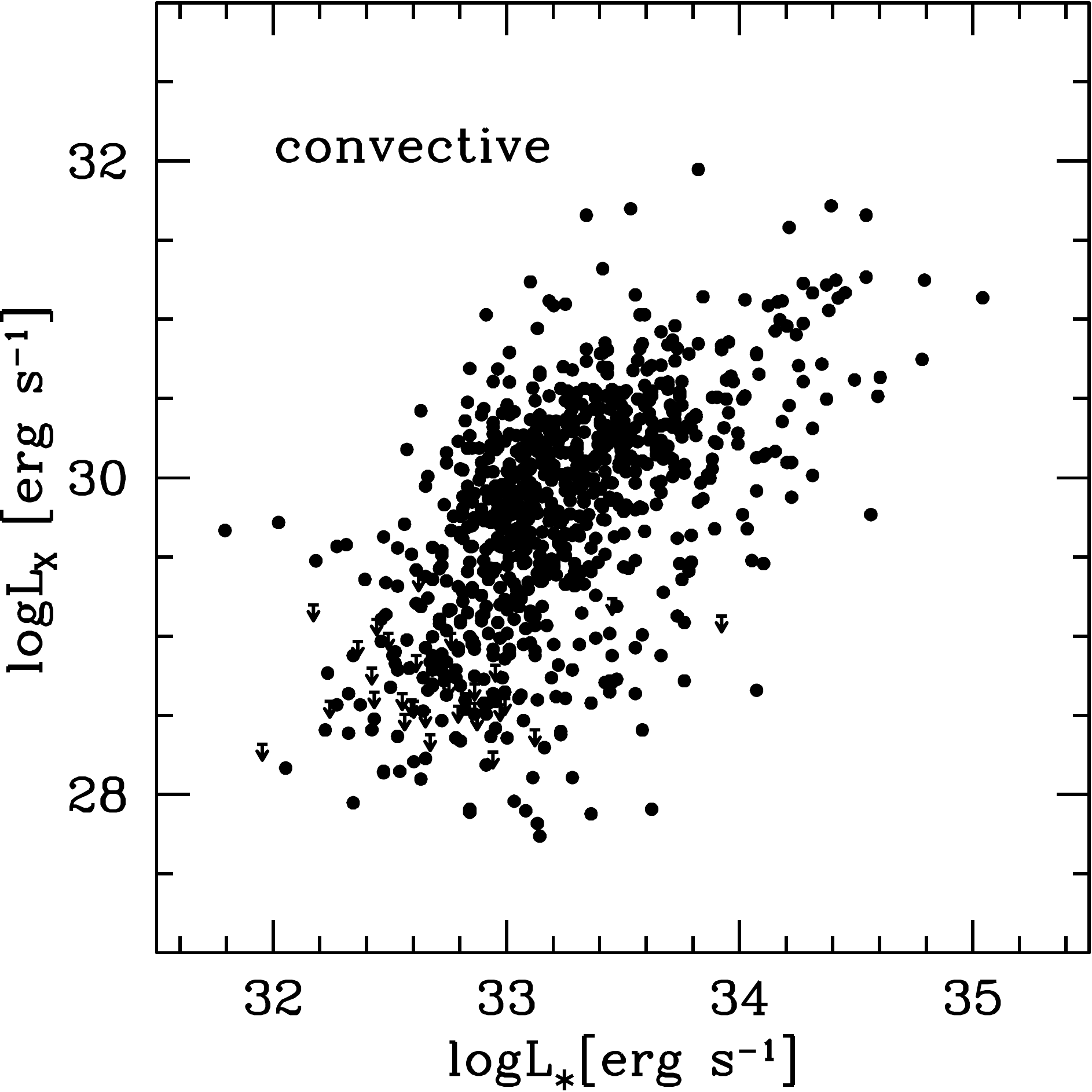}
  \includegraphics[width=0.35\textwidth]{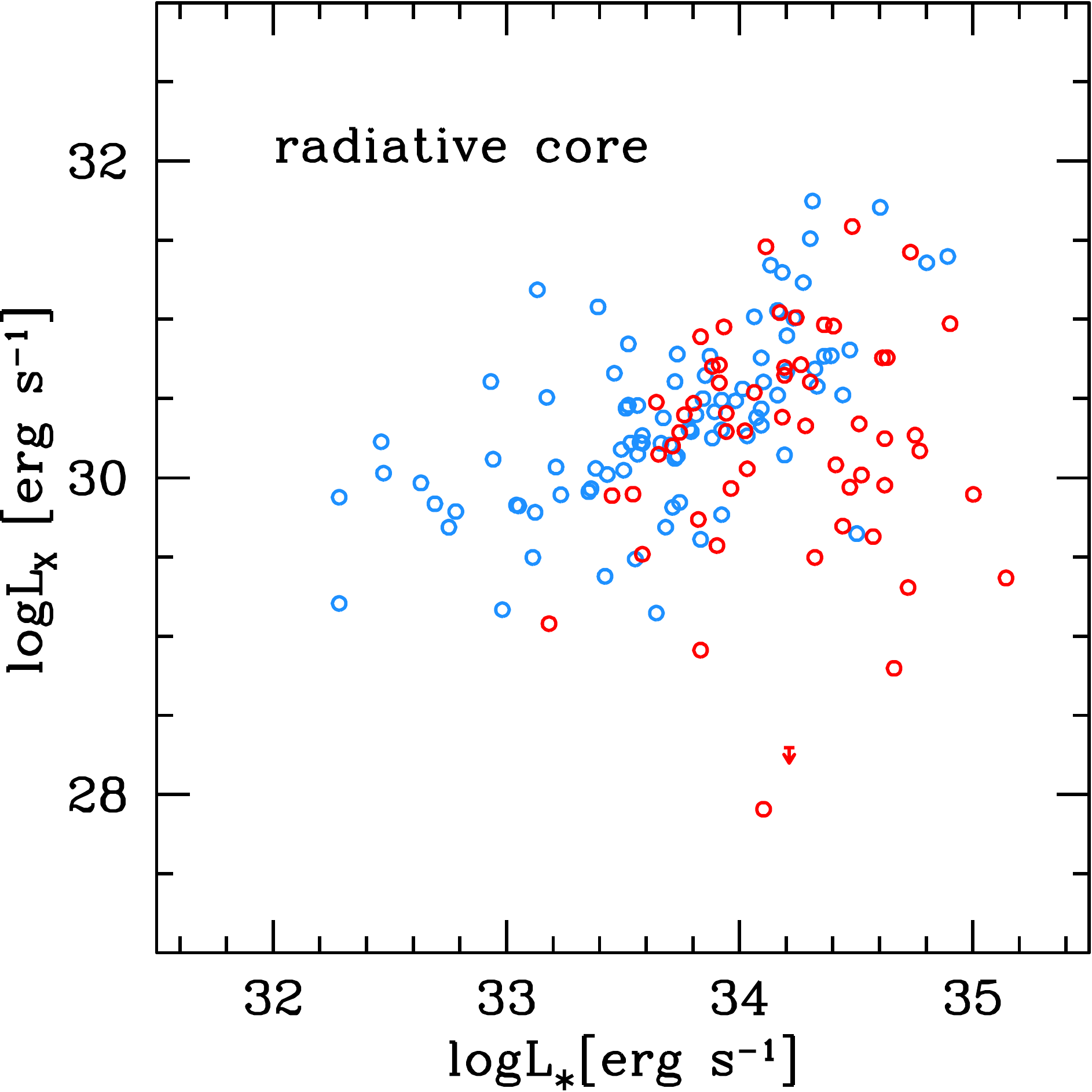}
   \includegraphics[width=0.35\textwidth]{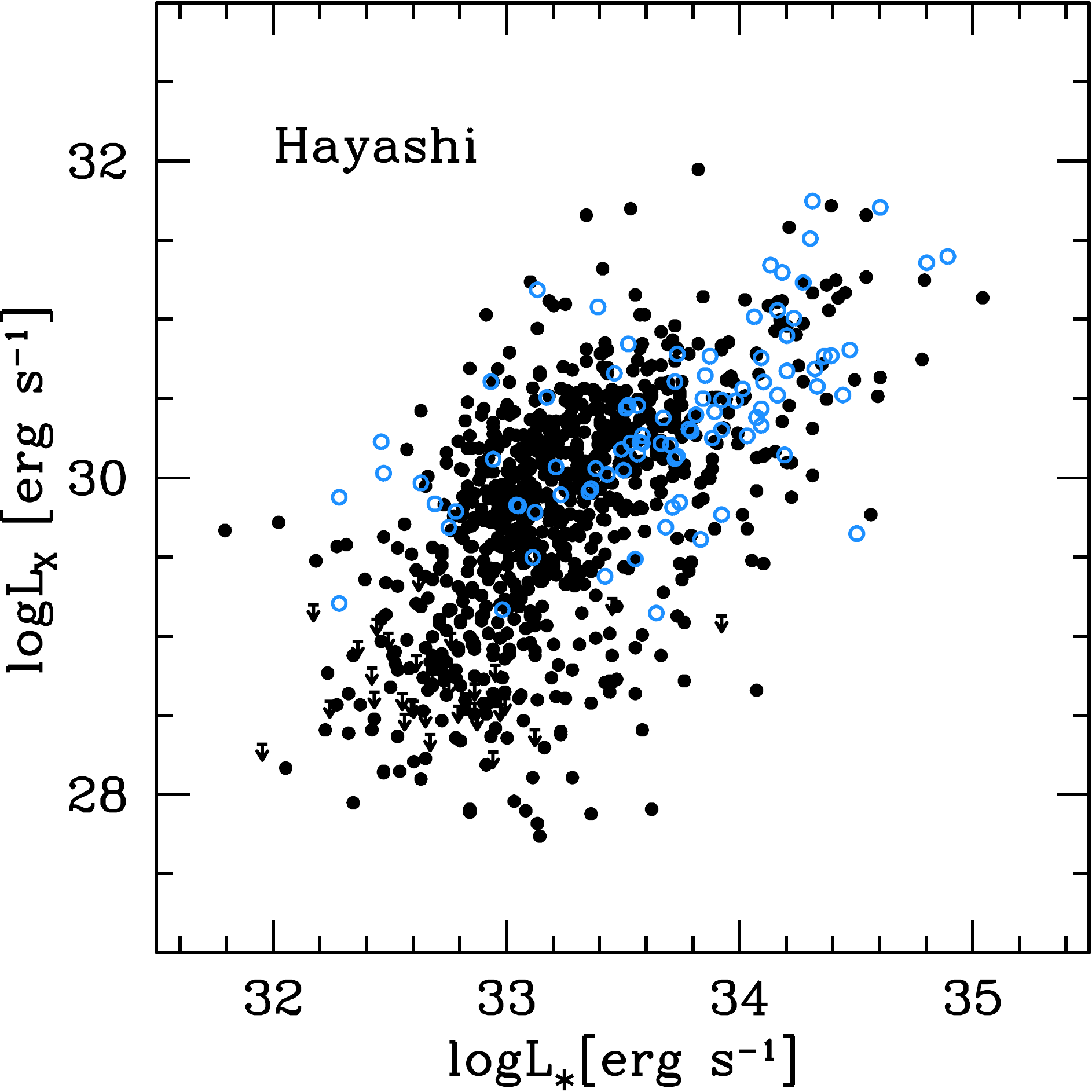}
    \includegraphics[width=0.35\textwidth]{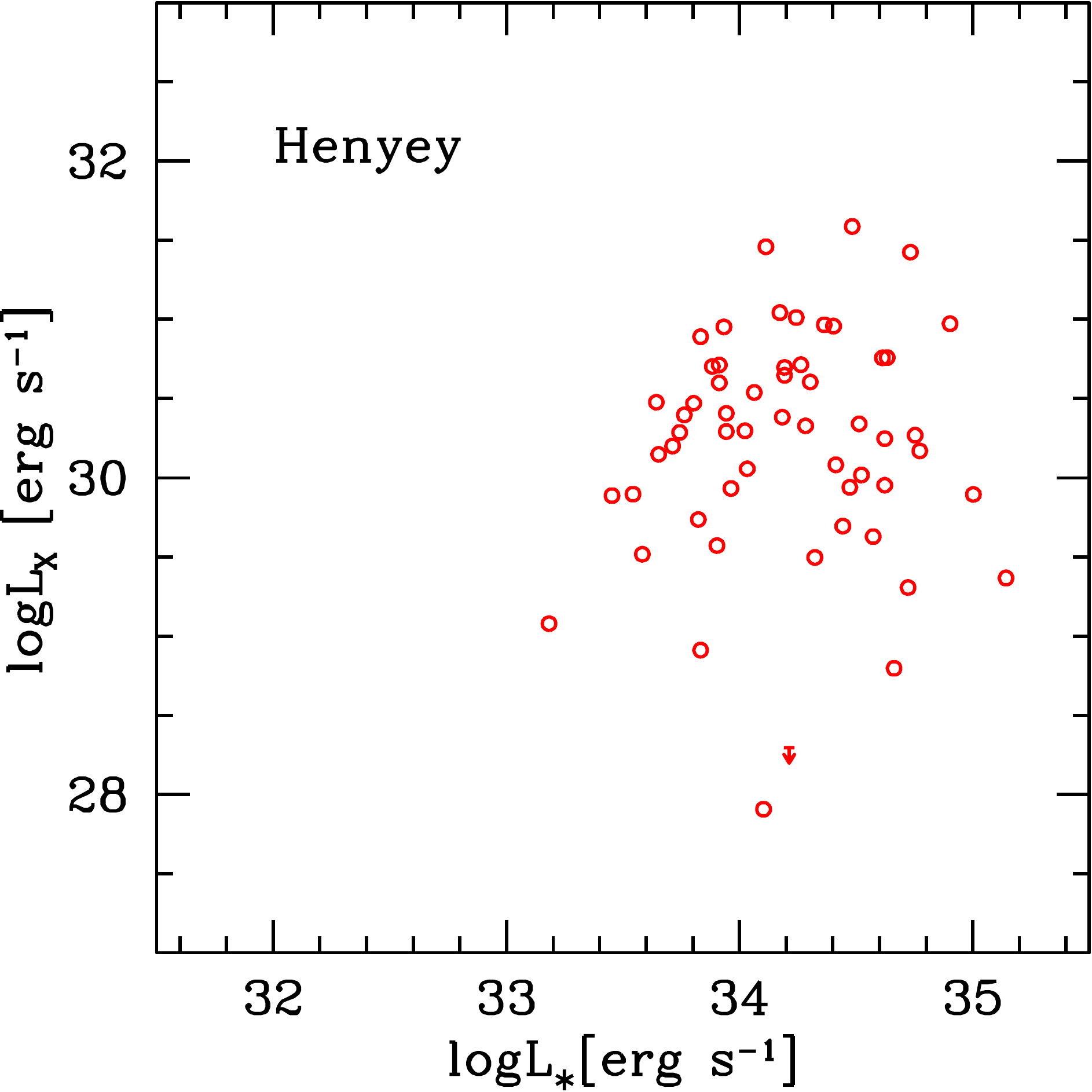}
   \caption{$\log{L_{\rm X}}$ vs $\log{L_\ast}$ for stars in the five star forming regions combined, split into the fully convective (top left), radiative core (top right), Hayashi (bottom left), and Henyey (bottom right) samples.  Fully convective stars are shown as black symbols and partially convective stars as blue/red symbols.  Black and blue symbols combined are stars on Hayashi tracks; those shown as red symbols have evolved onto Henyey tracks.  Stars with $L_{\rm X}$ upper limits (all from IC~348) are shown as downward pointing arrows.     
   }
   \label{logLX_logLstar}
\end{figure*}

We note in passing that, on average, partially convective PMS stars have larger $L_\ast$ and larger $L_{\rm X}$ compared to fully convective stars (likewise for stars on Henyey tracks compared to those on Hayashi tracks, where the difference is greater\footnote{The stellar population of NGC~6530 is an exception to this trend for $L_{\rm X}$ as a result of an observational bias (the cluster has a lack of low-mass PMS stars that have estimated spectral types available in the literature, see Appendix \ref{appendix_logLX}).}).  This is exactly as expected, as fully convective stars are located in the lower right of the PMS of the H-R diagram, see Figure \ref{hrd}, whereas partially convective stars are predominantly in the upper left.  X-ray luminosity is known to correlate with stellar luminosity (e.g. \citealt{tel07} as well as Figure \ref{logLX_logLstar} and the discussion below) and therefore the partially convective stars are also more X-ray luminous than those which are fully convective.  Likewise, $L_{\rm X}$ is correlated with stellar mass (e.g. \citealt{fla03b,pre05b}), and as the partially convective stars are typically more massive than the fully convective stars (evident from their H-R diagram positions alone), it is not a surprise that we find them to be brighter in X-rays.         
   
Of particular interest is the comparison between $\log(L_{\rm X}/L_\ast)$ for fully convective stars and those with radiative cores (and for stars on Hayashi and Henyey tracks), and the relationship between $L_{\rm X}$ and $L_\ast$ for stars with different internal structure.  Previous studies have reported that PMS stars on radiative tracks have lower $\log(L_{\rm X}/L_\ast)$ compared to those on convective tracks in the H-R diagram (e.g. \citealt{fei03,fla03,reb06,cur09,may10}). The most quantitative study of this effect, that of \citet{reb06}, report a factor of $\sim$10 reduction in $\log(L_{\rm X}/L_\ast)$ for partially convective compared to fully convective PMS stars in the mass range $1-2\,{\rm M}_\odot$.  The \citet{reb06} sample of stars, from NGC~2264 (and from the Orion Flanking Fields (FFs) a region we do not consider in this paper), contained many X-ray luminosity upper limits which were carefully accounted for.  Thanks to the improved X-ray data analysis methods of \citet{get10}, see section \ref{ngc2264LX}, these upper limits have been replaced by firm detections.  This, combined with our larger sample from multiple star forming regions\footnote{\citet{reb06} consider a sample of 317 detections plus 139 $L_{\rm X}$ upper limits from NGC~2264 (and 250 detections plus 80 upper limits from the Orion FFs). About 28\% of their sample are $L_{\rm X}$ upper limits.  For comparison, our total sample consists of 950 detections and 34 upper limits, about 3\%.  There are, however, important difference between our data sets, the most important being that \citet{reb06} include stars without spectral type estimates, dereddening the photometry of such stars using ``the most likely reddening in the direction of the cluster''.  Our sample size would increase had we adopted a similar approach rather than selecting only stars with spectral types.} and updated stellar luminosity/mass/age estimates, warrants a detailed comparison of $\log(L_{\rm X}/L_\ast)$ for PMS stars of differing internal structure, as well as the correlation between $L_{\rm X}$ and $L_\ast$.

\begin{table*}
  \caption{Linear regression fits to $\log{L_{\rm X}}$ vs $\log{L_\ast}$ shown in Figure $\ref{logLX_logLstar}$, with
    $\log{L_{\rm X}}=b+a\log{L_\ast}$ (i.e. $L_{\rm X}\propto L_\ast^a$), calculated using the EM algorithm with ASURV. The number in parenthesis in the $N_{\rm stars}$ column is the number of $L_{\rm X}$ upper limits that contribute to the total number of stars in that category. The upper limits are accounted for in the fits. The fifth column gives the standard deviation, and the final column lists the probability of there not being a correlation calculated using a generalised Kendall's $\tau$ test. There is no correlation for Henyey track stars.}
  \begin{tabular}{cccccc}
  \hline
  & $N_{\rm stars}$ & $b$ & $a$ & std. dev. & prob. \\
  \hline
  all stars       &  984 (34)  &   2.82$\pm$1.23     &   0.81$\pm$0.04   & 0.62 & $<$5e-5  \\
  fully convective           &  836 (33)  &   -1.06$\pm$1.49    &   0.93$\pm$0.04   & 0.61 & $<$5e-5  \\
  radiative core     &  148 (1)    &   19.1$\pm$3.0   &   0.33$\pm$0.09   & 0.60 & $<$5e-5  \\
  Hayashi track     &  927 (33)  &   -0.65$\pm$1.32    &   0.92$\pm$0.04   & 0.60 & $<$5e-5  \\
  Henyey track      &  57 (1)     &   25.6$\pm$8.0    &   0.13$\pm$0.23   & 0.73 & 0.518     \\
  \hline
  Hayashi track with radiative core &  91 (0)     &   9.91$\pm$2.82    &   0.61$\pm$0.08   & 0.44 & $<$5e-5    \\
  \hline
\end{tabular}
\label{table_logLX_logLstar}
\end{table*}

\begin{figure*}
   \centering
     \includegraphics[width=0.33\textwidth]{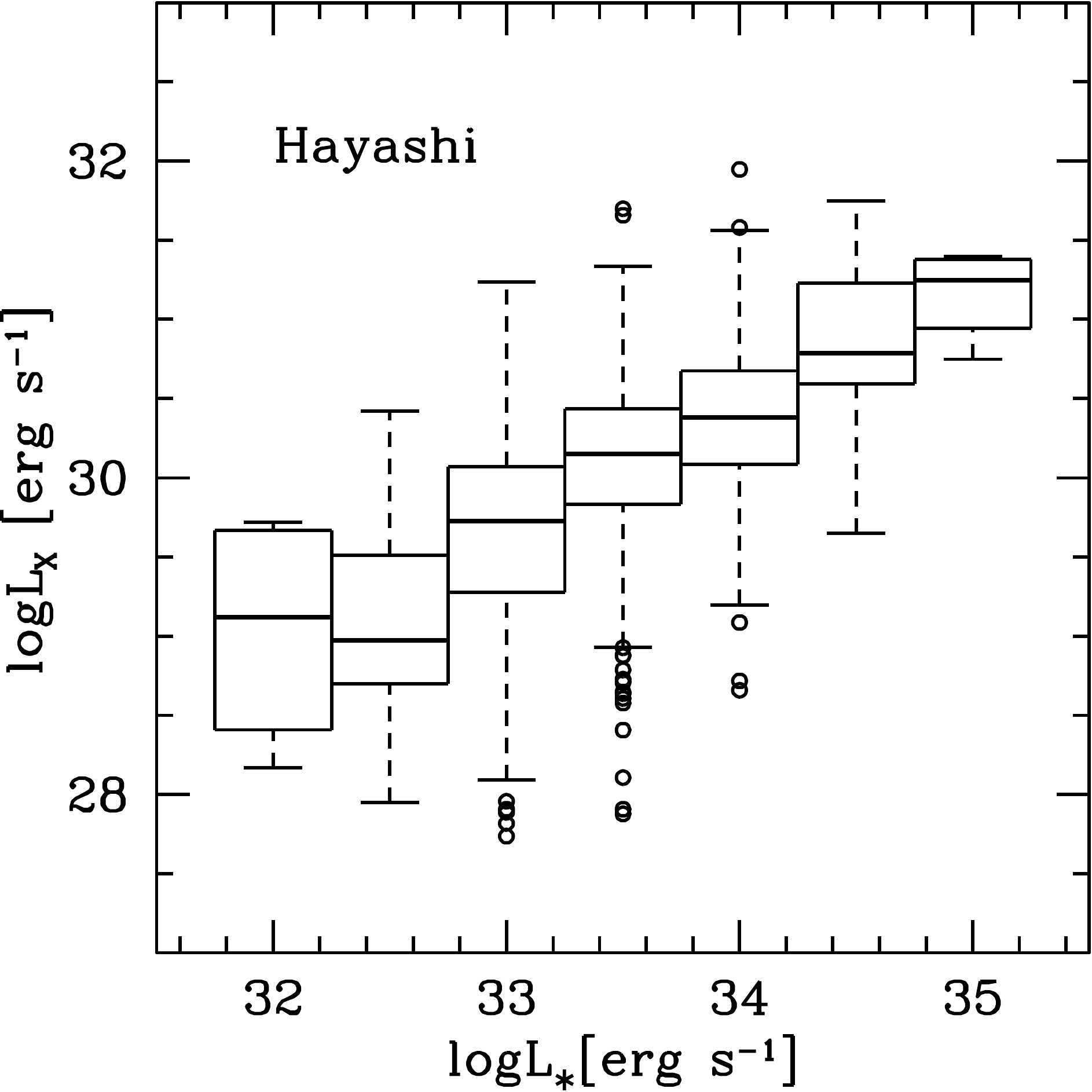}  
      \includegraphics[width=0.33\textwidth]{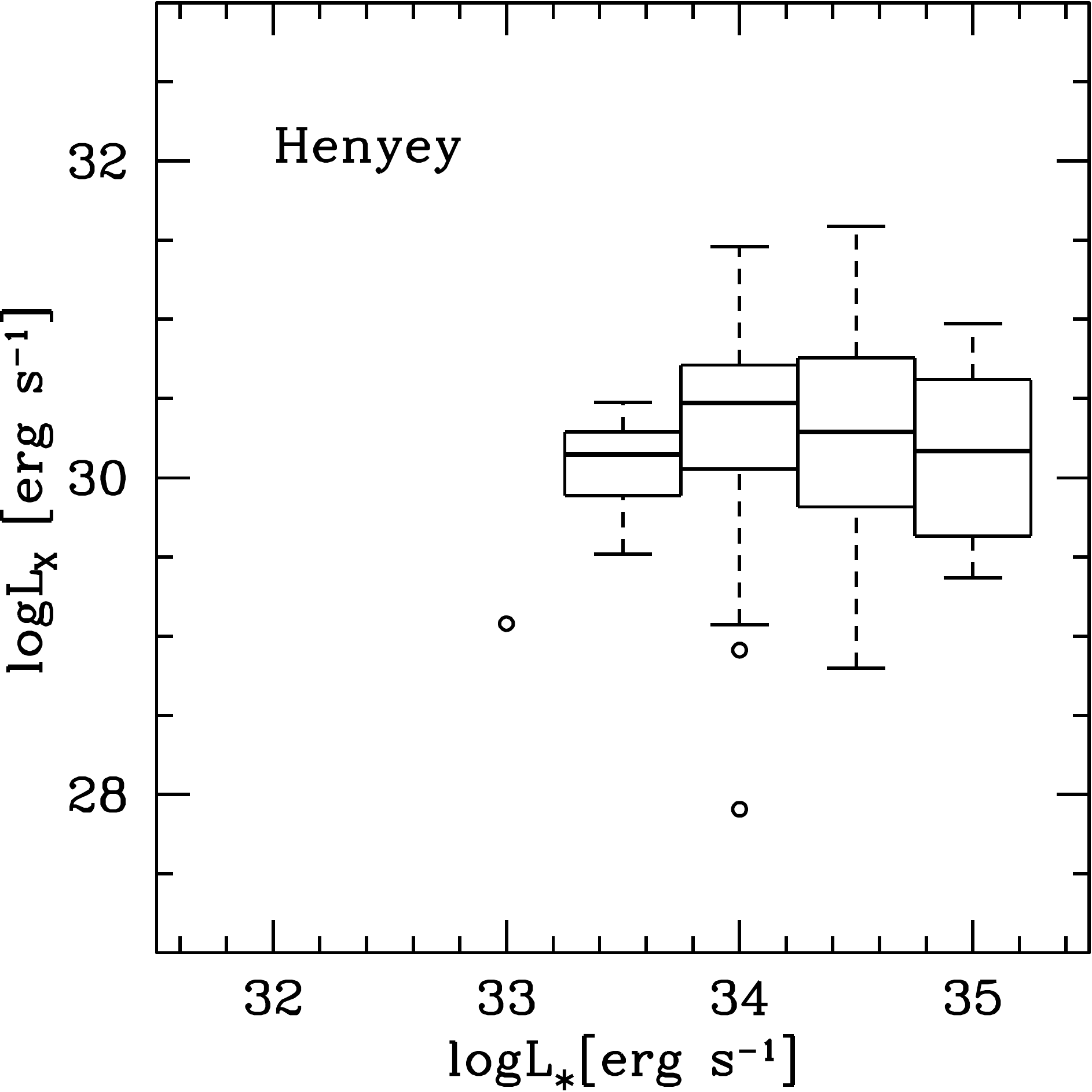} 
       \includegraphics[width=0.33\textwidth]{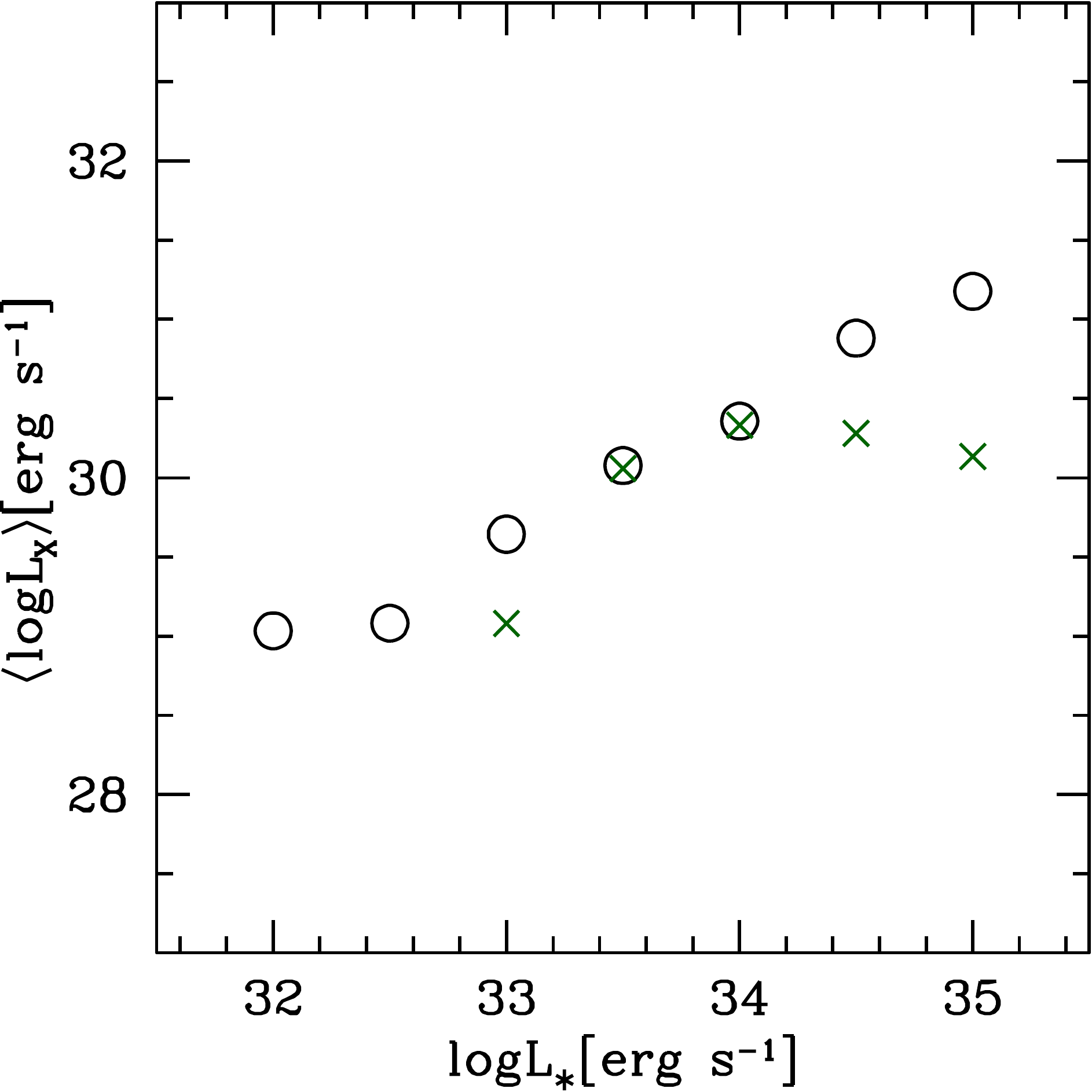} 
   \caption{Boxplots of $\log L_{\rm X}$ in $\log L_\ast$ bins for the Hayashi sample (left; constructed from the black and blue points in Figure \ref{logLX_logLstar}) and the Henyey sample (middle; constructed from the red points in Figure \ref{logLX_logLstar}). The right panel shows the mean value of $\log L_{\rm X}$ in each $\log L_\ast$ bin (plotted with an ordinate value equal to the midpoint of each box in the left/middle panels) with black circles/green crosses for the Hayashi/Henyey sample. Examination of the middle panel reveals that the green cross at $\log L_\ast \approx 33\,{\rm ergs^{-1}}$, $\langle\log L_{\rm X}\rangle \approx 29\,{\rm ergs^{-1}}$ (left most green cross in the right-hand panel) is generated by a single star, which invalidates a comparison between the mean value of $\log{L_{\rm X}}$ for Hayashi and Henyey track stars in that particular $\log{L_\ast}$ bin. The relation between $\log{L_{\rm X}}$ and $\log{L_\ast}$ flattens at large stellar luminosities - this is driven by the evolution of $L_{\rm X}$ of partially convective stars (see section \ref{sectionevolve}).   
   }
   \label{logLX_logLstar_boxes}
\end{figure*}

Considering our whole sample of stars we find only a slight reduction in $\langle\log(L_{\rm X}/L_\ast)\rangle$ for radiative core stars, of 0.13~dex, when comparing them to fully convective stars (Figure \ref{regions_hist_logLXLbol}, upper left, and Table \ref{tablestats}).  The difference is larger when comparing the Henyey and Hayashi sample (Figure \ref{regions_hist_logLXLbol}, lower left, and Table \ref{tablestats2}) with stars on Henyey tracks having $\langle\log(L_{\rm X}/L_\ast)\rangle$ of 0.52~dex less than those on Hayashi tracks. It is immediately apparent when we consider the individual star forming regions why we find little difference in $\langle\log(L_{\rm X}/L_\ast)\rangle$ between fully convective stars and stars with radiative cores when considering all five star regions combined.  The partially convective sample of stars have lower $\langle\log(L_{\rm X}/L_\ast)\rangle$ compared to fully convective stars in each region except the ONC, where the trend is the opposite way around. $\langle\log(L_{\rm X}/L_\ast)\rangle$ is, however, less for Henyey track compared to Hayashi track stars in the ONC.  The ONC is one of the youngest star forming regions, and an examination of the H-R diagrams in Figure \ref{hrd} reveals that few partially convective stars in our ONC sample have yet evolved onto Henyey tracks (about $\sim$80\% of the partially convective stars in the ONC sample are still on Hayashi tracks).  In the other regions the partially convective stars have evolved further towards the ZAMS. This immediately indicates that there is something happening in the evolution of $\log(L_{\rm X}/L_\ast)$ as PMS develop radiative cores, with the decrease becoming greater as stars evolve onto Henyey tracks. If the ONC is removed, and the other four regions are considered together, we find a reduction in $\langle\log(L_{\rm X}/L_\ast)\rangle$ of 0.38~(0.64)~dex when comparing fully convective to radiative core (Hayashi to Henyey track) stars, a larger difference than when the ONC is included.

The mean value of $\log(L_{\rm X}/L_\ast)$ can be written as,
\begin{equation}
\langle\log\left( L_{\rm X}/L_\ast\right) \rangle = \langle\log L_{\rm X}\rangle - \langle\log\left(L_\ast/L_\odot\right)\rangle - \log L_\odot,
\label{Lav} 
\end{equation}
where $\langle\log L_{\rm X}\rangle$ and $\langle\log\left(L_\ast/L_\odot\right)\rangle$ are the mean values of $\log L_{\rm X}$ and $\log\left(L_\ast/L_\odot\right)$ respectively.  If the mean logarithmic X-ray and bolometric luminosities of stars on Hayashi tracks are denoted $\langle\log L_{\rm X}\rangle_{\rm HAY}$ and $\langle\log L_\ast \rangle_{\rm HAY}$ then the equivalent for stars on Henyey tracks can be written as $\langle\log L_{\rm X} \rangle_{\rm HEN} = \langle\log L_{\rm X}\rangle_{\rm HAY} + \alpha$ and $\langle\log L_{\ast} \rangle_{\rm HEN} = \langle\log L_{\ast}\rangle_{\rm HAY} + \beta$; where $\alpha,\beta>0$ to ensure that $\langle\log L_{\rm X}\rangle_{\rm HEN} > \langle\log L_{\rm X}\rangle_{\rm HAY}$ and $\langle\log L_{\ast}\rangle_{\rm HEN} > \langle\log L_{\ast}\rangle_{\rm HAY}$ (as required since Henyey track stars are more luminous than Hayashi track stars, on average, see Appendix \ref{appendix_logLX}).  Using equation (\ref{Lav}) the fractional X-ray luminosities are then related as,
\begin{equation}
\langle\log(L_{\rm_X}/L_\ast)\rangle_{\rm HEN} = \langle\log(L_{\rm_X}/L_\ast)\rangle_{\rm HAY} + (\alpha-\beta).
\end{equation}
Therefore, in order to explain why $\langle\log(L_{\rm_X}/L_\ast)\rangle$ is smaller for Henyey track stars compared to Hayashi track stars, we require that $\alpha<\beta$. In other words, the increase in 
$\langle \log L_{\rm_X}\rangle$ for Henyey track stars when comparing them to Hayashi track stars must be less than the corresponding increase in $\langle \log L_{\ast}\rangle$ in order to ensure that 
$\langle\log(L_{\rm X}/L_\ast)\rangle_{\rm HEN} < \langle\log(L_{\rm X}/L_{\ast})\rangle_{\rm HAY}$. This smaller increase is readily apparent when considering the correlation between $\log L_\ast$ and $\log L_{\rm X}$ for the complete sample.   Figure \ref{logLX_logLstar} plots $\log L_{\rm X}$ versus $\log L_\ast$ with the results of linear regression fits given in Table \ref{table_logLX_logLstar}.  

For the linear regression fits we used the EM (expectation maximisation) algorithm of the ASURV (Astronomy SURVival analysis) package, as described by \citet*{iso86}.   To calculate the probabilities that correlations are present we used generalized Kendall's $\tau$ tests within the ASURV package. ASURV allows censored data, $L_{\rm X}$ upper limits in our case, to be properly accounted for in the analysis, alongside detections. The EM algorithm is a parametric technique that yields the linear regression coefficients (the gradient and the ordinate intercept) assuming a normal distribution for the residuals.  In the absence of censored data the EM algorithm provides an ordinary least squares fit.  In Table \ref{table_logLX_logLstar}, and throughout, we present the values of the regression coefficients to two decimal places to be consistent with the literature; notice, however, that the uncertainties are sometimes large enough that the second decimal place is not fully justified.

Examination Figure \ref{logLX_logLstar} and Table \ref{table_logLX_logLstar} reveals that fully and partially convective stars behave differently in terms of their X-ray emission, with the effect becoming more pronounced when comparing stars on Hayashi and Henyey tracks. For fully convective PMS stars we find an almost linear relationship between the stellar and X-ray luminosities, $L_{\rm X}\propto L_\ast^{0.93\pm0.04}$, as has been reported by other authors (e.g. \citealt{tel07}). The power law dependency is weaker for PMS stars that have developed radiative cores, $L_{\rm X}\propto L_\ast^{0.33\pm0.09}$.  If we consider the Hayashi sample, which includes all of the fully convective stars as well as some that are partially convective (those which still have $L_\ast$ reducing with increasing age) the almost linear relationship is maintained, $L_{\rm X}\propto L_\ast^{0.92\pm0.04}$. However, this hides the fact that the Hayashi sample is dominated by the behaviour of the 836 fully convective stars, compared to the 91 Hayashi track partially convective stars.  If we consider only the 91 stars, the blue points in Figure \ref{logLX_logLstar}, we find a weaker power law dependence, $L_{\rm X}\propto L_\ast^{0.61\pm0.08}$, than what is obtained for the fully convective stars alone. A two-sided Kolmogorov-Smirnov test gives a probability of 1e-8 that the distribution of $\log L_{\rm X}$ for partially convective Hayashi track stars is drawn from the same parent population as that for fully convective Hayashi track stars. 
For PMS stars on Henyey tracks we find no correlation between $\log{L_{\rm X}}$ and $\log{L_\ast}$,  with a probability of 0.52 from a generalised Kendall's $\tau$ test. Taken together, this is evidence for an evolution in the X-ray emission from PMS stars as they evolve across the H-R diagram and develop radiative cores.  The X-ray luminosity appears to begin to decrease once a radiative core develops, with the effect becoming more pronounced for stars that have evolved onto Henyey tracks and which have substantially radiative interiors.

The difference in X-ray emission properties of Hayashi and Henyey track stars is further illustrated in Figure \ref{logLX_logLstar_boxes}.  The figure is constructed from the Hayashi and Henyey plots in Figure \ref{logLX_logLstar}.  The correlation between $L_{\rm X}$ and $L_\ast$ is clear for Hayashi track stars, with $\log{L_{\rm X}}$ increasing uniformly with $\log{L_\ast}$.  For Henyey track stars, as we increase in $L_\ast$, there is no corresponding increase in $L_{\rm X}$.  The mean values in each $\log{L_\ast}$ bin are shown in the right panel of Figure \ref{logLX_logLstar_boxes}.  As we increase in $\log{L_\ast}$ the mean values of $\log{L_{\rm X}}$ for Hayashi (black circles) and Henyey (green crosses) track stars are initially well matched. The most luminous PMS stars on Hayashi tracks are, however, more X-ray luminous than those of comparable $L_\ast$ on Henyey tracks.
The reduction in $\langle\log(L_{\rm X}/L_\ast)\rangle$ for Henyey track compared to Hayashi track PMS stars is therefore directly related to how the X-ray emission evolves with radiative core growth.  In the next section we compare X-ray luminosities with the length of time that stars have spent with partially convective interiors.


\section{X-ray luminosity and radiative core development}\label{sectionevolve}
\subsection{The evolution of the X-ray luminosity of partially convective PMS stars}
\label{dashedline}
In this section we demonstrate that for partially convective PMS stars the stellar X-ray luminosity is anti-correlated with the length of time that stars have spent with partially convective interiors.   

For the partially convective PMS stars we can estimate how long it has been since they developed a radiative core, $t_{\rm since}$, 
\begin{equation}
t_{\rm since}\approx t - \left(\frac{1.494}{M_\ast/{\rm M}_\odot}\right)^{2.364},
\label{tsince}
\end{equation}
where $t_{\rm since}$ and the stellar age $t$ are in units of Myr, and the expression is valid for $0.35<M_\ast/M_\odot\le3$. The second term is an approximation for when a star of specified mass ends the fully convective phase of evolution, derived by \citet{gre12} using the same \citet{sie00} models as we adopt here.\footnote{For a fully convective star with $M_\ast>0.35\,{\rm M}_\odot$ equation (\ref{tsince}) yields a negative number, the modulus of which is the number of Myr that the star still has to evolve before developing a radiative core.  Likewise, for a star on the Hayashi track with $M_\ast\gtrsim0.63\,{\rm M}_\odot$, equation (\ref{tsincehen}) yields a negative number, the modulus of which is the number of Myr that the star still have to evolve before transitioning onto its Henyey track.}  

In a similar fashion, we can derive a rough estimate from the \citet{sie00} models for how long it has been since a PMS star transitioned onto the Henyey track,
\begin{equation}
t_{\rm sincehen} \approx t - \left(\frac{3.04}{M_\ast/{\rm M}_\odot}\right)^{2.42},
\label{tsincehen}
\end{equation}
where both $t_{\rm sincehen}$ and $t$ are in units of Myr, and the expression is valid for $0.63\lesssim M_\ast/M_\odot\le3$.  Stars of mass $M_\ast\lesssim0.63\,{\rm M}_\odot$ never transition onto Henyey tracks; they evolve directly from their Hayashi tracks to the ZAMS.  

\begin{figure}
   \centering
   \includegraphics[width=0.35\textwidth]{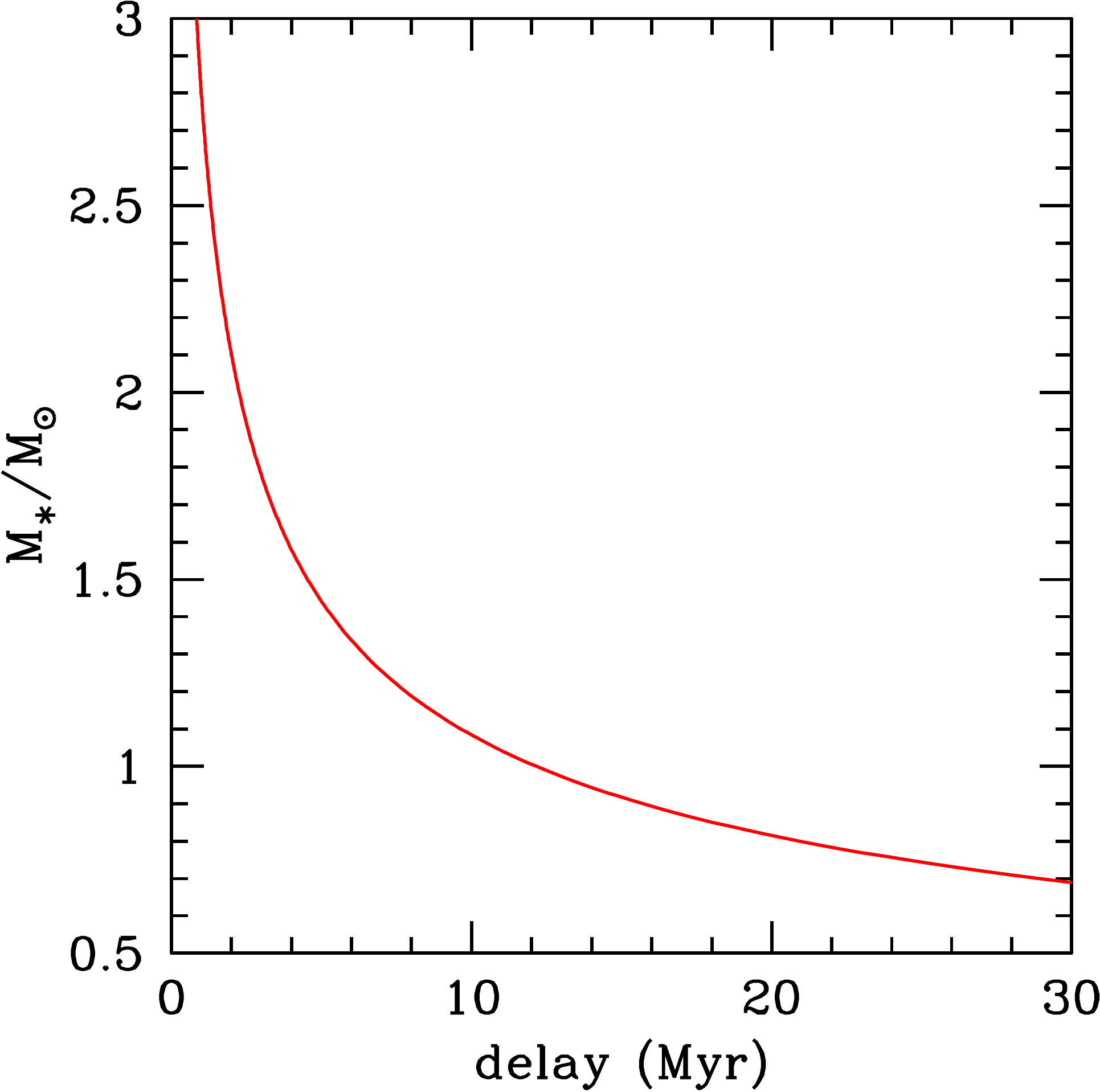}
   \caption{The delay in Myr between stars developing a radiative core and transitioning from their Hayashi tracks (where $L_\ast$ decreases with increasing age) to their Henyey tracks (where $L_\ast$ increases with increasing age) as estimated from the \citet{sie00} models, using equations (\ref{tsince}) and (\ref{tsincehen}).}
   \label{coredelay}
\end{figure}

By subtracting equation (\ref{tsince}) from (\ref{tsincehen}) we can estimate the delay, in Myr, between when a star develops a radiative core until it subsequently transitions from its Hayashi track to its Henyey track in the H-R diagram.  As shown in Figure \ref{coredelay}, a solar mass star requires $\sim$12$\,{\rm Myr}$ of evolution from ending the fully convective phase to transitioning onto its Henyey tracks [a radiative core begins to grow within a solar mass star after $\sim$2.6$\,{\rm Myr}$, equation (\ref{tsince}) while it takes $\sim$14.7$\,{\rm Myr}$ to reach the Henyey track, equation (\ref{tsincehen})].  As expected, the delay is less/more for higher/lower mass stars.

\begin{figure}
   \centering
     \includegraphics[width=0.35\textwidth]{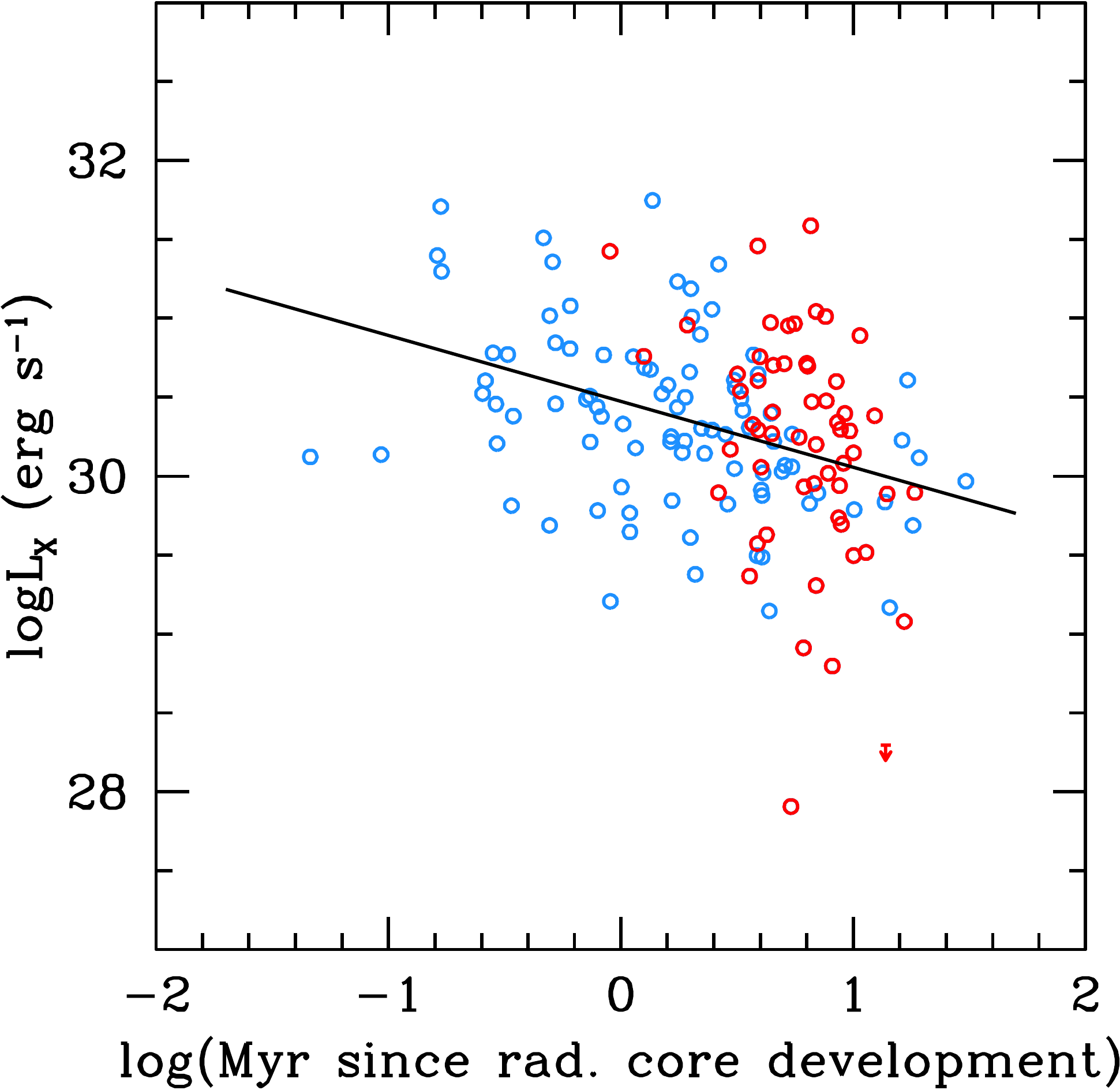}  
   \caption{The X-ray luminosity of partially convective PMS stars as a function of the number of Myr since radiative core development, $t_{\rm since}$, calculated using equation (\ref{tsince}). Blue/red circles represent stars on Hayashi/Henyey tracks in the H-R diagram.  The linear regression fit (solid line), calculated using the EM algorithm within ASURV, indicates that $L_{\rm X}\propto t_{\rm since}^{-2/5}$,
while the probability of there not being a correlation determined from a generalised Kendall's $\tau$ test is $P(0)<5\times10^{-5}$, indicating the presence of a strong correlation. The longer a PMS star has spent with a radiative core, the weaker its X-ray luminosity.}
   \label{logLX_tsince}
\end{figure}

Figure \ref{logLX_tsince} plots $\log{L_{\rm X}}$ versus the time, in Myr, since the beginning of radiative core development, $t_{\rm since}$, estimated from equation (\ref{tsince}).  Using the EM algorithm in ASURV we find that,
\begin{equation}
\log L_{\rm X}=(-0.42\pm0.09)\log t_{\rm since} + (30.47\pm0.06),
\label{tsince_relation}
\end{equation}
which (neglecting the errors) is the equation of the black line in Figure \ref{logLX_tsince}. The standard deviation is 0.59 and the probability of there not being a correlation from a generalized Kendall's $\tau$ test is $<5\times10^{-5}$, indicating the presence of a strong correlation.  PMS stars that have spent longer with radiative cores have lower X-ray luminosities, albeit with a large scatter in $\log L_{\rm X}$ values at a given $\log t_{\rm since}$.  Such scatter in $\log L_{\rm X}$ is inherent to all of the reported X-ray correlations, those with stellar mass, age, and bolometric luminosity (e.g. \citealt{pre05b}, section \ref{xraycompare}, and Appendix \ref{appendix_correlations}), and here with time since radiative core development.  Magnetospheric accretion may contribute to some of this scatter, as we will address in future work. For now, we note that if star forming regions are considered as a whole, then accreting PMS stars are observed to be a factor of 2-to-3 times less luminous in X-rays than the non-accretors (e.g. \citealt{ste01}; \citealt*{fla03c}; \citealt{pre05b}; \citealt{tel07}).  We thus expect the differences between the accreting and non-accreting stars to give rise to a scatter of $\sim$0.4$\,~$dex in $\log L_{\rm X}$ (see \citealt{pre05a}).  This is far less than the observed scatter of $\sim$2.5$\,~$dex apparent from Figure \ref{logLX_tsince}.  Accretion alone cannot explain the observed scatter in $\log L_{\rm X}$.

If we only consider the Hayashi track partially convective stars, the blue points in Figure \ref{logLX_tsince} we find,
\begin{equation}
\log L_{\rm X}=(-0.39\pm0.10)\log t_{\rm since} + (30.44\pm0.06),
\end{equation}
with a standard deviation of 0.51 and with the probability of there not being a correlation of $1\times10^{-4}$.  This is almost identical to that found when considering all of the partially convective stars [equation (\ref{tsince_relation}].  Considering only Henyey track stars, the red points in Figure \ref{logLX_tsince} gives, 
\begin{equation}
\log L_{\rm X}=(-1.10\pm0.36)\log t_{\rm since} + (31.05\pm0.30),
\end{equation}
with a standard deviation of 0.68 and with the probability of there not being a correlation of $9\times10^{-3}$. Tentatively, this perhaps indicates that the drop in stellar X-ray emission quickens once stars have evolved onto Henyey tracks in the H-R diagram.  However, the probability of there being a correlation is marginal for this subsample. A power law decay of the form $L_{\rm X}\propto t_{\rm since}^{a}$ ($a<0$) may not sufficiently approximate the true behaviour of a star's X-ray emission with time since radiative core development for the entirety of its subsequent PMS evolution. 

Is the strong correlation $L_{\rm X}\propto t_{\rm since}^{-2/5}$, equation (\ref{tsince_relation}), a consequence of the correlations between $L_{\rm X}$ and stellar mass $M_\ast$, and between $L_{\rm X}$ and age $t$ (see Appendix \ref{appendix_correlations})?  We do not believe that it is, although $t_{\rm since}$ does depend on both quantities, see equation (\ref{tsince}). Imagine that we had a sample of approximately coeval PMS stars.  As we consider progressively more massive stars in the sample there would come a point where the stars transitioned from fully convective to partially convective objects. As the sample is nearly coeval, the more massive the star, the longer it would have spent with a radiative core; and as $L_{\rm X}$ is positively correlated with $M_\ast$, we would also expect them to be more X-ray luminous, on average, than the lower mass stars in the sample.  Thus we would expect stars which have spent longer with radiative cores to have larger $L_{\rm X}$, and therefore there would be a positive correlation between $L_{\rm X}$ and $t_{\rm since}$.  We observe the opposite, see Fig. \ref{logLX_tsince}.  

Although X-ray luminosities are positively correlated with stellar mass, many of the partially convective stars, and in particular those on Henyey tracks, fall below the main body of the $L_{\rm X}-M_\ast$ correlation (see Appendix \ref{appendix_mass} and Fig. \ref{logLX_logMstar}).  Are these stars the ones that are the cause of the $L_{\rm X}-t_{\rm since}$ correlation?  To test this, we removed the partially convective stars that fall on or below the dashed line in Fig. \ref{logLX_logMstar}, which has the same gradient as the linear regression fit to the entire sample (see Table \ref{table_logLX_logMstar}).  Although this cut is arbitrary, the results below are the same if we choose a slightly different gradient and intercept.  With the stars removed the $L_{\rm X}-t_{\rm since}$ correlation is maintained but with a weaker decay: exponent $-0.27\pm0.08$ and a probability of 0.0004 [see equation (\ref{tsince_relation})].  The decrease in $L_{\rm X}$ with time since radiative core development is therefore apparent even for the stars within the main body of the $L_{\rm X}-M_\ast$ correlation.       

Our sample is also not coeval. A significant age spread does exist in all of the star forming regions, evident from the H-R diagram shown in Fig. \ref{hrd}.  Also noticeable, by eye from the H-R diagrams, is that the earlier spectral type stars (those of large $T_{\rm eff}$) appear to be older, on average, than those of later spectral types.  This has been noted before. \citet*{hil08} argue that PMS evolutionary models under predict the age of low mass stars with, for example, G-type stars being $\sim$2-5 times older than K-types \citep{her15}.  If we focus only on the partially convective PMS stars that form the correlation in Fig. \ref{logLX_tsince} there is less of an age difference.  We separated this subsample into early and late spectral types and found the biggest age difference when using a spectral type cut of earlier than K0 and K0 \& later, but of only $\approx$0.9$\,{\rm Myr}$.    

\begin{figure}
   \centering
     \includegraphics[width=0.35\textwidth]{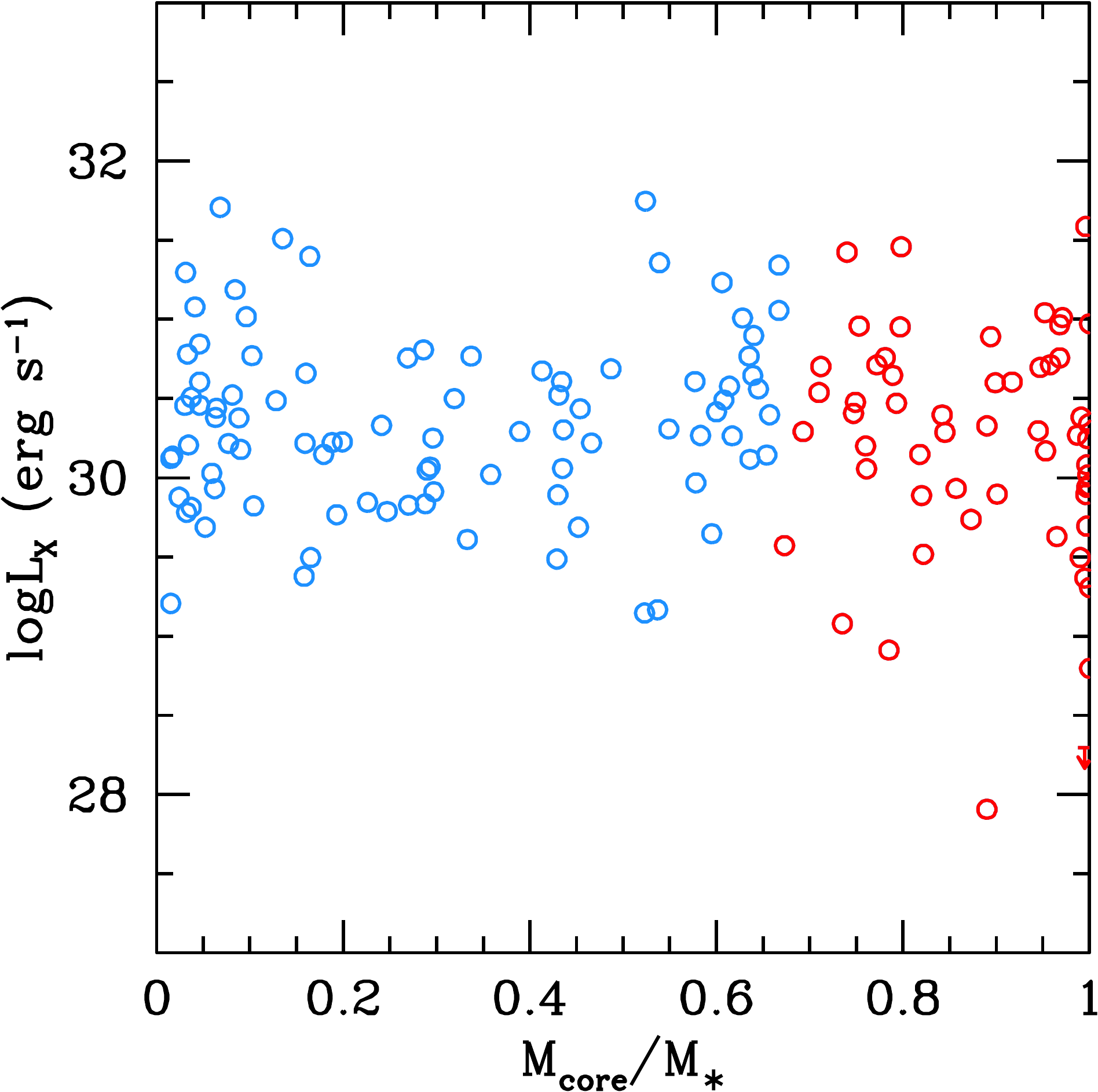}  
      \includegraphics[width=0.35\textwidth]{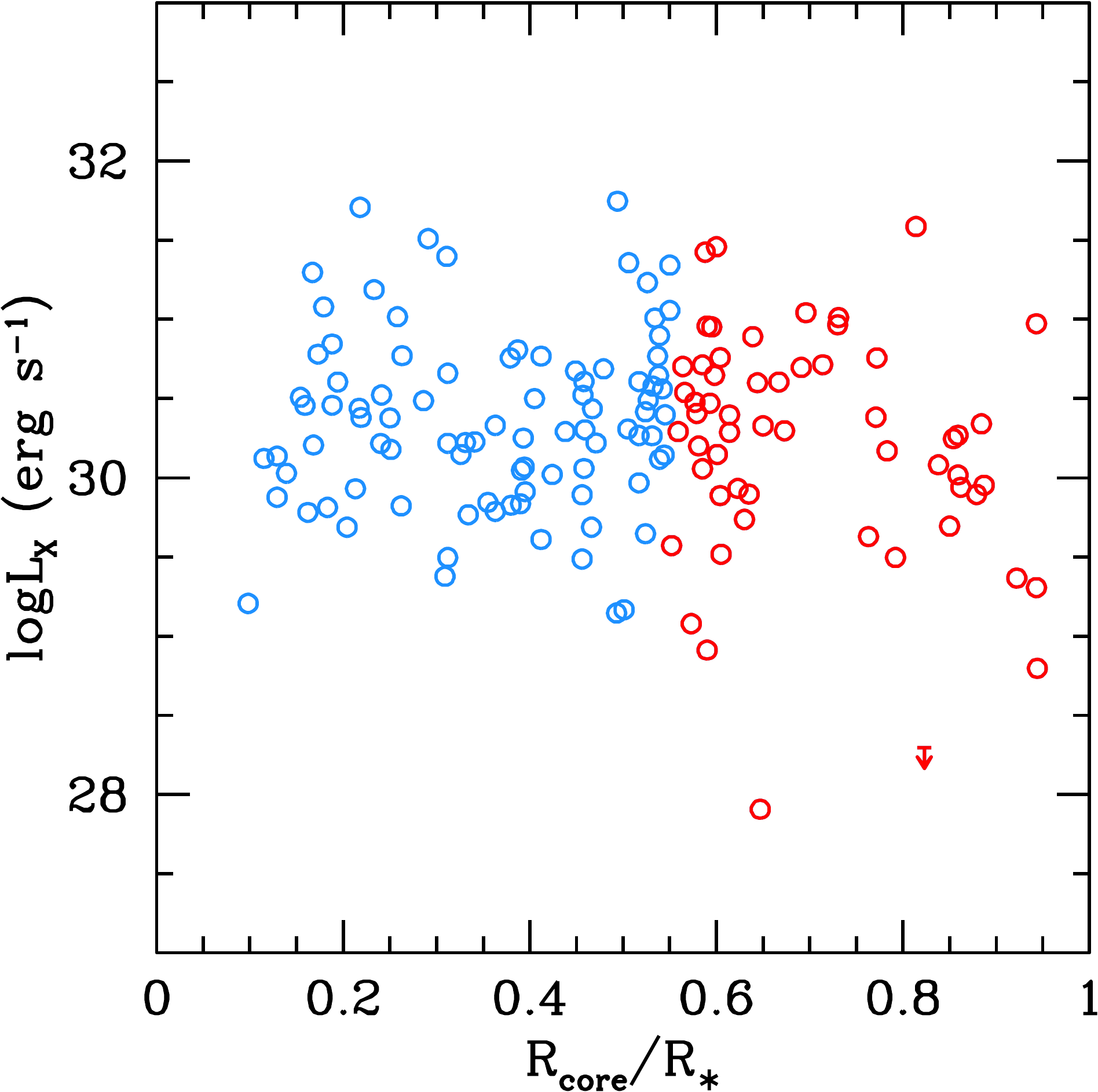} 
   \caption{The X-ray luminosity versus the ratio of the radiative core mass to the stellar mass (upper) and the ratio of radiative core radius to the stellar radius (lower). Points are coloured as in Figure \ref{logLX_tsince}. There is no correlation between $L_{\rm X}$ and $M_{\rm core}/M_\ast$, nor between $L_{\rm X}$ and $R_{\rm core}/R_\ast$.
   }
   \label{LX_mcore}
\end{figure}

To examine if this had any bearing on the correlation, we increased the age of the KM-type stars in Fig. \ref{logLX_tsince} by a constant such that their mean age matched the mean age of the FG-type stars.  This shifted 18 stars from Hayashi to Henyey tracks.  The $L_{\rm X}-t_{\rm since}$ correlation was maintained although with a steeper decay than that from equation (\ref{tsince_relation}) with an exponent of $-0.68\pm0.14$ and a probability of $<$5e-5.  However, increasing the age of KM-types stars results in additional stars becoming partially convective.  Accounting for this adds 29 stars to the partially convective sample and results in slower decay with an exponent of $-0.27\pm0.08$ and a probability of 0.0001.  For completeness, we also tried decreasing the age of the FG-type stars such that their mean age matched the mean age of the partially convective KM-type stars even though this is the opposite of what the work of \citet{hil08} suggests, where it is the late types that have under predicted ages. This resulted in a smaller sample of 120 partially convective stars, with 28 moving to portions of their mass tracks where they have fully convective interiors.  Again the $L_{\rm X}-t_{\rm since}$ correlation was maintained with an exponent of $-0.26\pm0.10$ and a probability of 0.0023.  We can therefore conclude that any age difference between early and late spectral type stars inherent in our adopted PMS evolutionary model has no bearing on the basic result.  We also note that our results are unaffected by the choice of PMS evolutionary model (see Appendix \ref{appendix_comparison}). On average, the longer a PMS star has spent with a partially convective interior the less X-ray luminous it becomes.


\subsection{X-ray luminosity as a function of core mass and radius}
As shown in Figure \ref{LX_mcore}, we find no simple correlation between $L_{\rm X}$ and radiative core mass, nor between $L_{\rm X}$ and radiative core radius.  This lack of a clear relationship between radiative core size and $L_{\rm X}$, with $\sim$2 dex variation in $\log L_{\rm X}$ at any given core mass or radius, likely contributes to the scatter in the other correlations e.g. between $L_{\rm X}$ and time since core development. We do find a strong correlation ($P(0)<5\times10^{-5}$) between $L_{\rm X}$ and radiative core density, $\rho_{\rm core}=3M_{\rm core}/(4\pi R_{\rm core}^3)$, with, $\log L_{\rm X} = (30.28\pm0.05) + (-0.36\pm0.09)\log \rho_{\rm core}$. However, this is a consequence of the correlation between $L_{\rm X}$ and $L_\ast$ (see section \ref{xraycompare}), as $\rho_{\rm core}$ scales inversely with $L_\ast$ \citep{sie00}.


\section{Discussion \& conclusions}\label{conclusion}
We have presented a thorough examination of the differences in, and the evolution of, the coronal X-ray emission properties of fully and partially convective PMS stars.  We considered the most up-to-date spectral types, X-ray luminosities, and photometry from 
the literature for PMS stars in five of the best studied star forming regions: the ONC, IC~348, NGC~2264, NGC~2362, and NGC~6530. The spectral type of each star was estimated from spectra and not by comparing de-reddened colours to (spectral type dependent) intrinsic colours.  After de-reddening the observed colours, we used the most modern PMS stars intrinsic colour/bolometric correction/effective temperature scale of \citet{pec13} to position stars in the H-R diagram and the models of \citet{sie00} to estimate stellar mass, age, and internal structure for our sample of close to 1000 stars.   

We find a (slightly) sub-linear relationship between stellar luminosity and X-ray luminosity for fully convective PMS stars $L_{\rm X}\propto L_\ast^{0.93\pm0.04}$.  The exponent is reduced to 0.61$\pm$0.08 for stars that have developed radiative cores but which are still on their Hayashi tracks in the H-R diagram (where $L_\ast$ is decreasing with age), with a two-sided Kolmogorov-Smirnov test revealing that they are drawn from a different population than the fully convective stars (see section \ref{xraycompare}). Considering all partially convective PMS stars including those on Henyey tracks ($L_\ast$ increasing with age), which have mostly radiative interiors, the exponent is further reduced to 0.33$\pm$0.09.  There is no correlation between $L_{\rm X}$ and $L_\ast$ for PMS stars on Henyey tracks in the H-R diagram (see Figures \ref{logLX_logLstar} and \ref{logLX_logLstar_boxes}).  We note that there are only 148 stars in the radiative core sample compared with 836 fully convective stars, and only 57 Henyey track stars compared to 927 Hayashi track stars.  This is as expected since our PMS clusters are young and there will always be a greater proportion of low-mass compared to high-mass stars.  In future work, adding data for older PMS clusters, where more stars will have developed radiative cores, and a greater fraction will have completed the Hayashi-to-Henyey track transition, will help to increase the sample size.

The evolution of $L_{\rm X}-L_\ast$ correlation is clearly driven by the stellar structure transition from fully convective to partially convective with an inner radiative core and outer convective envelop.  By comparing the mean value of $\log{L_{\rm X}}$ in bins of increasing $\log{L_\ast}$ we demonstrated that the $L_{\rm X}-L_\ast$ correlation flattens at large stellar luminosities.  This is because, on average, partially convective PMS stars are less X-ray luminous than fully convective ones at the same $L_\ast$.  PMS stars with radiative cores have lower fractional X-ray luminosities, $\log(L_{\rm X}/L_\ast)$, on average, than their fully convective counterparts. This effect is greater when comparing Henyey track to Hayashi track stars, with $\langle \log(L_{\rm X}/L_\ast) \rangle_{\rm HAY} - \langle \log(L_{\rm X}/L_\ast) \rangle_{\rm HEN} \approx$ 0.52~dex when considering our entire sample of X-ray detections.     

The drop off in X-ray luminosity for partially convective PMS stars is encapsulated in the correlations between $L_{\rm X}$ and age, $t$. \citet{pre05a} studied the age evolution of PMS star X-ray emission, reporting a weak decrease of approximately $L_{\rm X}\propto t^{-1/3}$ when dividing their stars into mass stratified subsamples.  We find the same correlation but only for mass bins that are dominated by fully convective PMS stars.  As a side note, $L_{\rm X}\propto t^{-1/3}$ is slower than we would expect from simple theoretical considerations.  There is an approximately linear relationship between $L_{\rm X}$ and $L_\ast$ for fully convective PMS stars which contract roughly as $R_\ast\propto t^{-1/3}$ (e.g. \citealt{lam05}; \citealt{dav14}).  As such we would expect $L_{\rm X}\propto L_\ast\propto R_\ast^2 T_{\rm eff}^4 \propto t^{-2/3}$, as $T_{\rm eff}$ is approximately constant during this phase. This theoretical result represents a faster decrease than is observed.  

With our larger sample of stars we have firmly established that $L_{\rm X}$ decays more rapidly with age for higher mass bins which contain mostly partially convective PMS stars.  For example, $L_{\rm X} \propto t^{-0.53\pm0.10}$ for 1-1.5$\,{\rm M}_\odot$ stars, with the exponent dropping to $-0.86\pm0.19$ and then to $-1.19\pm0.35$ for 1.5-2 and 2-3$\,{\rm M}_\odot$ stars.  Across these three mass bins the fraction of partially convective stars in our sample is 59\%, 88\% and 100\%, further emphasising that $L_{\rm X}$ decays once convective zones start to vanish within stellar interiors.  

This quickening of the decay in $L_{\rm X}$ with age for higher mass bins is driven by the $L_{\rm X}$ decay once stars have developed radiative cores. We find that, on average, the longer PMS stars have spent with radiative cores the less X-ray luminous that they become with $L_{\rm X}\propto t_{\rm since}^{-2/5}$.  The correlation is robust, with the probability of there not being a correlation from a generalised Kendall's $\tau$ test of $<$5e-5.  The basic result that $L_{\rm X}$ decays with time since radiative core development is not influenced by any age differences between early and late spectral type stars that are inherent to PMS evolutionary models (e.g. \citealt{hil08}). There is, however, no direct link between the fraction of the stellar mass contained with the radiative core and the X-ray luminosity, nor between the core radius relative to the stellar radius and $L_{\rm X}$ (equivalently between convective zone depth and $L_{\rm X}$).   

The decay of $L_{\rm X}$ with time since radiative core development can be linked to the observed increase in the complexity of the magnetic topology of PMS stars as they evolve across the H-R diagram \citep{gre12,gre14}.  Zeeman-Doppler imaging studies of accreting PMS stars (e.g. \citealt{don08bptau,hus09,don11,don12}) have revealed those which are fully convective, at least those above $\sim$0.5$\,{\rm M}_\odot$, and those with small radiative cores host large-scale magnetic fields that are dominantly axisymmetric, with kilo-Gauss octupole or dipole components slightly tilted with respect to the stellar rotation axis \citep{gre11}, with the octupole becoming more dominant relative to the dipole with age \citep{gre14}.  In contrast, PMS stars with large radiative cores host complex, multipolar, and dominantly non-axisymmetric large-scale magnetic fields.  $L_{\rm X}$ is approximately the product of a radiative loss function $\Lambda(T_{\rm X})$ and the volume emission measure, $L_{\rm X}\approx \Lambda(T_{\rm X})EM$ (e.g. \citealt{gue09}), where $\Lambda(T_{\rm X})$ is a piecewise function of the coronal temperature $T_{\rm X}$ (e.g. \citealt*{asc08}). If we approximate the emission measure as $EM\approx n^2 V$, where $n$ is the number density of the coronal plasma and $V$ the emitting volume, then the X-ray luminosity $L_{\rm X}\propto V$.  An increase in the magnetic field complexity as PMS stars evolve from fully convective objects, to hosting substantial radiative cores will result in a decrease of the available coronal emitting volume, and therefore a decrease in $L_{\rm X}$.  For example, the magnetic field of a PMS star with a strong dipole component would be able to contain coronal plasma out to a large radius from the stellar surface.  A PMS star with a magnetic field dominated by high order and non-axisymmetric components, where the field strength would decay faster with height above the star, would have a more compact corona, a smaller emitting volume, and therefore a lower X-ray luminosity.  These qualitative ideas will be developed further in a companion paper, where semi-analytic models of the evolution of the coronal X-ray emission from PMS stars with multipolar magnetic fields will be presented.  

The change in magnetic field topology as PMS evolve may affect other X-ray signatures as well.  For example, compact, multipolar coronae would be more likely to produce rotationally modulated X-ray light curves compared to more extended coronae (e.g. \citealt{gre06}; \citealt{hus09}; \citealt{joh14}).  We also speculate that the flaring properties of Hayashi and Henyey track PMS stars may differ, although a detailed comparison of flares in X-ray light curves as a function of the stellar internal structure has not yet been undertaken. 

Finally, we end by noting that the partially convective mid-K to F-type PMS stars in our sample will evolve to mid-G to A-type stars once they arrive on the main sequence.  A-type main sequence stars, in particular, lack outer convective envelopes. Only $\sim$10-15\% of them are detected in X-rays, with this emission thought to come from later spectral type binary companions \citep{schr07}.  Our work in this paper has shown that we can already observe the decay of X-ray emission during the first few Myr of the PMS evolution of stars that will become A-type main sequence stars, as their radiative zones are beginning to grow.  
Additionally, the disappearance of highly luminous X-ray coronae as PMS stars become largely, and fully, radiative, may be linked to the low ($\sim$5-10\%) detection rate of significant surface magnetic fields in Herbig AeBe stars \citep{ale13}.  Newly initiated surveys of PMS stars of mass $1.5-3\,{\rm M}_\odot$ will help to further complete our understanding of the evolution of stellar magnetism, and of coronal X-ray emission, with the change in stellar internal structure \citep{hus14}.


\section*{Acknowledgements}
SGG acknowledges support from the Science \& Technology Facilities Council (STFC) via an Ernest Rutherford Fellowship [ST/J003255/1], and A. Aarnio, C. Argiroffi, M. Audard, P. Bodenheimer, A. M. Cody, J.-F. Donati, E. D. Feigelson, K. V. Getman, M. G\"{u}del, L. A. Hillenbrand, V. R. Holzwarth, K. Horne, N. Mayne, M. Pecaut \& B. Stelzer for useful discussions. CLD acknowledges support from STFC via a PhD studentship and additional funding via the STFC Studentship Enhancement Programme [ST/J500744/1], and support from the ERC Starting Grant "ImagePlanetFormDiscs" (Grant Agreement No. 639889). We thank the referee for their constructive and detailed report. This research has made use of NASA's Astrophysics Data System. This research has made use of the VizieR catalogue access tool, CDS, Strasbourg, France. This research has made use of the SIMBAD database, operated at CDS, Strasbourg. France. Statistical tests and fits were made using {\bf ASURV} Rev 1.2 \citep*{lav92}, which implements the methods presented in \citet{fei85} and \citet{iso86}.This publication makes use of data products from the Two Micron All Sky Survey, which is a joint project of the University of Massachusetts and the Infrared Processing and Analysis Center/California Institute of Technology, funded by the National Aeronautics and Space Administration and the National Science Foundation. This research has made use of the International Variable Star Index (VSX) database, operated at AAVSO, Cambridge, Massachusetts, USA.




\bibliographystyle{mnras}
\bibliography{xray_v3} 

\begin{thebibliography}{}
\makeatletter
\relax
\def\mn@urlcharsother{\let\do\@makeother \do\$\do\&\do\#\do\^\do\_\do\%\do\~}
\def\mn@doi{\begingroup\mn@urlcharsother \@ifnextchar [ {\mn@doi@}
  {\mn@doi@[]}}
\def\mn@doi@[#1]#2{\def\@tempa{#1}\ifx\@tempa\@empty \href
  {http://dx.doi.org/#2} {doi:#2}\else \href {http://dx.doi.org/#2} {#1}\fi
  \endgroup}
\def\mn@eprint#1#2{\mn@eprint@#1:#2::\@nil}
\def\mn@eprint@arXiv#1{\href {http://arxiv.org/abs/#1} {{\tt arXiv:#1}}}
\def\mn@eprint@dblp#1{\href {http://dblp.uni-trier.de/rec/bibtex/#1.xml}
  {dblp:#1}}
\def\mn@eprint@#1:#2:#3:#4\@nil{\def\@tempa {#1}\def\@tempb {#2}\def\@tempc
  {#3}\ifx \@tempc \@empty \let \@tempc \@tempb \let \@tempb \@tempa \fi \ifx
  \@tempb \@empty \def\@tempb {arXiv}\fi \@ifundefined
  {mn@eprint@\@tempb}{\@tempb:\@tempc}{\expandafter \expandafter \csname
  mn@eprint@\@tempb\endcsname \expandafter{\@tempc}}}

\bibitem[\protect\citeauthoryear{{Aarnio}, {Stassun}  \& {Matt}}{{Aarnio}
  et~al.}{2010}]{aar10}
{Aarnio} A.~N.,  {Stassun} K.~G.,   {Matt} S.~P.,  2010, \mn@doi [\apj]
  {10.1088/0004-637X/717/1/93}, \href
  {http://adsabs.harvard.edu/abs/2010ApJ...717...93A} {717, 93}

\bibitem[\protect\citeauthoryear{{Alecian} et~al.,}{{Alecian}
  et~al.}{2013}]{ale13}
{Alecian} E.,  et~al., 2013, \mn@doi [\mnras] {10.1093/mnras/sts383}, \href
  {http://adsabs.harvard.edu/abs/2013MNRAS.429.1001A} {429, 1001}

\bibitem[\protect\citeauthoryear{{Alexander} \& {Preibisch}}{{Alexander} \&
  {Preibisch}}{2012}]{ale12}
{Alexander} F.,  {Preibisch} T.,  2012, \mn@doi [\aap]
  {10.1051/0004-6361/201118100}, \href
  {http://adsabs.harvard.edu/abs/2012A%26A...539A..64A} {539, A64}

\bibitem[\protect\citeauthoryear{{Arias}, {Barb{\'a}}, {Ma{\'{\i}}z
  Apell{\'a}niz}, {Morrell}  \& {Rubio}}{{Arias} et~al.}{2006}]{ari06}
{Arias} J.~I.,  {Barb{\'a}} R.~H.,  {Ma{\'{\i}}z Apell{\'a}niz} J.,  {Morrell}
  N.~I.,   {Rubio} M.,  2006, \mn@doi [\mnras]
  {10.1111/j.1365-2966.2005.09829.x}, \href
  {http://adsabs.harvard.edu/abs/2006MNRAS.366..739A} {366, 739}

\bibitem[\protect\citeauthoryear{{Arias}, {Barb{\'a}}  \& {Morrell}}{{Arias}
  et~al.}{2007}]{ari07}
{Arias} J.~I.,  {Barb{\'a}} R.~H.,   {Morrell} N.~I.,  2007, \mn@doi [\mnras]
  {10.1111/j.1365-2966.2006.11217.x}, \href
  {http://adsabs.harvard.edu/abs/2007MNRAS.374.1253A} {374, 1253}

\bibitem[\protect\citeauthoryear{{Aschwanden}, {Stern}  \&
  {G{\"u}del}}{{Aschwanden} et~al.}{2008}]{asc08}
{Aschwanden} M.~J.,  {Stern} R.~A.,   {G{\"u}del} M.,  2008, \mn@doi [\apj]
  {10.1086/523926}, \href {http://adsabs.harvard.edu/abs/2008ApJ...672..659A}
  {672, 659}

\bibitem[\protect\citeauthoryear{{Balona} \& {Laney}}{{Balona} \&
  {Laney}}{1996}]{bal96}
{Balona} L.~A.,  {Laney} C.~D.,  1996, \mnras, \href
  {http://adsabs.harvard.edu/abs/1996MNRAS.281.1341B} {281, 1341}

\bibitem[\protect\citeauthoryear{{Balona}, {Krisciunas}  \& {Cousins}}{{Balona}
  et~al.}{1994}]{bal94}
{Balona} L.~A.,  {Krisciunas} K.,   {Cousins} A.~W.~J.,  1994, \mnras, \href
  {http://adsabs.harvard.edu/abs/1994MNRAS.270..905B} {270, 905}

\bibitem[\protect\citeauthoryear{{Baraffe}, {Homeier}, {Allard}  \&
  {Chabrier}}{{Baraffe} et~al.}{2015}]{bar15}
{Baraffe} I.,  {Homeier} D.,  {Allard} F.,   {Chabrier} G.,  2015, \mn@doi
  [\aap] {10.1051/0004-6361/201425481}, \href
  {http://adsabs.harvard.edu/abs/2015A%26A...577A..42B} {577, A42}

\bibitem[\protect\citeauthoryear{{Batygin} \& {Adams}}{{Batygin} \&
  {Adams}}{2013}]{bat13}
{Batygin} K.,  {Adams} F.~C.,  2013, \mn@doi [\apj]
  {10.1088/0004-637X/778/2/169}, \href
  {http://adsabs.harvard.edu/abs/2013ApJ...778..169B} {778, 169}

\bibitem[\protect\citeauthoryear{{Baxter}, {Covey}, {Muench}, {F{\H
  u}r{\'e}sz}, {Rebull}  \& {Szentgyorgyi}}{{Baxter} et~al.}{2009}]{bax09}
{Baxter} E.~J.,  {Covey} K.~R.,  {Muench} A.~A.,  {F{\H u}r{\'e}sz} G.,
  {Rebull} L.,   {Szentgyorgyi} A.~H.,  2009, \mn@doi [\aj]
  {10.1088/0004-6256/138/3/963}, \href
  {http://adsabs.harvard.edu/abs/2009AJ....138..963B} {138, 963}

\bibitem[\protect\citeauthoryear{{Bessell}}{{Bessell}}{1979}]{bes79}
{Bessell} M.~S.,  1979, \mn@doi [\pasp] {10.1086/130542}, \href
  {http://adsabs.harvard.edu/abs/1979PASP...91..589B} {91, 589}

\bibitem[\protect\citeauthoryear{{Bodenheimer}}{{Bodenheimer}}{2011}]{bod11}
{Bodenheimer} P.~H.,  2011, {Principles of Star Formation}

\bibitem[\protect\citeauthoryear{{Broos} et~al.,}{{Broos} et~al.}{2013}]{bro13}
{Broos} P.~S.,  et~al., 2013, \mn@doi [\apjs] {10.1088/0067-0049/209/2/32},
  \href {http://adsabs.harvard.edu/abs/2013ApJS..209...32B} {209, 32}

\bibitem[\protect\citeauthoryear{{Cargile}, {Stassun}  \& {Mathieu}}{{Cargile}
  et~al.}{2008}]{car08}
{Cargile} P.~A.,  {Stassun} K.~G.,   {Mathieu} R.~D.,  2008, \mn@doi [\apj]
  {10.1086/524346}, \href {http://adsabs.harvard.edu/abs/2008ApJ...674..329C}
  {674, 329}

\bibitem[\protect\citeauthoryear{{Chabrier} \& {Baraffe}}{{Chabrier} \&
  {Baraffe}}{1997}]{cha97}
{Chabrier} G.,  {Baraffe} I.,  1997, \aap, \href
  {http://adsabs.harvard.edu/abs/1997A%26A...327.1039C} {327, 1039}

\bibitem[\protect\citeauthoryear{{Chambers}, {Cleveland}, {Kleiner}  \&
  {Tukey}}{{Chambers} et~al.}{1983}]{cha83}
{Chambers} J.~M.,  {Cleveland} W.~S.,  {Kleiner} B.,   {Tukey} P.~A.,  1983,
  {Graphical Methods for Data Analysis}

\bibitem[\protect\citeauthoryear{{Cody} et~al.,}{{Cody} et~al.}{2014}]{cod14}
{Cody} A.~M.,  et~al., 2014, \mn@doi [\aj] {10.1088/0004-6256/147/4/82}, \href
  {http://adsabs.harvard.edu/abs/2014AJ....147...82C} {147, 82}

\bibitem[\protect\citeauthoryear{{Cohen} \& {Kuhi}}{{Cohen} \&
  {Kuhi}}{1979}]{coh79}
{Cohen} M.,  {Kuhi} L.~V.,  1979, \mn@doi [\apjs] {10.1086/190641}, \href
  {http://adsabs.harvard.edu/abs/1979ApJS...41..743C} {41, 743}

\bibitem[\protect\citeauthoryear{{Correia} et~al.,}{{Correia}
  et~al.}{2013}]{cor13}
{Correia} S.,  et~al., 2013, \mn@doi [\aap] {10.1051/0004-6361/201220681},
  \href {http://adsabs.harvard.edu/abs/2013A%26A...557A..63C} {557, A63}

\bibitem[\protect\citeauthoryear{{Currie}, {Evans}, {Spitzbart}, {Irwin},
  {Wolk}, {Hernandez}, {Kenyon}  \& {Pasachoff}}{{Currie} et~al.}{2009}]{cur09}
{Currie} T.,  {Evans} N.~R.,  {Spitzbart} B.~D.,  {Irwin} J.,  {Wolk} S.~J.,
  {Hernandez} J.,  {Kenyon} S.~J.,   {Pasachoff} J.~M.,  2009, \mn@doi [\aj]
  {10.1088/0004-6256/137/2/3210}, \href
  {http://adsabs.harvard.edu/abs/2009AJ....137.3210C} {137, 3210}

\bibitem[\protect\citeauthoryear{{Cutri} et~al.,}{{Cutri} et~al.}{2003}]{cut03}
{Cutri} R.~M.,  et~al., 2003, VizieR Online Data Catalog, \href
  {http://adsabs.harvard.edu/abs/2003yCat.2246....0C} {2246, 0}

\bibitem[\protect\citeauthoryear{{Cvetkovic}, {Vince}  \&
  {Ninkovic}}{{Cvetkovic} et~al.}{2009}]{cve09}
{Cvetkovic} Z.,  {Vince} I.,   {Ninkovic} S.,  2009, Publications de
  l'Observatoire Astronomique de Beograd, \href
  {http://adsabs.harvard.edu/abs/2009POBeo..86..331C} {86, 331}

\bibitem[\protect\citeauthoryear{{Daemgen}, {Correia}  \&
  {Petr-Gotzens}}{{Daemgen} et~al.}{2012}]{dae12}
{Daemgen} S.,  {Correia} S.,   {Petr-Gotzens} M.~G.,  2012, \mn@doi [\aap]
  {10.1051/0004-6361/201118314}, \href
  {http://adsabs.harvard.edu/abs/2012A%26A...540A..46D} {540, A46}

\bibitem[\protect\citeauthoryear{{Dahm}}{{Dahm}}{2005}]{dah05b}
{Dahm} S.~E.,  2005, \mn@doi [\aj] {10.1086/433178}, \href
  {http://adsabs.harvard.edu/abs/2005AJ....130.1805D} {130, 1805}

\bibitem[\protect\citeauthoryear{{Dahm}}{{Dahm}}{2008a}]{dah08c}
{Dahm} S.~E.,  2008a, {NGC 2362: The Terminus of Star Formation}.
p.~26

\bibitem[\protect\citeauthoryear{{Dahm}}{{Dahm}}{2008b}]{dah08}
{Dahm} S.~E.,  2008b, {The Young Cluster and Star Forming Region NGC 2264}.
p.~966

\bibitem[\protect\citeauthoryear{{Dahm}}{{Dahm}}{2008c}]{dah08b}
{Dahm} S.~E.,  2008c, \mn@doi [\aj] {10.1088/0004-6256/136/2/521}, \href
  {http://adsabs.harvard.edu/abs/2008AJ....136..521D} {136, 521}

\bibitem[\protect\citeauthoryear{{Dahm} \& {Hillenbrand}}{{Dahm} \&
  {Hillenbrand}}{2007}]{dah07b}
{Dahm} S.~E.,  {Hillenbrand} L.~A.,  2007, \mn@doi [\aj] {10.1086/512156},
  \href {http://adsabs.harvard.edu/abs/2007AJ....133.2072D} {133, 2072}

\bibitem[\protect\citeauthoryear{{Dahm} \& {Simon}}{{Dahm} \&
  {Simon}}{2005}]{dah05}
{Dahm} S.~E.,  {Simon} T.,  2005, \mn@doi [\aj] {10.1086/426326}, \href
  {http://adsabs.harvard.edu/abs/2005AJ....129..829D} {129, 829}

\bibitem[\protect\citeauthoryear{{Dahm}, {Simon}, {Proszkow}  \&
  {Patten}}{{Dahm} et~al.}{2007}]{dah07}
{Dahm} S.~E.,  {Simon} T.,  {Proszkow} E.~M.,   {Patten} B.~M.,  2007, \mn@doi
  [\aj] {10.1086/519954}, \href
  {http://adsabs.harvard.edu/abs/2007AJ....134..999D} {134, 999}

\bibitem[\protect\citeauthoryear{{Damiani}, {Flaccomio}, {Micela}, {Sciortino},
  {Harnden}  \& {Murray}}{{Damiani} et~al.}{2004}]{dam04}
{Damiani} F.,  {Flaccomio} E.,  {Micela} G.,  {Sciortino} S.,  {Harnden} Jr.
  F.~R.,   {Murray} S.~S.,  2004, \mn@doi [\apj] {10.1086/420779}, \href
  {http://adsabs.harvard.edu/abs/2004ApJ...608..781D} {608, 781}

\bibitem[\protect\citeauthoryear{{Damiani}, {Micela}, {Sciortino},
  {Hu{\'e}lamo}, {Moitinho}, {Harnden}  \& {Murray}}{{Damiani}
  et~al.}{2006}]{dam06}
{Damiani} F.,  {Micela} G.,  {Sciortino} S.,  {Hu{\'e}lamo} N.,  {Moitinho} A.,
   {Harnden} Jr. F.~R.,   {Murray} S.~S.,  2006, \mn@doi [\aap]
  {10.1051/0004-6361:20065011}, \href
  {http://adsabs.harvard.edu/abs/2006A%26A...460..133D} {460, 133}

\bibitem[\protect\citeauthoryear{{Davies}, {Gregory}  \& {Greaves}}{{Davies}
  et~al.}{2014}]{dav14}
{Davies} C.~L.,  {Gregory} S.~G.,   {Greaves} J.~S.,  2014, \mn@doi [\mnras]
  {10.1093/mnras/stu1488}, \href
  {http://adsabs.harvard.edu/abs/2014MNRAS.444.1157D} {444, 1157}

\bibitem[\protect\citeauthoryear{{Delgado}, {Gonz{\'a}lez-Mart{\'{\i}}n},
  {Alfaro}  \& {Yun}}{{Delgado} et~al.}{2006}]{del06}
{Delgado} A.~J.,  {Gonz{\'a}lez-Mart{\'{\i}}n} O.,  {Alfaro} E.~J.,   {Yun} J.,
   2006, \mn@doi [\apj] {10.1086/504828}, \href
  {http://adsabs.harvard.edu/abs/2006ApJ...646..269D} {646, 269}

\bibitem[\protect\citeauthoryear{{Donati} et~al.,}{{Donati}
  et~al.}{2008a}]{don08bptau}
{Donati} J.-F.,  et~al., 2008a, \mn@doi [\mnras]
  {10.1111/j.1365-2966.2008.13111.x}, \href
  {http://ukads.nottingham.ac.uk/abs/2008MNRAS.386.1234D} {386, 1234}

\bibitem[\protect\citeauthoryear{{Donati} et~al.,}{{Donati}
  et~al.}{2008b}]{don08}
{Donati} J.-F.,  et~al., 2008b, \mn@doi [\mnras]
  {10.1111/j.1365-2966.2008.13799.x}, \href
  {http://adsabs.harvard.edu/abs/2008MNRAS.390..545D} {390, 545}

\bibitem[\protect\citeauthoryear{{Donati} et~al.,}{{Donati}
  et~al.}{2011}]{don11}
{Donati} J.-F.,  et~al., 2011, \mn@doi [\mnras]
  {10.1111/j.1365-2966.2011.19288.x}, \href
  {http://adsabs.harvard.edu/abs/2011MNRAS.417..472D} {417, 472}

\bibitem[\protect\citeauthoryear{{Donati} et~al.,}{{Donati}
  et~al.}{2012}]{don12}
{Donati} J.-F.,  et~al., 2012, \mn@doi [\mnras]
  {10.1111/j.1365-2966.2012.21482.x}, \href
  {http://adsabs.harvard.edu/abs/2012MNRAS.425.2948D} {425, 2948}

\bibitem[\protect\citeauthoryear{{Duch{\^e}ne}, {Bouvier}  \&
  {Simon}}{{Duch{\^e}ne} et~al.}{1999}]{duc99}
{Duch{\^e}ne} G.,  {Bouvier} J.,   {Simon} T.,  1999, \aap, \href
  {http://adsabs.harvard.edu/abs/1999A%26A...343..831D} {343, 831}

\bibitem[\protect\citeauthoryear{{Duncan}}{{Duncan}}{1993}]{dun93}
{Duncan} D.~K.,  1993, \mn@doi [\apj] {10.1086/172428}, \href
  {http://adsabs.harvard.edu/abs/1993ApJ...406..172D} {406, 172}

\bibitem[\protect\citeauthoryear{{Dunham} \& {Vorobyov}}{{Dunham} \&
  {Vorobyov}}{2012}]{dun12}
{Dunham} M.~M.,  {Vorobyov} E.~I.,  2012, \mn@doi [\apj]
  {10.1088/0004-637X/747/1/52}, \href
  {http://adsabs.harvard.edu/abs/2012ApJ...747...52D} {747, 52}

\bibitem[\protect\citeauthoryear{{Dzib}, {Loinard}, {Rodr{\'{\i}}guez}  \&
  {Galli}}{{Dzib} et~al.}{2014}]{dzi14}
{Dzib} S.~A.,  {Loinard} L.,  {Rodr{\'{\i}}guez} L.~F.,   {Galli} P.,  2014,
  \mn@doi [\apj] {10.1088/0004-637X/788/2/162}, \href
  {http://adsabs.harvard.edu/abs/2014ApJ...788..162D} {788, 162}

\bibitem[\protect\citeauthoryear{{Edwards} et~al.,}{{Edwards}
  et~al.}{1993}]{edw93}
{Edwards} S.,  et~al., 1993, \mn@doi [\aj] {10.1086/116646}, \href
  {http://adsabs.harvard.edu/abs/1993AJ....106..372E} {106, 372}

\bibitem[\protect\citeauthoryear{{F{\H u}r{\'e}sz} et~al.,}{{F{\H u}r{\'e}sz}
  et~al.}{2006}]{fur06}
{F{\H u}r{\'e}sz} G.,  et~al., 2006, \mn@doi [\apj] {10.1086/506140}, \href
  {http://adsabs.harvard.edu/abs/2006ApJ...648.1090F} {648, 1090}

\bibitem[\protect\citeauthoryear{{F{\H u}r{\'e}sz}, {Hartmann}, {Megeath},
  {Szentgyorgyi}  \& {Hamden}}{{F{\H u}r{\'e}sz} et~al.}{2008}]{fur08}
{F{\H u}r{\'e}sz} G.,  {Hartmann} L.~W.,  {Megeath} S.~T.,  {Szentgyorgyi}
  A.~H.,   {Hamden} E.~T.,  2008, \mn@doi [\apj] {10.1086/525844}, \href
  {http://adsabs.harvard.edu/abs/2008ApJ...676.1109F} {676, 1109}

\bibitem[\protect\citeauthoryear{{Favata}, {Flaccomio}, {Reale}, {Micela},
  {Sciortino}, {Shang}, {Stassun}  \& {Feigelson}}{{Favata}
  et~al.}{2005}]{fav05}
{Favata} F.,  {Flaccomio} E.,  {Reale} F.,  {Micela} G.,  {Sciortino} S.,
  {Shang} H.,  {Stassun} K.~G.,   {Feigelson} E.~D.,  2005, \mn@doi [\apjs]
  {10.1086/432542}, \href {http://adsabs.harvard.edu/abs/2005ApJS..160..469F}
  {160, 469}

\bibitem[\protect\citeauthoryear{{Feigelson} \& {Nelson}}{{Feigelson} \&
  {Nelson}}{1985}]{fei85}
{Feigelson} E.~D.,  {Nelson} P.~I.,  1985, \mn@doi [\apj] {10.1086/163225},
  \href {http://adsabs.harvard.edu/abs/1985ApJ...293..192F} {293, 192}

\bibitem[\protect\citeauthoryear{{Feigelson}, {Gaffney}, {Garmire},
  {Hillenbrand}  \& {Townsley}}{{Feigelson} et~al.}{2003}]{fei03}
{Feigelson} E.~D.,  {Gaffney} III J.~A.,  {Garmire} G.,  {Hillenbrand} L.~A.,
  {Townsley} L.,  2003, \mn@doi [\apj] {10.1086/345811}, \href
  {http://adsabs.harvard.edu/abs/2003ApJ...584..911F} {584, 911}

\bibitem[\protect\citeauthoryear{{Feigelson} et~al.,}{{Feigelson}
  et~al.}{2005}]{fei05}
{Feigelson} E.~D.,  et~al., 2005, \mn@doi [\apjs] {10.1086/432512}, \href
  {http://adsabs.harvard.edu/abs/2005ApJS..160..379F} {160, 379}

\bibitem[\protect\citeauthoryear{{Feigelson} et~al.,}{{Feigelson}
  et~al.}{2013}]{fei13}
{Feigelson} E.~D.,  et~al., 2013, \mn@doi [\apjs] {10.1088/0067-0049/209/2/26},
  \href {http://adsabs.harvard.edu/abs/2013ApJS..209...26F} {209, 26}

\bibitem[\protect\citeauthoryear{{Flaccomio}, {Micela}  \&
  {Sciortino}}{{Flaccomio} et~al.}{2003a}]{fla03c}
{Flaccomio} E.,  {Micela} G.,   {Sciortino} S.,  2003a, \mn@doi [\aap]
  {10.1051/0004-6361:20021484}, \href
  {http://ukads.nottingham.ac.uk/abs/2003A%26A...397..611F} {397, 611}

\bibitem[\protect\citeauthoryear{{Flaccomio}, {Micela}  \&
  {Sciortino}}{{Flaccomio} et~al.}{2003b}]{fla03b}
{Flaccomio} E.,  {Micela} G.,   {Sciortino} S.,  2003b, \mn@doi [\aap]
  {10.1051/0004-6361:20030203}, \href
  {http://adsabs.harvard.edu/abs/2003A%26A...402..277F} {402, 277}

\bibitem[\protect\citeauthoryear{{Flaccomio}, {Damiani}, {Micela}, {Sciortino},
  {Harnden}, {Murray}  \& {Wolk}}{{Flaccomio} et~al.}{2003c}]{fla03}
{Flaccomio} E.,  {Damiani} F.,  {Micela} G.,  {Sciortino} S.,  {Harnden} Jr.
  F.~R.,  {Murray} S.~S.,   {Wolk} S.~J.,  2003c, \mn@doi [\apj]
  {10.1086/344536}, \href {http://adsabs.harvard.edu/abs/2003ApJ...582..398F}
  {582, 398}

\bibitem[\protect\citeauthoryear{{Flaccomio}, {Micela}  \&
  {Sciortino}}{{Flaccomio} et~al.}{2006}]{fla06}
{Flaccomio} E.,  {Micela} G.,   {Sciortino} S.,  2006, \mn@doi [\aap]
  {10.1051/0004-6361:20065084}, \href
  {http://adsabs.harvard.edu/abs/2006A%26A...455..903F} {455, 903}

\bibitem[\protect\citeauthoryear{{Forbrich}, {Osten}  \& {Wolk}}{{Forbrich}
  et~al.}{2011}]{for11}
{Forbrich} J.,  {Osten} R.~A.,   {Wolk} S.~J.,  2011, \mn@doi [\apj]
  {10.1088/0004-637X/736/1/25}, \href
  {http://adsabs.harvard.edu/abs/2011ApJ...736...25F} {736, 25}

\bibitem[\protect\citeauthoryear{{Getman} et~al.,}{{Getman}
  et~al.}{2005}]{get05}
{Getman} K.~V.,  et~al., 2005, \mn@doi [\apjs] {10.1086/432092}, \href
  {http://adsabs.harvard.edu/abs/2005ApJS..160..319G} {160, 319}

\bibitem[\protect\citeauthoryear{{Getman}, {Feigelson}, {Micela}, {Jardine},
  {Gregory}  \& {Garmire}}{{Getman} et~al.}{2008}]{get08}
{Getman} K.~V.,  {Feigelson} E.~D.,  {Micela} G.,  {Jardine} M.~M.,  {Gregory}
  S.~G.,   {Garmire} G.~P.,  2008, \mn@doi [\apj] {10.1086/592034}, \href
  {http://adsabs.harvard.edu/abs/2008ApJ...688..437G} {688, 437}

\bibitem[\protect\citeauthoryear{{Getman}, {Feigelson}, {Broos}, {Townsley}  \&
  {Garmire}}{{Getman} et~al.}{2010}]{get10}
{Getman} K.~V.,  {Feigelson} E.~D.,  {Broos} P.~S.,  {Townsley} L.~K.,
  {Garmire} G.~P.,  2010, \mn@doi [\apj] {10.1088/0004-637X/708/2/1760}, \href
  {http://adsabs.harvard.edu/abs/2010ApJ...708.1760G} {708, 1760}

\bibitem[\protect\citeauthoryear{{Gillen} et~al.,}{{Gillen}
  et~al.}{2014}]{gil14}
{Gillen} E.,  et~al., 2014, \mn@doi [\aap] {10.1051/0004-6361/201322493}, \href
  {http://adsabs.harvard.edu/abs/2014A%26A...562A..50G} {562, A50}

\bibitem[\protect\citeauthoryear{{Greenstein} \& {Struve}}{{Greenstein} \&
  {Struve}}{1946}]{gre46}
{Greenstein} J.~L.,  {Struve} O.,  1946, \mn@doi [\pasp] {10.1086/125874},
  \href {http://adsabs.harvard.edu/abs/1946PASP...58..366G} {58, 366}

\bibitem[\protect\citeauthoryear{{Gregory} \& {Donati}}{{Gregory} \&
  {Donati}}{2011}]{gre11}
{Gregory} S.~G.,  {Donati} J.-F.,  2011, \mn@doi [Astronomische Nachrichten]
  {10.1002/asna.201111621}, \href
  {http://adsabs.harvard.edu/abs/2011AN....332.1027G} {332, 1027}

\bibitem[\protect\citeauthoryear{{Gregory}, {Jardine}, {Collier Cameron}  \&
  {Donati}}{{Gregory} et~al.}{2006}]{gre06}
{Gregory} S.~G.,  {Jardine} M.,  {Collier Cameron} A.,   {Donati} J.-F.,  2006,
  \mn@doi [\mnras] {10.1111/j.1365-2966.2006.11086.x}, \href
  {http://ukads.nottingham.ac.uk/abs/2006MNRAS.373..827G} {373, 827}

\bibitem[\protect\citeauthoryear{{Gregory}, {Donati}, {Morin}, {Hussain},
  {Mayne}, {Hillenbrand}  \& {Jardine}}{{Gregory} et~al.}{2012}]{gre12}
{Gregory} S.~G.,  {Donati} J.-F.,  {Morin} J.,  {Hussain} G.~A.~J.,  {Mayne}
  N.~J.,  {Hillenbrand} L.~A.,   {Jardine} M.,  2012, \mn@doi [\apj]
  {10.1088/0004-637X/755/2/97}, \href
  {http://adsabs.harvard.edu/abs/2012ApJ...755...97G} {755, 97}

\bibitem[\protect\citeauthoryear{{Gregory}, {Donati}, {Morin}, {Hussain},
  {Mayne}, {Hillenbrand}  \& {Jardine}}{{Gregory} et~al.}{2014}]{gre14}
{Gregory} S.~G.,  {Donati} J.-F.,  {Morin} J.,  {Hussain} G.~A.~J.,  {Mayne}
  N.~J.,  {Hillenbrand} L.~A.,   {Jardine} M.,  2014, in IAU Symposium. pp
  40--43 (\mn@eprint {arXiv} {1309.7556}), \mn@doi{10.1017/S1743921314001677}

\bibitem[\protect\citeauthoryear{{G{\"u}del}}{{G{\"u}del}}{2009}]{gue09}
{G{\"u}del} M.,  2009, in {Cargill} P.,  {Vlahos} L.,  eds,  Lecture Notes in
  Physics, Berlin Springer Verlag Vol. 778, Turbulence in Space Plasmas.
  p.~269, \mn@doi{10.1007/978-3-642-00210-6_8}

\bibitem[\protect\citeauthoryear{{Harris}, {Morgan}  \& {Roman}}{{Harris}
  et~al.}{1954}]{har54}
{Harris} D.~L.,  {Morgan} W.~W.,   {Roman} N.~G.,  1954, \mn@doi [\apj]
  {10.1086/145865}, \href {http://cdsads.u-strasbg.fr/abs/1954ApJ...119..622H}
  {119, 622}

\bibitem[\protect\citeauthoryear{{Henderson} \& {Stassun}}{{Henderson} \&
  {Stassun}}{2012}]{hen12}
{Henderson} C.~B.,  {Stassun} K.~G.,  2012, \mn@doi [\apj]
  {10.1088/0004-637X/747/1/51}, \href
  {http://adsabs.harvard.edu/abs/2012ApJ...747...51H} {747, 51}

\bibitem[\protect\citeauthoryear{{Herbig}}{{Herbig}}{1978}]{her78}
{Herbig} G.~H.,  1978, {Can Post-T Tauri Stars Be Found?}.
p.~171

\bibitem[\protect\citeauthoryear{{Herbig}}{{Herbig}}{1998}]{her98}
{Herbig} G.~H.,  1998, \mn@doi [\apj] {10.1086/305500}, \href
  {http://adsabs.harvard.edu/abs/1998ApJ...497..736H} {497, 736}

\bibitem[\protect\citeauthoryear{{Herbig} \& {Bell}}{{Herbig} \&
  {Bell}}{1988}]{her88}
{Herbig} G.~H.,  {Bell} K.~R.,  1988, {Third Catalog of Emission-Line Stars of
  the Orion Population : 3 : 1988}

\bibitem[\protect\citeauthoryear{{Herbig} \& {Terndrup}}{{Herbig} \&
  {Terndrup}}{1986}]{her86}
{Herbig} G.~H.,  {Terndrup} D.~M.,  1986, \mn@doi [\apj] {10.1086/164447},
  \href {http://cdsads.u-strasbg.fr/abs/1986ApJ...307..609H} {307, 609}

\bibitem[\protect\citeauthoryear{{Herbst}}{{Herbst}}{2008}]{her08}
{Herbst} W.,  2008, {Star Formation in IC 348}.
p.~372

\bibitem[\protect\citeauthoryear{{Herczeg} \& {Hillenbrand}}{{Herczeg} \&
  {Hillenbrand}}{2014}]{her14}
{Herczeg} G.~J.,  {Hillenbrand} L.~A.,  2014, \mn@doi [\apj]
  {10.1088/0004-637X/786/2/97}, \href
  {http://adsabs.harvard.edu/abs/2014ApJ...786...97H} {786, 97}

\bibitem[\protect\citeauthoryear{{Herczeg} \& {Hillenbrand}}{{Herczeg} \&
  {Hillenbrand}}{2015}]{her15}
{Herczeg} G.~J.,  {Hillenbrand} L.~A.,  2015, \mn@doi [\apj]
  {10.1088/0004-637X/808/1/23}, \href
  {http://adsabs.harvard.edu/abs/2015ApJ...808...23H} {808, 23}

\bibitem[\protect\citeauthoryear{{Hillenbrand}}{{Hillenbrand}}{1997}]{hil97}
{Hillenbrand} L.~A.,  1997, \mn@doi [\aj] {10.1086/118389}, \href
  {http://adsabs.harvard.edu/abs/1997AJ....113.1733H} {113, 1733}

\bibitem[\protect\citeauthoryear{{Hillenbrand}, {Bauermeister}  \&
  {White}}{{Hillenbrand} et~al.}{2008}]{hil08}
{Hillenbrand} L.~A.,  {Bauermeister} A.,   {White} R.~J.,  2008, in {van Belle}
  G.,  ed.,  Astronomical Society of the Pacific Conference Series Vol. 384,
  14th Cambridge Workshop on Cool Stars, Stellar Systems, and the Sun. p.~200
  (\mn@eprint {} {astro-ph/0703642})

\bibitem[\protect\citeauthoryear{{Hillenbrand}, {Hoffer}  \&
  {Herczeg}}{{Hillenbrand} et~al.}{2013}]{hil13}
{Hillenbrand} L.~A.,  {Hoffer} A.~S.,   {Herczeg} G.~J.,  2013, \mn@doi [\aj]
  {10.1088/0004-6256/146/4/85}, \href
  {http://adsabs.harvard.edu/abs/2013AJ....146...85H} {146, 85}

\bibitem[\protect\citeauthoryear{{Hirota} et~al.,}{{Hirota}
  et~al.}{2007}]{hir07}
{Hirota} T.,  et~al., 2007, \mn@doi [\pasj] {10.1093/pasj/59.5.897}, \href
  {http://adsabs.harvard.edu/abs/2007PASJ...59..897H} {59, 897}

\bibitem[\protect\citeauthoryear{{Hoff}, {Henning}  \& {Pfau}}{{Hoff}
  et~al.}{1998}]{hof98}
{Hoff} W.,  {Henning} T.,   {Pfau} W.,  1998, \aap, \href
  {http://adsabs.harvard.edu/abs/1998A%26A...336..242H} {336, 242}

\bibitem[\protect\citeauthoryear{{Hussain} \& {Alecian}}{{Hussain} \&
  {Alecian}}{2014}]{hus14}
{Hussain} G.~A.~J.,  {Alecian} E.,  2014, in {Petit} P.,  {Jardine} M.,
  {Spruit} H.~C.,  eds,  IAU Symposium Vol. 302, IAU Symposium. pp 25--37
  (\mn@eprint {arXiv} {1402.7130}), \mn@doi{10.1017/S1743921314001653}

\bibitem[\protect\citeauthoryear{{Hussain} et~al.,}{{Hussain}
  et~al.}{2009}]{hus09}
{Hussain} G.~A.~J.,  et~al., 2009, \mn@doi [\mnras]
  {10.1111/j.1365-2966.2009.14881.x}, \href
  {http://ukads.nottingham.ac.uk/abs/2009MNRAS.398..189H} {398, 189}

\bibitem[\protect\citeauthoryear{{Iben}}{{Iben}}{2013}]{ibe13}
{Iben} Jr. I.,  2013, {Stellar Evolution Physics, Volume 1: Physical Processes
  in Stellar Interiors}

\bibitem[\protect\citeauthoryear{{Isobe}, {Feigelson}  \& {Nelson}}{{Isobe}
  et~al.}{1986}]{iso86}
{Isobe} T.,  {Feigelson} E.~D.,   {Nelson} P.~I.,  1986, \mn@doi [\apj]
  {10.1086/164359}, \href {http://adsabs.harvard.edu/abs/1986ApJ...306..490I}
  {306, 490}

\bibitem[\protect\citeauthoryear{{Jeffries}}{{Jeffries}}{2007}]{jef07}
{Jeffries} R.~D.,  2007, \mn@doi [\mnras] {10.1111/j.1365-2966.2007.11471.x},
  \href {http://adsabs.harvard.edu/abs/2007MNRAS.376.1109J} {376, 1109}

\bibitem[\protect\citeauthoryear{{Johnson}}{{Johnson}}{1965}]{joh65}
{Johnson} H.~M.,  1965, \mn@doi [\apj] {10.1086/148365}, \href
  {http://adsabs.harvard.edu/abs/1965ApJ...142..964J} {142, 964}

\bibitem[\protect\citeauthoryear{{Johnstone}, {Jardine}, {Gregory}, {Donati}
  \& {Hussain}}{{Johnstone} et~al.}{2014}]{joh14}
{Johnstone} C.~P.,  {Jardine} M.,  {Gregory} S.~G.,  {Donati} J.-F.,
  {Hussain} G.,  2014, \mn@doi [\mnras] {10.1093/mnras/stt2107}, \href
  {http://ukads.nottingham.ac.uk/abs/2014MNRAS.437.3202J} {437, 3202}

\bibitem[\protect\citeauthoryear{{Jung} \& {Kim}}{{Jung} \&
  {Kim}}{2007}]{jun07}
{Jung} Y.~K.,  {Kim} Y.-C.,  2007, Journal of Astronomy and Space Sciences,
  \href {http://adsabs.harvard.edu/abs/2007JASS...24....1J} {24, 1}

\bibitem[\protect\citeauthoryear{{Kamezaki} et~al.,}{{Kamezaki}
  et~al.}{2014}]{kam14}
{Kamezaki} T.,  et~al., 2014, \mn@doi [\apjs] {10.1088/0067-0049/211/2/18},
  \href {http://adsabs.harvard.edu/abs/2014ApJS..211...18K} {211, 18}

\bibitem[\protect\citeauthoryear{{Karnath}, {Prato}, {Wasserman}, {Torres},
  {Skiff}  \& {Mathieu}}{{Karnath} et~al.}{2013}]{kar13}
{Karnath} N.,  {Prato} L.,  {Wasserman} L.~H.,  {Torres} G.,  {Skiff} B.~A.,
  {Mathieu} R.~D.,  2013, \mn@doi [\aj] {10.1088/0004-6256/146/6/149}, \href
  {http://adsabs.harvard.edu/abs/2013AJ....146..149K} {146, 149}

\bibitem[\protect\citeauthoryear{{Kaye}, {Handler}, {Krisciunas}, {Poretti}  \&
  {Zerbi}}{{Kaye} et~al.}{1999}]{kay99}
{Kaye} A.~B.,  {Handler} G.,  {Krisciunas} K.,  {Poretti} E.,   {Zerbi} F.~M.,
  1999, \mn@doi [\pasp] {10.1086/316399}, \href
  {http://adsabs.harvard.edu/abs/1999PASP..111..840K} {111, 840}

\bibitem[\protect\citeauthoryear{{Kenyon} \& {Hartmann}}{{Kenyon} \&
  {Hartmann}}{1995}]{ken95}
{Kenyon} S.~J.,  {Hartmann} L.,  1995, \mn@doi [\apjs] {10.1086/192235}, \href
  {http://adsabs.harvard.edu/abs/1995ApJS..101..117K} {101, 117}

\bibitem[\protect\citeauthoryear{{Kharchenko}, {Piskunov}, {Schilbach},
  {R{\"o}ser}  \& {Scholz}}{{Kharchenko} et~al.}{2013}]{kha13}
{Kharchenko} N.~V.,  {Piskunov} A.~E.,  {Schilbach} E.,  {R{\"o}ser} S.,
  {Scholz} R.-D.,  2013, \mn@doi [\aap] {10.1051/0004-6361/201322302}, \href
  {http://cdsads.u-strasbg.fr/abs/2013A%26A...558A..53K} {558, A53}

\bibitem[\protect\citeauthoryear{{Kim} et~al.,}{{Kim} et~al.}{2008}]{kim08}
{Kim} M.~K.,  et~al., 2008, \mn@doi [\pasj] {10.1093/pasj/60.5.991}, \href
  {http://adsabs.harvard.edu/abs/2008PASJ...60..991K} {60, 991}

\bibitem[\protect\citeauthoryear{{Klagyivik} et~al.,}{{Klagyivik}
  et~al.}{2013}]{kla13}
{Klagyivik} P.,  et~al., 2013, \mn@doi [\apj] {10.1088/0004-637X/773/1/54},
  \href {http://adsabs.harvard.edu/abs/2013ApJ...773...54K} {773, 54}

\bibitem[\protect\citeauthoryear{{Klimchuk} \& {Bradshaw}}{{Klimchuk} \&
  {Bradshaw}}{2014}]{kli14}
{Klimchuk} J.~A.,  {Bradshaw} S.~J.,  2014, \mn@doi [\apj]
  {10.1088/0004-637X/791/1/60}, \href
  {http://adsabs.harvard.edu/abs/2014ApJ...791...60K} {791, 60}

\bibitem[\protect\citeauthoryear{{Koch} \& {Perry}}{{Koch} \&
  {Perry}}{1974}]{koc74}
{Koch} R.~H.,  {Perry} P.~M.,  1974, \mn@doi [\aj] {10.1086/111554}, \href
  {http://adsabs.harvard.edu/abs/1974AJ.....79..379K} {79, 379}

\bibitem[\protect\citeauthoryear{{Kraus} et~al.,}{{Kraus} et~al.}{2007}]{kra07}
{Kraus} S.,  et~al., 2007, \mn@doi [\aap] {10.1051/0004-6361:20066965}, \href
  {http://adsabs.harvard.edu/abs/2007A%26A...466..649K} {466, 649}

\bibitem[\protect\citeauthoryear{{Kumar}, {Sagar}, {Sanwal}  \&
  {Bessell}}{{Kumar} et~al.}{2004}]{kum04}
{Kumar} B.,  {Sagar} R.,  {Sanwal} B.~B.,   {Bessell} M.~S.,  2004, \mn@doi
  [\mnras] {10.1111/j.1365-2966.2004.08130.x}, \href
  {http://adsabs.harvard.edu/abs/2004MNRAS.353..991K} {353, 991}

\bibitem[\protect\citeauthoryear{{Lamm}, {Bailer-Jones}, {Mundt}, {Herbst}  \&
  {Scholz}}{{Lamm} et~al.}{2004}]{lam04}
{Lamm} M.~H.,  {Bailer-Jones} C.~A.~L.,  {Mundt} R.,  {Herbst} W.,   {Scholz}
  A.,  2004, \mn@doi [\aap] {10.1051/0004-6361:20035588}, \href
  {http://adsabs.harvard.edu/abs/2004A%26A...417..557L} {417, 557}

\bibitem[\protect\citeauthoryear{{Lamm}, {Mundt}, {Bailer-Jones}  \&
  {Herbst}}{{Lamm} et~al.}{2005}]{lam05}
{Lamm} M.~H.,  {Mundt} R.,  {Bailer-Jones} C.~A.~L.,   {Herbst} W.,  2005,
  \mn@doi [\aap] {10.1051/0004-6361:20040492}, \href
  {http://adsabs.harvard.edu/abs/2005A%26A...430.1005L} {430, 1005}

\bibitem[\protect\citeauthoryear{{Laughlin}, {Bodenheimer}  \&
  {Adams}}{{Laughlin} et~al.}{1997}]{lau97}
{Laughlin} G.,  {Bodenheimer} P.,   {Adams} F.~C.,  1997, \apj, \href
  {http://adsabs.harvard.edu/abs/1997ApJ...482..420L} {482, 420}

\bibitem[\protect\citeauthoryear{{Lavalley}, {Isobe}  \&
  {Feigelson}}{{Lavalley} et~al.}{1992}]{lav92}
{Lavalley} M.~P.,  {Isobe} T.,   {Feigelson} E.~D.,  1992, in Bulletin of the
  American Astronomical Society. pp 839--840

\bibitem[\protect\citeauthoryear{{Lucas}, {Roche}, {Allard}  \&
  {Hauschildt}}{{Lucas} et~al.}{2001}]{luc01}
{Lucas} P.~W.,  {Roche} P.~F.,  {Allard} F.,   {Hauschildt} P.~H.,  2001,
  \mn@doi [\mnras] {10.1046/j.1365-8711.2001.04666.x}, \href
  {http://adsabs.harvard.edu/abs/2001MNRAS.326..695L} {326, 695}

\bibitem[\protect\citeauthoryear{{Luhman}}{{Luhman}}{1999}]{luh99}
{Luhman} K.~L.,  1999, \mn@doi [\apj] {10.1086/307902}, \href
  {http://cdsads.u-strasbg.fr/abs/1999ApJ...525..466L} {525, 466}

\bibitem[\protect\citeauthoryear{{Luhman}, {Rieke}, {Lada}  \& {Lada}}{{Luhman}
  et~al.}{1998}]{luh98}
{Luhman} K.~L.,  {Rieke} G.~H.,  {Lada} C.~J.,   {Lada} E.~A.,  1998, \mn@doi
  [\apj] {10.1086/306393}, \href
  {http://cdsads.u-strasbg.fr/abs/1998ApJ...508..347L} {508, 347}

\bibitem[\protect\citeauthoryear{{Luhman}, {Rieke}, {Young}, {Cotera}, {Chen},
  {Rieke}, {Schneider}  \& {Thompson}}{{Luhman} et~al.}{2000}]{luh00}
{Luhman} K.~L.,  {Rieke} G.~H.,  {Young} E.~T.,  {Cotera} A.~S.,  {Chen} H.,
  {Rieke} M.~J.,  {Schneider} G.,   {Thompson} R.~I.,  2000, \mn@doi [\apj]
  {10.1086/309365}, \href {http://adsabs.harvard.edu/abs/2000ApJ...540.1016L}
  {540, 1016}

\bibitem[\protect\citeauthoryear{{Luhman}, {Stauffer}, {Muench}, {Rieke},
  {Lada}, {Bouvier}  \& {Lada}}{{Luhman} et~al.}{2003}]{luh03a}
{Luhman} K.~L.,  {Stauffer} J.~R.,  {Muench} A.~A.,  {Rieke} G.~H.,  {Lada}
  E.~A.,  {Bouvier} J.,   {Lada} C.~J.,  2003, \mn@doi [\apj] {10.1086/376594},
  \href {http://adsabs.harvard.edu/abs/2003ApJ...593.1093L} {593, 1093}

\bibitem[\protect\citeauthoryear{{Luhman}, {Lada}, {Muench}  \&
  {Elston}}{{Luhman} et~al.}{2005}]{luh05}
{Luhman} K.~L.,  {Lada} E.~A.,  {Muench} A.~A.,   {Elston} R.~J.,  2005,
  \mn@doi [\apj] {10.1086/426052}, \href
  {http://adsabs.harvard.edu/abs/2005ApJ...618..810L} {618, 810}

\bibitem[\protect\citeauthoryear{{Maeder}}{{Maeder}}{2009}]{mae09}
{Maeder} A.,  2009, {Physics, Formation and Evolution of Rotating Stars},
  \mn@doi{10.1007/978-3-540-76949-1.
}

\bibitem[\protect\citeauthoryear{{Mamajek}}{{Mamajek}}{2012}]{mam12}
{Mamajek} E.~E.,  2012, \mn@doi [\apjl] {10.1088/2041-8205/754/2/L20}, \href
  {http://adsabs.harvard.edu/abs/2012ApJ...754L..20M} {754, L20}

\bibitem[\protect\citeauthoryear{{Mari{\~n}as}, {Lada}, {Teixeira}  \&
  {Lada}}{{Mari{\~n}as} et~al.}{2013}]{mar13}
{Mari{\~n}as} N.,  {Lada} E.~A.,  {Teixeira} P.~S.,   {Lada} C.~J.,  2013,
  \mn@doi [\apj] {10.1088/0004-637X/772/2/81}, \href
  {http://adsabs.harvard.edu/abs/2013ApJ...772...81M} {772, 81}

\bibitem[\protect\citeauthoryear{{Marschall} \& {Mathieu}}{{Marschall} \&
  {Mathieu}}{1988}]{mar88}
{Marschall} L.~A.,  {Mathieu} R.~D.,  1988, \mn@doi [\aj] {10.1086/114942},
  \href {http://adsabs.harvard.edu/abs/1988AJ.....96.1956M} {96, 1956}

\bibitem[\protect\citeauthoryear{{Mathieu}}{{Mathieu}}{1994}]{mat94}
{Mathieu} R.~D.,  1994, \mn@doi [\araa] {10.1146/annurev.aa.32.090194.002341},
  \href {http://adsabs.harvard.edu/abs/1994ARA%26A..32..465M} {32, 465}

\bibitem[\protect\citeauthoryear{{Mayne}}{{Mayne}}{2010}]{may10}
{Mayne} N.~J.,  2010, \mn@doi [\mnras] {10.1111/j.1365-2966.2010.17245.x},
  \href {http://adsabs.harvard.edu/abs/2010MNRAS.408.1409M} {408, 1409}

\bibitem[\protect\citeauthoryear{{Mayne} \& {Naylor}}{{Mayne} \&
  {Naylor}}{2008}]{may08}
{Mayne} N.~J.,  {Naylor} T.,  2008, \mn@doi [\mnras]
  {10.1111/j.1365-2966.2008.13025.x}, \href
  {http://adsabs.harvard.edu/abs/2008MNRAS.386..261M} {386, 261}

\bibitem[\protect\citeauthoryear{{McGill}, {Tukey}  \& {Larsen}}{{McGill}
  et~al.}{1978}]{mcg78}
{McGill} R.,  {Tukey} J.~W.,   {Larsen} W.~A.,  1978, The American
  Statistician, 32, 12

\bibitem[\protect\citeauthoryear{{Menten}, {Reid}, {Forbrich}  \&
  {Brunthaler}}{{Menten} et~al.}{2007}]{men07}
{Menten} K.~M.,  {Reid} M.~J.,  {Forbrich} J.,   {Brunthaler} A.,  2007,
  \mn@doi [\aap] {10.1051/0004-6361:20078247}, \href
  {http://adsabs.harvard.edu/abs/2007A%26A...474..515M} {474, 515}

\bibitem[\protect\citeauthoryear{{Moitinho}, {Alves}, {Hu{\'e}lamo}  \&
  {Lada}}{{Moitinho} et~al.}{2001}]{moi01}
{Moitinho} A.,  {Alves} J.,  {Hu{\'e}lamo} N.,   {Lada} C.~J.,  2001, \mn@doi
  [\apjl] {10.1086/338503}, \href
  {http://adsabs.harvard.edu/abs/2001ApJ...563L..73M} {563, L73}

\bibitem[\protect\citeauthoryear{{Montmerle}, {Grosso}, {Tsuboi}  \&
  {Koyama}}{{Montmerle} et~al.}{2000}]{mon00}
{Montmerle} T.,  {Grosso} N.,  {Tsuboi} Y.,   {Koyama} K.,  2000, \mn@doi
  [\apj] {10.1086/308611}, \href
  {http://adsabs.harvard.edu/abs/2000ApJ...532.1097M} {532, 1097}

\bibitem[\protect\citeauthoryear{{Morales-Calder{\'o}n}
  et~al.,}{{Morales-Calder{\'o}n} et~al.}{2012}]{mor12}
{Morales-Calder{\'o}n} M.,  et~al., 2012, \mn@doi [\apj]
  {10.1088/0004-637X/753/2/149}, \href
  {http://adsabs.harvard.edu/abs/2012ApJ...753..149M} {753, 149}

\bibitem[\protect\citeauthoryear{{Morin} et~al.,}{{Morin} et~al.}{2008}]{mor08}
{Morin} J.,  et~al., 2008, \mn@doi [\mnras] {10.1111/j.1365-2966.2008.13809.x},
  \href {http://adsabs.harvard.edu/abs/2008MNRAS.390..567M} {390, 567}

\bibitem[\protect\citeauthoryear{{Muench}, {Lada}, {Luhman}, {Muzerolle}  \&
  {Young}}{{Muench} et~al.}{2007}]{mue07}
{Muench} A.~A.,  {Lada} C.~J.,  {Luhman} K.~L.,  {Muzerolle} J.,   {Young} E.,
  2007, \mn@doi [\aj] {10.1086/518560}, \href
  {http://adsabs.harvard.edu/abs/2007AJ....134..411M} {134, 411}

\bibitem[\protect\citeauthoryear{{Muench}, {Getman}, {Hillenbrand}  \&
  {Preibisch}}{{Muench} et~al.}{2008}]{mue08}
{Muench} A.,  {Getman} K.,  {Hillenbrand} L.,   {Preibisch} T.,  2008, {Star
  Formation in the Orion Nebula I: Stellar Content}.
p.~483

\bibitem[\protect\citeauthoryear{{Narain} \& {Ulmschneider}}{{Narain} \&
  {Ulmschneider}}{1996}]{nar96}
{Narain} U.,  {Ulmschneider} P.,  1996, \mn@doi [\ssr] {10.1007/BF00833341},
  \href {http://adsabs.harvard.edu/abs/1996SSRv...75..453N} {75, 453}

\bibitem[\protect\citeauthoryear{{Naylor}}{{Naylor}}{2009}]{nay09}
{Naylor} T.,  2009, \mn@doi [\mnras] {10.1111/j.1365-2966.2009.15295.x}, \href
  {http://adsabs.harvard.edu/abs/2009MNRAS.399..432N} {399, 432}

\bibitem[\protect\citeauthoryear{{Nordhagen}, {Herbst}, {Rhode}  \&
  {Williams}}{{Nordhagen} et~al.}{2006}]{nor06}
{Nordhagen} S.,  {Herbst} W.,  {Rhode} K.~L.,   {Williams} E.~C.,  2006,
  \mn@doi [\aj] {10.1086/506985}, \href
  {http://adsabs.harvard.edu/abs/2006AJ....132.1555N} {132, 1555}

\bibitem[\protect\citeauthoryear{{O'Dell}, {Muench}, {Smith}  \&
  {Zapata}}{{O'Dell} et~al.}{2008}]{ode08}
{O'Dell} C.~R.,  {Muench} A.,  {Smith} N.,   {Zapata} L.,  2008, {Star
  Formation in the Orion Nebula II: Gas, Dust, Proplyds and Outflows}.
p.~544

\bibitem[\protect\citeauthoryear{{Offner} \& {McKee}}{{Offner} \&
  {McKee}}{2011}]{off11}
{Offner} S.~S.~R.,  {McKee} C.~F.,  2011, \mn@doi [\apj]
  {10.1088/0004-637X/736/1/53}, \href
  {http://adsabs.harvard.edu/abs/2011ApJ...736...53O} {736, 53}

\bibitem[\protect\citeauthoryear{{Padgett} \& {Stapelfeldt}}{{Padgett} \&
  {Stapelfeldt}}{1994}]{pad94}
{Padgett} D.~L.,  {Stapelfeldt} K.~R.,  1994, \mn@doi [\aj] {10.1086/116890},
  \href {http://adsabs.harvard.edu/abs/1994AJ....107..720P} {107, 720}

\bibitem[\protect\citeauthoryear{{Palla} \& {Stahler}}{{Palla} \&
  {Stahler}}{1993}]{pal93}
{Palla} F.,  {Stahler} S.~W.,  1993, \mn@doi [\apj] {10.1086/173402}, \href
  {http://adsabs.harvard.edu/abs/1993ApJ...418..414P} {418, 414}

\bibitem[\protect\citeauthoryear{{Parenago}}{{Parenago}}{1954}]{par54}
{Parenago} P.~P.,  1954, Trudy Gosudarstvennogo Astronomicheskogo Instituta,
  \href {http://adsabs.harvard.edu/abs/1954TrSht..25....1P} {25, 1}

\bibitem[\protect\citeauthoryear{{Park}, {Sung}, {Bessell}  \& {Kang}}{{Park}
  et~al.}{2000}]{par00}
{Park} B.-G.,  {Sung} H.,  {Bessell} M.~S.,   {Kang} Y.~H.,  2000, \mn@doi
  [\aj] {10.1086/301459}, \href
  {http://adsabs.harvard.edu/abs/2000AJ....120..894P} {120, 894}

\bibitem[\protect\citeauthoryear{{Parker}}{{Parker}}{1988}]{par88}
{Parker} E.~N.,  1988, \mn@doi [\apj] {10.1086/166485}, \href
  {http://adsabs.harvard.edu/abs/1988ApJ...330..474P} {330, 474}

\bibitem[\protect\citeauthoryear{{Pecaut} \& {Mamajek}}{{Pecaut} \&
  {Mamajek}}{2013}]{pec13}
{Pecaut} M.~J.,  {Mamajek} E.~E.,  2013, \mn@doi [\apjs]
  {10.1088/0067-0049/208/1/9}, \href
  {http://adsabs.harvard.edu/abs/2013ApJS..208....9P} {208, 9}

\bibitem[\protect\citeauthoryear{{Penston}}{{Penston}}{1973}]{pen73}
{Penston} M.~V.,  1973, \mn@doi [\apj] {10.1086/152243}, \href
  {http://adsabs.harvard.edu/abs/1973ApJ...183..505P} {183, 505}

\bibitem[\protect\citeauthoryear{{Penston}, {Hunter}  \& {Oneill}}{{Penston}
  et~al.}{1975}]{pen75}
{Penston} M.~V.,  {Hunter} J.~K.,   {Oneill} A.,  1975, \mnras, \href
  {http://adsabs.harvard.edu/abs/1975MNRAS.171..219P} {171, 219}

\bibitem[\protect\citeauthoryear{{Prato}, {Simon}, {Mazeh}, {McLean}, {Norman}
  \& {Zucker}}{{Prato} et~al.}{2002}]{pra02}
{Prato} L.,  {Simon} M.,  {Mazeh} T.,  {McLean} I.~S.,  {Norman} D.,   {Zucker}
  S.,  2002, \mn@doi [\apj] {10.1086/339397}, \href
  {http://adsabs.harvard.edu/abs/2002ApJ...569..863P} {569, 863}

\bibitem[\protect\citeauthoryear{{Preibisch} \& {Feigelson}}{{Preibisch} \&
  {Feigelson}}{2005}]{pre05a}
{Preibisch} T.,  {Feigelson} E.~D.,  2005, \mn@doi [\apjs] {10.1086/432094},
  \href {http://adsabs.harvard.edu/abs/2005ApJS..160..390P} {160, 390}

\bibitem[\protect\citeauthoryear{{Preibisch} \& {Zinnecker}}{{Preibisch} \&
  {Zinnecker}}{2001}]{pre01}
{Preibisch} T.,  {Zinnecker} H.,  2001, \mn@doi [\aj] {10.1086/321177}, \href
  {http://adsabs.harvard.edu/abs/2001AJ....122..866P} {122, 866}

\bibitem[\protect\citeauthoryear{{Preibisch} \& {Zinnecker}}{{Preibisch} \&
  {Zinnecker}}{2002}]{pre02}
{Preibisch} T.,  {Zinnecker} H.,  2002, \mn@doi [\aj] {10.1086/338851}, \href
  {http://adsabs.harvard.edu/abs/2002AJ....123.1613P} {123, 1613}

\bibitem[\protect\citeauthoryear{{Preibisch} et~al.,}{{Preibisch}
  et~al.}{2005}]{pre05b}
{Preibisch} T.,  et~al., 2005, \mn@doi [\apjs] {10.1086/432891}, \href
  {http://adsabs.harvard.edu/abs/2005ApJS..160..401P} {160, 401}

\bibitem[\protect\citeauthoryear{{Prisinzano}, {Damiani}, {Micela}  \&
  {Sciortino}}{{Prisinzano} et~al.}{2005}]{pri05}
{Prisinzano} L.,  {Damiani} F.,  {Micela} G.,   {Sciortino} S.,  2005, \mn@doi
  [\aap] {10.1051/0004-6361:20040432}, \href
  {http://adsabs.harvard.edu/abs/2005A%26A...430..941P} {430, 941}

\bibitem[\protect\citeauthoryear{{Prisinzano}, {Damiani}, {Micela}  \&
  {Pillitteri}}{{Prisinzano} et~al.}{2007}]{pri07}
{Prisinzano} L.,  {Damiani} F.,  {Micela} G.,   {Pillitteri} I.,  2007, \mn@doi
  [\aap] {10.1051/0004-6361:20065623}, \href
  {http://adsabs.harvard.edu/abs/2007A%26A...462..123P} {462, 123}

\bibitem[\protect\citeauthoryear{{Prisinzano}, {Micela}, {Sciortino}, {Affer}
  \& {Damiani}}{{Prisinzano} et~al.}{2012}]{pri12}
{Prisinzano} L.,  {Micela} G.,  {Sciortino} S.,  {Affer} L.,   {Damiani} F.,
  2012, \mn@doi [\aap] {10.1051/0004-6361/201219853}, \href
  {http://adsabs.harvard.edu/abs/2012A%26A...546A...9P} {546, A9}

\bibitem[\protect\citeauthoryear{{Prosser}, {Stauffer}, {Hartmann},
  {Soderblom}, {Jones}, {Werner}  \& {McCaughrean}}{{Prosser}
  et~al.}{1994}]{pro94}
{Prosser} C.~F.,  {Stauffer} J.~R.,  {Hartmann} L.,  {Soderblom} D.~R.,
  {Jones} B.~F.,  {Werner} M.~W.,   {McCaughrean} M.~J.,  1994, \mn@doi [\apj]
  {10.1086/173668}, \href {http://cdsads.u-strasbg.fr/abs/1994ApJ...421..517P}
  {421, 517}

\bibitem[\protect\citeauthoryear{{Ram{\'{\i}}rez} et~al.,}{{Ram{\'{\i}}rez}
  et~al.}{2004}]{ram04}
{Ram{\'{\i}}rez} S.~V.,  et~al., 2004, \mn@doi [\aj] {10.1086/383290}, \href
  {http://adsabs.harvard.edu/abs/2004AJ....127.2659R} {127, 2659}

\bibitem[\protect\citeauthoryear{{Rauw}, {Naz{\'e}}, {Gosset}, {Stevens},
  {Blomme}, {Corcoran}, {Pittard}  \& {Runacres}}{{Rauw} et~al.}{2002}]{rau02}
{Rauw} G.,  {Naz{\'e}} Y.,  {Gosset} E.,  {Stevens} I.~R.,  {Blomme} R.,
  {Corcoran} M.~F.,  {Pittard} J.~M.,   {Runacres} M.~C.,  2002, \mn@doi [\aap]
  {10.1051/0004-6361:20021230}, \href
  {http://adsabs.harvard.edu/abs/2002A%26A...395..499R} {395, 499}

\bibitem[\protect\citeauthoryear{{Rebull} et~al.,}{{Rebull}
  et~al.}{2002}]{reb02}
{Rebull} L.~M.,  et~al., 2002, \mn@doi [\aj] {10.1086/338904}, \href
  {http://adsabs.harvard.edu/abs/2002AJ....123.1528R} {123, 1528}

\bibitem[\protect\citeauthoryear{{Rebull}, {Stauffer}, {Ramirez}, {Flaccomio},
  {Sciortino}, {Micela}, {Strom}  \& {Wolff}}{{Rebull} et~al.}{2006}]{reb06}
{Rebull} L.~M.,  {Stauffer} J.~R.,  {Ramirez} S.~V.,  {Flaccomio} E.,
  {Sciortino} S.,  {Micela} G.,  {Strom} S.~E.,   {Wolff} S.~C.,  2006, \mn@doi
  [\aj] {10.1086/503835}, \href
  {http://adsabs.harvard.edu/abs/2006AJ....131.2934R} {131, 2934}

\bibitem[\protect\citeauthoryear{{Rhode}, {Herbst}  \& {Mathieu}}{{Rhode}
  et~al.}{2001}]{rho01}
{Rhode} K.~L.,  {Herbst} W.,   {Mathieu} R.~D.,  2001, \mn@doi [\aj]
  {10.1086/324448}, \href {http://adsabs.harvard.edu/abs/2001AJ....122.3258R}
  {122, 3258}

\bibitem[\protect\citeauthoryear{{Rieke} \& {Lebofsky}}{{Rieke} \&
  {Lebofsky}}{1985}]{rie85}
{Rieke} G.~H.,  {Lebofsky} M.~J.,  1985, \mn@doi [\apj] {10.1086/162827}, \href
  {http://adsabs.harvard.edu/abs/1985ApJ...288..618R} {288, 618}

\bibitem[\protect\citeauthoryear{{Samus}, {Durlevich}  \& {et al.}}{{Samus}
  et~al.}{2009}]{sam09}
{Samus} N.~N.,  {Durlevich} O.~V.,   {et al.} 2009, VizieR Online Data Catalog,
  \href {http://adsabs.harvard.edu/abs/2009yCat....102025S} {1, 2025}

\bibitem[\protect\citeauthoryear{{Sandstrom}, {Peek}, {Bower}, {Bolatto}  \&
  {Plambeck}}{{Sandstrom} et~al.}{2007}]{san07}
{Sandstrom} K.~M.,  {Peek} J.~E.~G.,  {Bower} G.~C.,  {Bolatto} A.~D.,
  {Plambeck} R.~L.,  2007, \mn@doi [\apj] {10.1086/520922}, \href
  {http://adsabs.harvard.edu/abs/2007ApJ...667.1161S} {667, 1161}

\bibitem[\protect\citeauthoryear{{Schr{\"o}der} \& {Schmitt}}{{Schr{\"o}der} \&
  {Schmitt}}{2007}]{schr07}
{Schr{\"o}der} C.,  {Schmitt} J.~H.~M.~M.,  2007, \mn@doi [\aap]
  {10.1051/0004-6361:20077429}, \href
  {http://adsabs.harvard.edu/abs/2007A%26A...475..677S} {475, 677}

\bibitem[\protect\citeauthoryear{{Siess}, {Dufour}  \& {Forestini}}{{Siess}
  et~al.}{2000}]{sie00}
{Siess} L.,  {Dufour} E.,   {Forestini} M.,  2000, \aap, \href
  {http://adsabs.harvard.edu/abs/2000A%26A...358..593S} {358, 593}

\bibitem[\protect\citeauthoryear{{Skrutskie} et~al.,}{{Skrutskie}
  et~al.}{2006}]{skr06}
{Skrutskie} M.~F.,  et~al., 2006, \mn@doi [\aj] {10.1086/498708}, \href
  {http://adsabs.harvard.edu/abs/2006AJ....131.1163S} {131, 1163}

\bibitem[\protect\citeauthoryear{{Slesnick}, {Hillenbrand}  \&
  {Carpenter}}{{Slesnick} et~al.}{2004}]{sle04}
{Slesnick} C.~L.,  {Hillenbrand} L.~A.,   {Carpenter} J.~M.,  2004, \mn@doi
  [\apj] {10.1086/421898}, \href
  {http://adsabs.harvard.edu/abs/2004ApJ...610.1045S} {610, 1045}

\bibitem[\protect\citeauthoryear{{Stahler}}{{Stahler}}{1983}]{sta83}
{Stahler} S.~W.,  1983, \mn@doi [\apj] {10.1086/161495}, \href
  {http://adsabs.harvard.edu/abs/1983ApJ...274..822S} {274, 822}

\bibitem[\protect\citeauthoryear{{Stassun}, {van den Berg}, {Feigelson}  \&
  {Flaccomio}}{{Stassun} et~al.}{2006}]{sta06}
{Stassun} K.~G.,  {van den Berg} M.,  {Feigelson} E.,   {Flaccomio} E.,  2006,
  \mn@doi [\apj] {10.1086/506422}, \href
  {http://adsabs.harvard.edu/abs/2006ApJ...649..914S} {649, 914}

\bibitem[\protect\citeauthoryear{{Stelzer} \& {Neuh{\"a}user}}{{Stelzer} \&
  {Neuh{\"a}user}}{2001}]{ste01}
{Stelzer} B.,  {Neuh{\"a}user} R.,  2001, \mn@doi [\aap]
  {10.1051/0004-6361:20011093}, \href
  {http://ukads.nottingham.ac.uk/abs/2001A%26A...377..538S} {377, 538}

\bibitem[\protect\citeauthoryear{{Stelzer}, {Preibisch}, {Alexander},
  {Mucciarelli}, {Flaccomio}, {Micela}  \& {Sciortino}}{{Stelzer}
  et~al.}{2012}]{ste12}
{Stelzer} B.,  {Preibisch} T.,  {Alexander} F.,  {Mucciarelli} P.,  {Flaccomio}
  E.,  {Micela} G.,   {Sciortino} S.,  2012, \mn@doi [\aap]
  {10.1051/0004-6361/201118118}, \href
  {http://adsabs.harvard.edu/abs/2012A%26A...537A.135S} {537, A135}

\bibitem[\protect\citeauthoryear{{Strom}, {Strom}  \& {Carrasco}}{{Strom}
  et~al.}{1974}]{str74}
{Strom} S.~E.,  {Strom} K.~A.,   {Carrasco} L.,  1974, \mn@doi [\pasp]
  {10.1086/129676}, \href {http://cdsads.u-strasbg.fr/abs/1974PASP...86..798S}
  {86, 798}

\bibitem[\protect\citeauthoryear{{Sung} \& {Bessell}}{{Sung} \&
  {Bessell}}{2010}]{sun10}
{Sung} H.,  {Bessell} M.~S.,  2010, \mn@doi [\aj]
  {10.1088/0004-6256/140/6/2070}, \href
  {http://adsabs.harvard.edu/abs/2010AJ....140.2070S} {140, 2070}

\bibitem[\protect\citeauthoryear{{Sung}, {Bessell}  \& {Lee}}{{Sung}
  et~al.}{1997}]{sun97}
{Sung} H.,  {Bessell} M.~S.,   {Lee} S.-W.,  1997, \mn@doi [\aj]
  {10.1086/118674}, \href {http://adsabs.harvard.edu/abs/1997AJ....114.2644S}
  {114, 2644}

\bibitem[\protect\citeauthoryear{{Sung}, {Chun}  \& {Bessell}}{{Sung}
  et~al.}{2000}]{sun00}
{Sung} H.,  {Chun} M.-Y.,   {Bessell} M.~S.,  2000, \mn@doi [\aj]
  {10.1086/301450}, \href {http://adsabs.harvard.edu/abs/2000AJ....120..333S}
  {120, 333}

\bibitem[\protect\citeauthoryear{{Sung}, {Bessell}  \& {Chun}}{{Sung}
  et~al.}{2004}]{sun04}
{Sung} H.,  {Bessell} M.~S.,   {Chun} M.-Y.,  2004, \mn@doi [\aj]
  {10.1086/423440}, \href {http://adsabs.harvard.edu/abs/2004AJ....128.1684S}
  {128, 1684}

\bibitem[\protect\citeauthoryear{{Telleschi}, {G{\"u}del}, {Briggs}, {Audard}
  \& {Palla}}{{Telleschi} et~al.}{2007}]{tel07}
{Telleschi} A.,  {G{\"u}del} M.,  {Briggs} K.~R.,  {Audard} M.,   {Palla} F.,
  2007, \mn@doi [\aap] {10.1051/0004-6361:20066565}, \href
  {http://adsabs.harvard.edu/abs/2007A%26A...468..425T} {468, 425}

\bibitem[\protect\citeauthoryear{{Tobin}, {Hartmann}, {Furesz}, {Mateo}  \&
  {Megeath}}{{Tobin} et~al.}{2009}]{tob09}
{Tobin} J.~J.,  {Hartmann} L.,  {Furesz} G.,  {Mateo} M.,   {Megeath} S.~T.,
  2009, \mn@doi [\apj] {10.1088/0004-637X/697/2/1103}, \href
  {http://adsabs.harvard.edu/abs/2009ApJ...697.1103T} {697, 1103}

\bibitem[\protect\citeauthoryear{{Tobin}, {Hartmann}, {Furesz}, {Mateo}  \&
  {Megeath}}{{Tobin} et~al.}{2013}]{tob13}
{Tobin} J.~J.,  {Hartmann} L.,  {Furesz} G.,  {Mateo} M.,   {Megeath} S.~T.,
  2013, \mn@doi [\apj] {10.1088/0004-637X/773/1/81}, \href
  {http://adsabs.harvard.edu/abs/2013ApJ...773...81T} {773, 81}

\bibitem[\protect\citeauthoryear{{Tognelli}, {Prada Moroni}  \&
  {Degl'Innocenti}}{{Tognelli} et~al.}{2011}]{tog11}
{Tognelli} E.,  {Prada Moroni} P.~G.,   {Degl'Innocenti} S.,  2011, \mn@doi
  [\aap] {10.1051/0004-6361/200913913}, \href
  {http://adsabs.harvard.edu/abs/2011A%26A...533A.109T} {533, A109}

\bibitem[\protect\citeauthoryear{{Torres}}{{Torres}}{2010}]{tor10}
{Torres} G.,  2010, \mn@doi [\aj] {10.1088/0004-6256/140/5/1158}, \href
  {http://adsabs.harvard.edu/abs/2010AJ....140.1158T} {140, 1158}

\bibitem[\protect\citeauthoryear{{Torres}, {Quast}, {da Silva}, {de La Reza},
  {Melo}  \& {Sterzik}}{{Torres} et~al.}{2006}]{tor06}
{Torres} C.~A.~O.,  {Quast} G.~R.,  {da Silva} L.,  {de La Reza} R.,  {Melo}
  C.~H.~F.,   {Sterzik} M.,  2006, \mn@doi [\aap] {10.1051/0004-6361:20065602},
  \href {http://adsabs.harvard.edu/abs/2006A%26A...460..695T} {460, 695}

\bibitem[\protect\citeauthoryear{{Tothill}, {Gagn{\'e}}, {Stecklum}  \&
  {Kenworthy}}{{Tothill} et~al.}{2008}]{tot08}
{Tothill} N.~F.~H.,  {Gagn{\'e}} M.,  {Stecklum} B.,   {Kenworthy} M.~A.,
  2008, {The Lagoon Nebula and its Vicinity}.
p.~533

\bibitem[\protect\citeauthoryear{{Turner}}{{Turner}}{2012}]{tur12}
{Turner} D.~G.,  2012, \mn@doi [Astronomische Nachrichten]
  {10.1002/asna.201111643}, \href
  {http://adsabs.harvard.edu/abs/2012AN....333..174T} {333, 174}

\bibitem[\protect\citeauthoryear{{Vacca} \& {Sandell}}{{Vacca} \&
  {Sandell}}{2011}]{vac11}
{Vacca} W.~D.,  {Sandell} G.,  2011, \mn@doi [\apj]
  {10.1088/0004-637X/732/1/8}, \href
  {http://adsabs.harvard.edu/abs/2011ApJ...732....8V} {732, 8}

\bibitem[\protect\citeauthoryear{{Vuong}, {Montmerle}, {Grosso}, {Feigelson},
  {Verstraete}  \& {Ozawa}}{{Vuong} et~al.}{2003}]{vuo03}
{Vuong} M.~H.,  {Montmerle} T.,  {Grosso} N.,  {Feigelson} E.~D.,  {Verstraete}
  L.,   {Ozawa} H.,  2003, \mn@doi [\aap] {10.1051/0004-6361:20030942}, \href
  {http://adsabs.harvard.edu/abs/2003A%26A...408..581V} {408, 581}

\bibitem[\protect\citeauthoryear{{Walker}}{{Walker}}{1956}]{wal56}
{Walker} M.~F.,  1956, \mn@doi [\apjs] {10.1086/190026}, \href
  {http://adsabs.harvard.edu/abs/1956ApJS....2..365W} {2, 365}

\bibitem[\protect\citeauthoryear{{Walker}}{{Walker}}{1957}]{wal57}
{Walker} M.~F.,  1957, \mn@doi [\apj] {10.1086/146337}, \href
  {http://adsabs.harvard.edu/abs/1957ApJ...125..636W} {125, 636}

\bibitem[\protect\citeauthoryear{{Walker}}{{Walker}}{1961}]{wal61}
{Walker} M.~F.,  1961, \mn@doi [\apj] {10.1086/147113}, \href
  {http://adsabs.harvard.edu/abs/1961ApJ...133.1081W} {133, 1081}

\bibitem[\protect\citeauthoryear{{Walker}}{{Walker}}{1983}]{wal83}
{Walker} M.~F.,  1983, \mn@doi [\apj] {10.1086/161232}, \href
  {http://adsabs.harvard.edu/abs/1983ApJ...271..642W} {271, 642}

\bibitem[\protect\citeauthoryear{{Weights}, {Lucas}, {Roche}, {Pinfield}  \&
  {Riddick}}{{Weights} et~al.}{2009}]{wei09}
{Weights} D.~J.,  {Lucas} P.~W.,  {Roche} P.~F.,  {Pinfield} D.~J.,   {Riddick}
  F.,  2009, \mn@doi [\mnras] {10.1111/j.1365-2966.2008.14096.x}, \href
  {http://adsabs.harvard.edu/abs/2009MNRAS.392..817W} {392, 817}

\bibitem[\protect\citeauthoryear{{Windemuth} \& {Herbst}}{{Windemuth} \&
  {Herbst}}{2014}]{win14}
{Windemuth} D.,  {Herbst} W.,  2014, \mn@doi [\aj] {10.1088/0004-6256/147/1/9},
  \href {http://adsabs.harvard.edu/abs/2014AJ....147....9W} {147, 9}

\bibitem[\protect\citeauthoryear{{Wolff}, {Strom}  \& {Hillenbrand}}{{Wolff}
  et~al.}{2004}]{wol04}
{Wolff} S.~C.,  {Strom} S.~E.,   {Hillenbrand} L.~A.,  2004, \mn@doi [\apj]
  {10.1086/380503}, \href {http://adsabs.harvard.edu/abs/2004ApJ...601..979W}
  {601, 979}

\bibitem[\protect\citeauthoryear{{Wolk}, {Harnden}, {Flaccomio}, {Micela},
  {Favata}, {Shang}  \& {Feigelson}}{{Wolk} et~al.}{2005}]{wol05}
{Wolk} S.~J.,  {Harnden} Jr. F.~R.,  {Flaccomio} E.,  {Micela} G.,  {Favata}
  F.,  {Shang} H.,   {Feigelson} E.~D.,  2005, \mn@doi [\apjs]
  {10.1086/432099}, \href {http://adsabs.harvard.edu/abs/2005ApJS..160..423W}
  {160, 423}

\bibitem[\protect\citeauthoryear{{Wright}, {Drake}, {Mamajek}  \&
  {Henry}}{{Wright} et~al.}{2011}]{wri11}
{Wright} N.~J.,  {Drake} J.~J.,  {Mamajek} E.~E.,   {Henry} G.~W.,  2011,
  \mn@doi [\apj] {10.1088/0004-637X/743/1/48}, \href
  {http://adsabs.harvard.edu/abs/2011ApJ...743...48W} {743, 48}

\bibitem[\protect\citeauthoryear{{van den Ancker}, {The}, {Feinstein},
  {Vazquez}, {de Winter}  \& {Perez}}{{van den Ancker} et~al.}{1997}]{van97}
{van den Ancker} M.~E.,  {The} P.~S.,  {Feinstein} A.,  {Vazquez} R.~A.,  {de
  Winter} D.,   {Perez} M.~R.,  1997, \mn@doi [\aaps] {10.1051/aas:1997306},
  \href {http://adsabs.harvard.edu/abs/1997A%26AS..123...63V} {123, 63}

\makeatother
\end{thebibliography}



\appendix

\section{Distance estimates}\label{appendix_distances}
Extensive discussion of literature estimates of the distance to the five star forming regions that we consider can be found in the papers of the Handbook of Star Forming Regions, ONC \citep{mue08}, NGC~2264 \citep{dah08}, IC~348 \citep{her08}, NGC~2362 \citep{dah08c}, and NGC~6530 \citep{tot08}.  We therefore limit any detailed discussion below to distance estimates published subsequently.  Our adopted distances are listed in Table \ref{table_distances}. 
 
\subsection{Orion Nebula Cluster (ONC)}
Historically, the distance to the ONC was subject to large uncertainty, but modern estimates have converged on $\sim400\,{\rm pc}$ (e.g. \citealt{jef07}; \citealt{kra07}; \citealt{hir07}; \citealt{san07}; \citealt{men07}; \citealt{may08}; \citealt{kim08}).  We assume $d=414\,{\rm pc}$, the value obtained from trigonometric parallax measurements made using the Very Long Baseline Array \citep{men07}. This corresponds to a distance modulus of $\mu=8.085$.

\subsection{NGC 2264}
The distance to NGC~2264 remains somewhat uncertain, with modern estimates ranging from $\sim$400$\,{\rm pc}$ \citep{dzi14} to $913\pm110\pm40\,{\rm pc}$ (systematic and sampling errors respectively; \citealt{bax09}).  The former was derived from a re-analysis of Hipparcos satellite parallax measurements; future, higher precession parallax observations to-be-made by the Gaia satellite will test the robustness of such a small cluster distance, $\sim$400$\,{\rm pc}$ being a factor of two below almost all other estimates (see below). The latter was estimated by calculating $\sin i$ for cluster members, where $i$ is the stellar inclination, using knowledge of the stellar rotation period, measured projected rotational velocities ($v_\ast\sin i$), and the stellar radius for an assumed cluster distance ($R_\ast$ depends on the stellar luminosity and therefore the cluster distance), $\sin i=P_{\rm rot}v_\ast\sin i/(2\pi R_\ast)$.  By then modelling the $\sin i$ distribution obtained assuming that the stellar inclinations are randomly oriented, \citet{bax09} varied the cluster distance until the best match with the observed distribution was obtained.                  

Most distance estimates fall in the range $\sim$700-780$\,{\rm pc}$ ($759\pm87\,{\rm pc}$, \citealt{sun97} with \citealt{par00} finding the same value but with a smaller error; $711<748<802\,{\rm pc}$, \citealt{may08}; $705\,{\rm pc}$, \citealt{nay09}; $750\,{\rm pc}$, \citealt*{cve09}; $815\pm95\,{\rm pc}$, \citealt{sun10}; $777\pm12\,{\rm pc}$, \citealt{tur12}; $756\pm96\,{\rm pc}$, \citealt{gil14}; and $738+57-50\,{\rm pc}$, \citealt{kam14}).  As the value derived by \citet{gil14} was obtained from an analysis of CoRoT satellite photometry of an eclipsing binary system, and is therefore arguably the most accurate, we use their value of $756\,{\rm pc}$ in this paper ($\mu=9.393$).   

\subsection{IC 348}
In additional to the Handbook of Star Forming Regions paper of \citet{her08}, an extensive comparison of distances estimates to IC~348 can be found in \citet{her98}, who settled on $316\,{\rm pc}$, the value we adopt here ($\mu=7.498$). 

\subsection{NGC 2362}
Historically, there was more than a kilo-parsec variation in derived distances to NGC~2362 (see \citealt{dah05b,dah08c}), but modern estimates have settled on $\sim$1.4-1.5$\,{\rm kpc}$ \citep{bal96,moi01,may08,kha13}. Many of the early type stars lie on the ZAMS, allowing the cluster distance to be derived by fitting the ZAMS in colour-magnitude diagrams.  We adopt a distance of $1480\,{\rm pc}$, the value derived by \citet{moi01} by fitting the B-star sequence in the $V$ vs $U-B$ plane ($\mu=10.851$).       

\begin{figure*}
   \centering
     \includegraphics[width=0.3\textwidth]{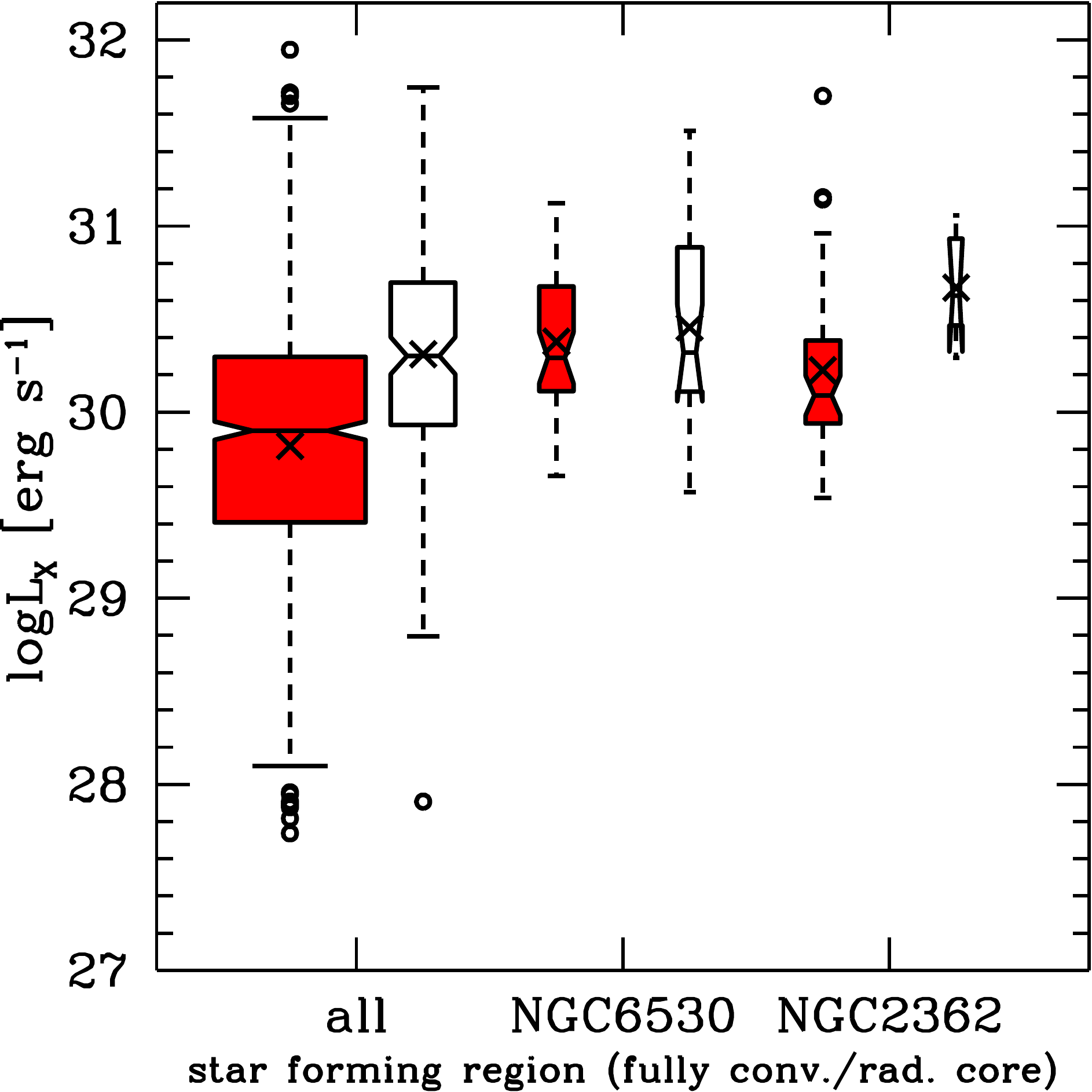}  
      \includegraphics[width=0.3\textwidth]{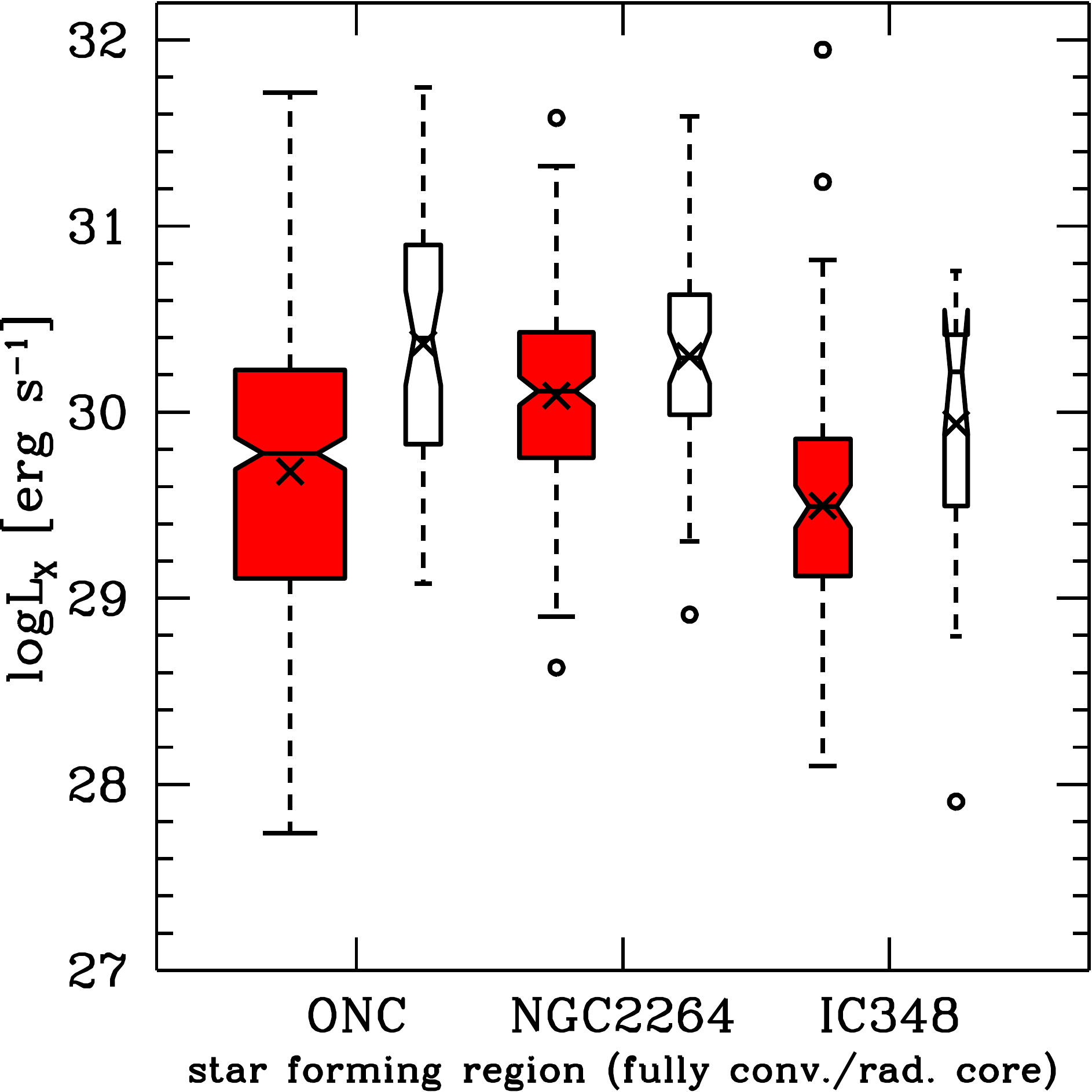} \\ 
           \includegraphics[width=0.3\textwidth]{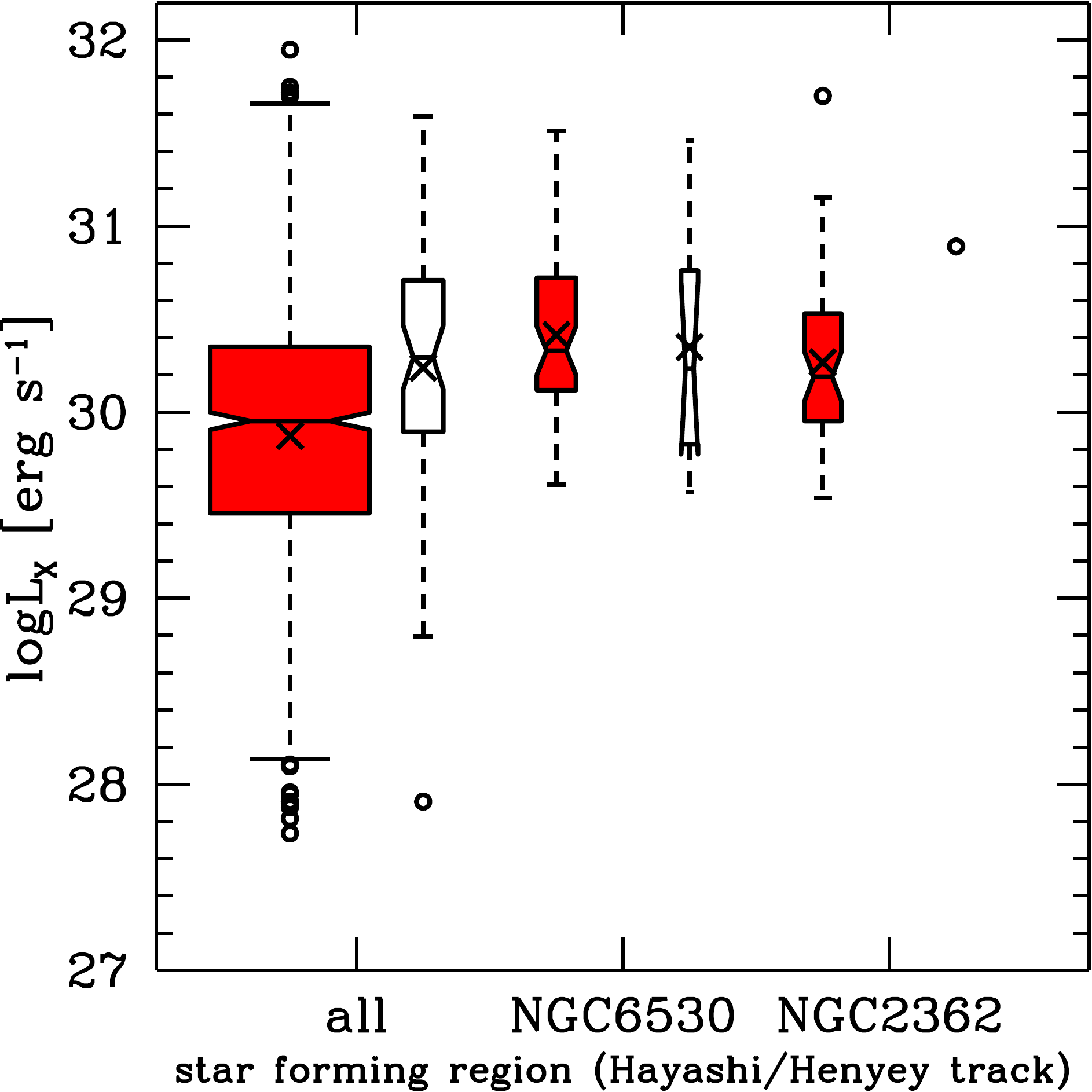}  
      \includegraphics[width=0.3\textwidth]{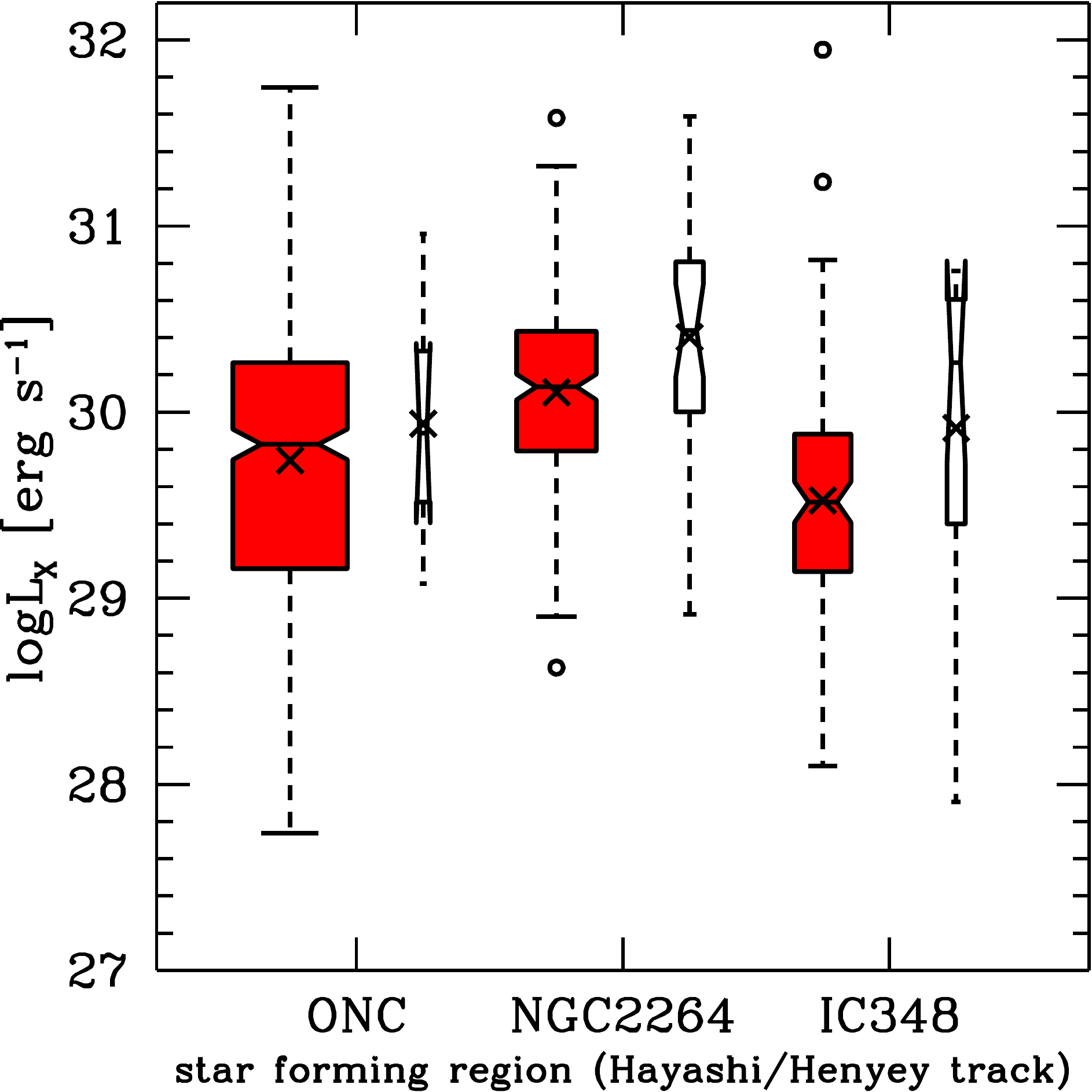}  
   \caption{As Figure \ref{regions_hist_logLXLbol} but for $\log L_{\rm X}$.}
   \label{regions_hist_logLX}
\end{figure*}

\begin{figure*}
   \centering
      \includegraphics[width=0.3\textwidth]{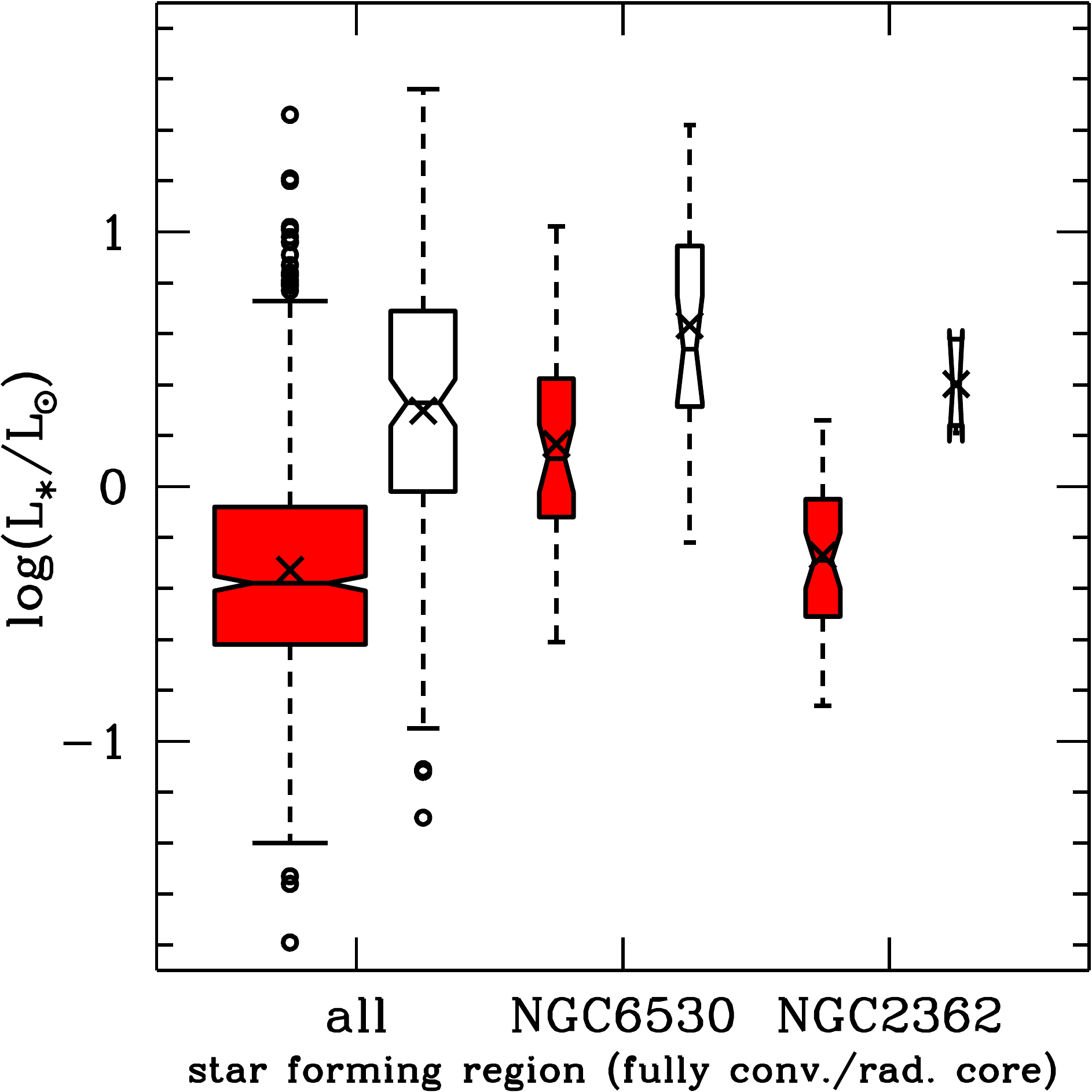} 
      \includegraphics[width=0.3\textwidth]{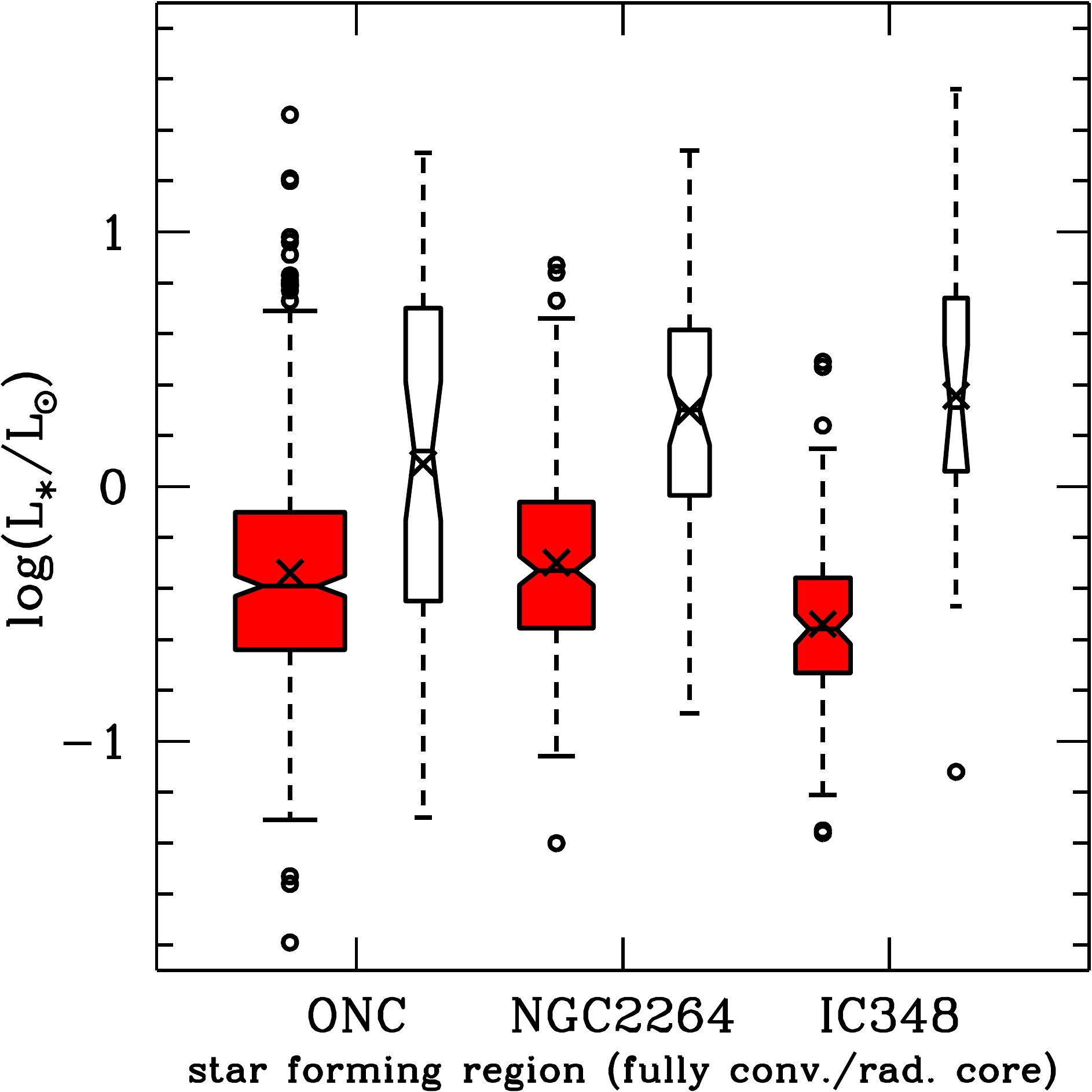}\\ 
           \includegraphics[width=0.3\textwidth]{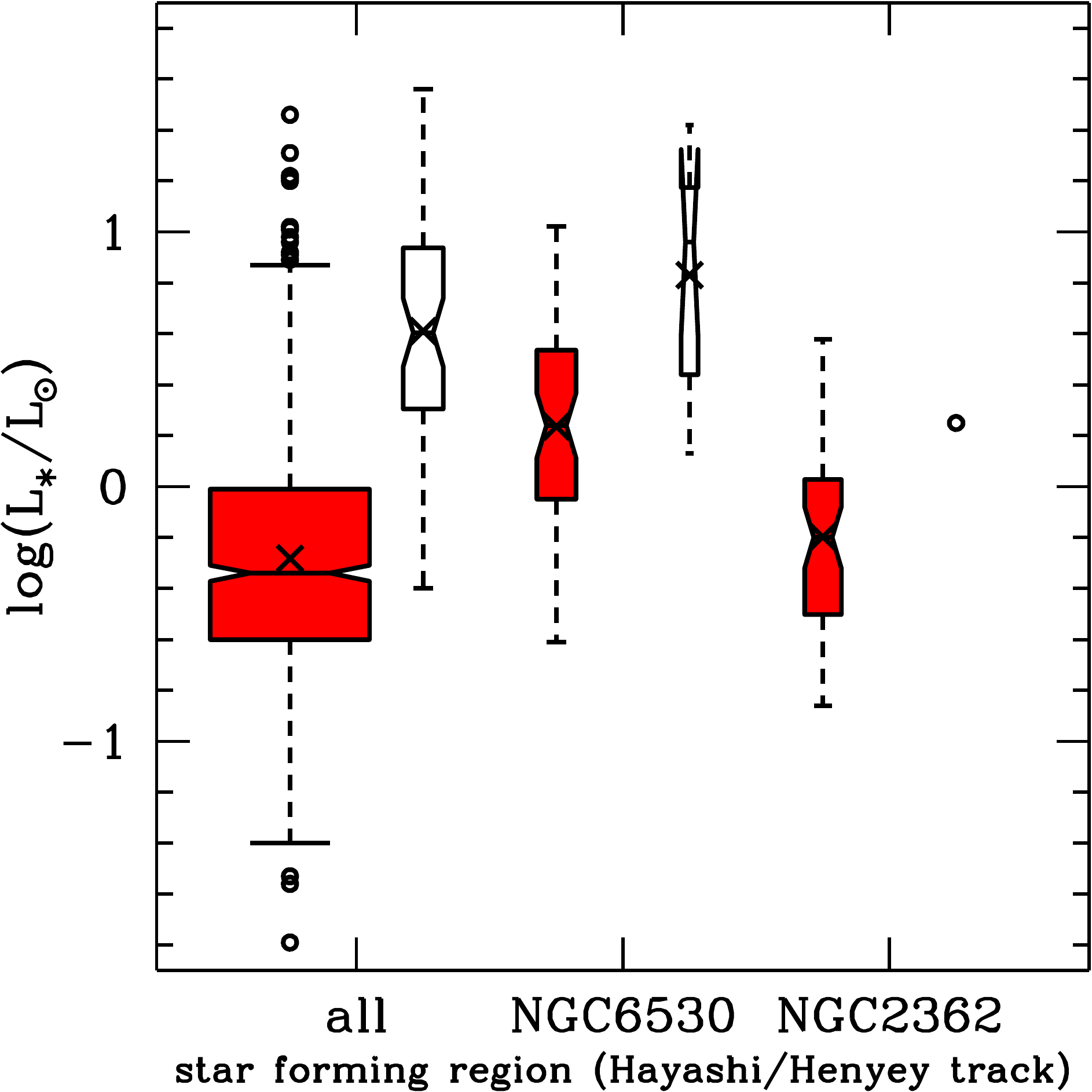}  
      \includegraphics[width=0.3\textwidth]{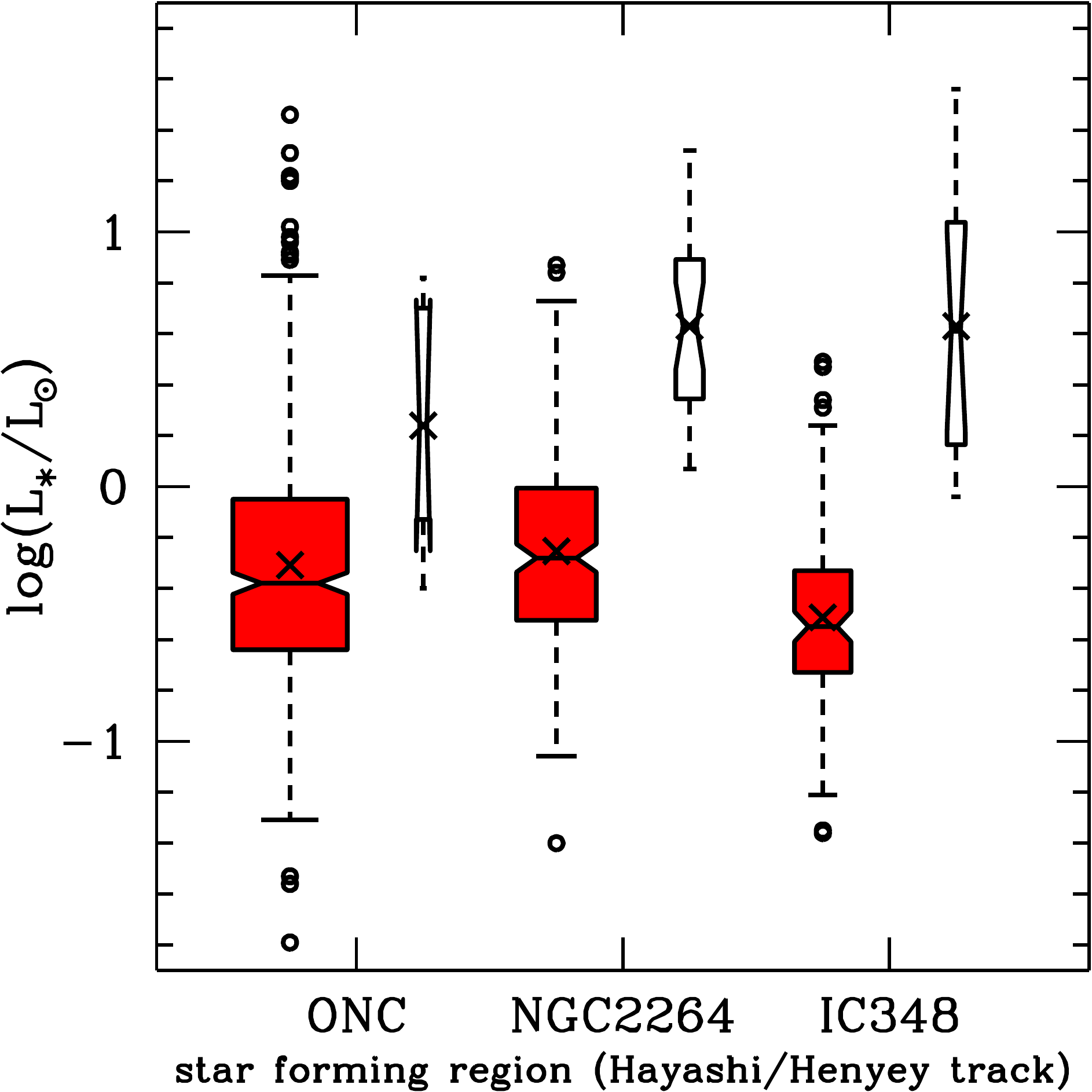}  
   \caption{As Figure \ref{regions_hist_logLXLbol} but for $\log(L_\ast/L_\odot)$.}
   \label{regions_hist_logLstar}
\end{figure*}

\subsection{NGC 6530}
Extensive discussion of the distance to NGC~6530 can be found in both \citet{tot08} and \citet{pri05}. Literature estimates range from $\sim$1.3-2$\,{\rm kpc}$, with most studies deriving either $\sim$1.3$\,{\rm kpc}$ (e.g. \citealt{pri05,ari06,may08,kha13}) or $\sim$1.8$\,{\rm kpc}$ (e.g. \citealt{van97,sun00}).  As the more recent estimates agree on the former, we adopt $1250\,{\rm pc}$, the value derived by \citet{pri05}, corresponding to a distance modulus of $\mu=10.485$.     


\section{A comparison of $\log L_{\rm X}$ and $\log(L_\ast/L_\odot)$ for fully and partially convective PMS stars}\label{appendix_logLX}
Box plots of $\log L_{\rm X}$ and $\log(L_\ast/L_\odot)$ for the entire sample and all regions combined, are shown in Figs.~\ref{regions_hist_logLX} \& \ref{regions_hist_logLstar} respectively.  For brevity we present only the box plots and not full breakdowns analogous to Tables~\ref{tablestats} \& \ref{tablestats2}.  Partially convective stars are, on average, more luminous than fully convective stars; likewise for Henyey compared to Hayashi track stars.  There is a single exception - Hayashi track stars in NGC~6530 are more X-ray luminous on average than Henyey track stars (see Fig.~\ref{regions_hist_logLX} lower left panel).  This is a result of an observational bias: the incompleteness of spectroscopic surveys of that region.  In this work we have selected only PMS stars where spectral type have been assigned from spectra (and not stars with an estimated spectral type from a comparison of dereddened photometry with spectral type-dependent intrinsic colours).  There are very few late-type stars with assigned spectral types in NGC~6530 (see \citealt{tot08}). In our sample, there are only 14 M-type stars and, of these, one is classified as M1 and the rest are earlier M-types.  The lack of low-mass stars in the region with spectral type assignments is evident from the H-R diagram for NGC~6530, Fig. \ref{hrd}, where there are no stars with $M_\ast\lesssim0.45\,{\rm M}_\odot$.  These low-mass stars, which would all be on Hayashi tracks, are weaker X-ray emitters than higher mass PMS stars, see Fig.~\ref{logLX_logMstar}.  Thus, had they been spectral typed and therefore included in our sample, we would expect to have found that $\langle\log L_{\rm X}\rangle_{\rm HEN}>\langle\log L_{\rm X}\rangle_{\rm HAY}$ for NGC~6530, as with the other four star forming regions that we consider. 


\section{Additional correlations}\label{appendix_correlations}
For PMS stars, correlations between X-ray luminosity and stellar mass, and between X-ray luminosity and stellar age, are well known (e.g. \citealt{pre05a,pre05b}).  In light of (i) the substantial updates to spectral type catalogs; (ii) the improved mass and age estimates derived from PMS star calibrated effective temperature and intrinsic colours that we present in this work; (iii) the large sample size of PMS stars from five star forming regions that we consider; and (iv) our division of the sample into fully/partially convective stars and Hayashi/Henyey track stars, in this Appendix we briefly re-examine the known correlations between $L_{\rm X}$ and stellar mass/age.  

\begin{table*}
  \caption{As Table \ref{table_logLX_logLstar} but for linear regression fits to $\log{L_{\rm X}}$ vs $\log(M_\ast/{\rm M}_\odot)$ shown in Figure $\ref{logLX_logMstar}$, with $\log{L_{\rm X}}=d+c\log(M_\ast/{\rm M}_\odot)$ [i.e. $L_{\rm X}\propto M_\ast^c$].}
  \begin{tabular}{cccccc}
  \hline
  & $N_{\rm stars}$ & $d$ & $c$ & std. dev. & prob. \\
  \hline
  all stars       &  984 (34)  &   30.34$\pm$0.03    &   1.45$\pm$0.06   & 0.60 & $<$5e-5  \\
  fully convective           &  836 (33)  &   30.55$\pm$0.04    &   1.84$\pm$0.08   & 0.59 & $<$5e-5  \\
  radiative core     &  148 (1)    &   30.11$\pm$0.06    &   1.32$\pm$0.28   & 0.59 & $<$5e-5  \\
  Hayashi track     &  927 (33)  &   30.44$\pm$0.03    &   1.64$\pm$0.07   & 0.58 & $<$5e-5  \\
  Henyey track      &  57 (1)     &   29.72$\pm$0.24     &   2.12$\pm$0.98   & 0.71 & 0.041       \\
  \hline
    Hayashi track with radiative core      &  91 (0)     &   30.20$\pm$0.05     &   1.69$\pm$0.26   & 0.45 & $<$5e-5       \\
all stars ($0.1\le M_\ast/{\rm M}_\odot\le 2$) &  963 (34)  &   30.35$\pm$0.03    &   1.47$\pm$0.06   & 0.60 & $<$5e-5  \\
ONC ($0.1\le M_\ast/{\rm M}_\odot\le 2$) &  455 (0)  &   30.38$\pm$0.05    &   1.61$\pm$0.10   & 0.62 & $<$5e-5  \\
  \hline
\end{tabular}
\label{table_logLX_logMstar}
\end{table*}


\subsection{$L_{\rm X}$ versus $M_\ast$}\label{appendix_mass}

\begin{figure}
   \centering
      \includegraphics[width=0.35\textwidth]{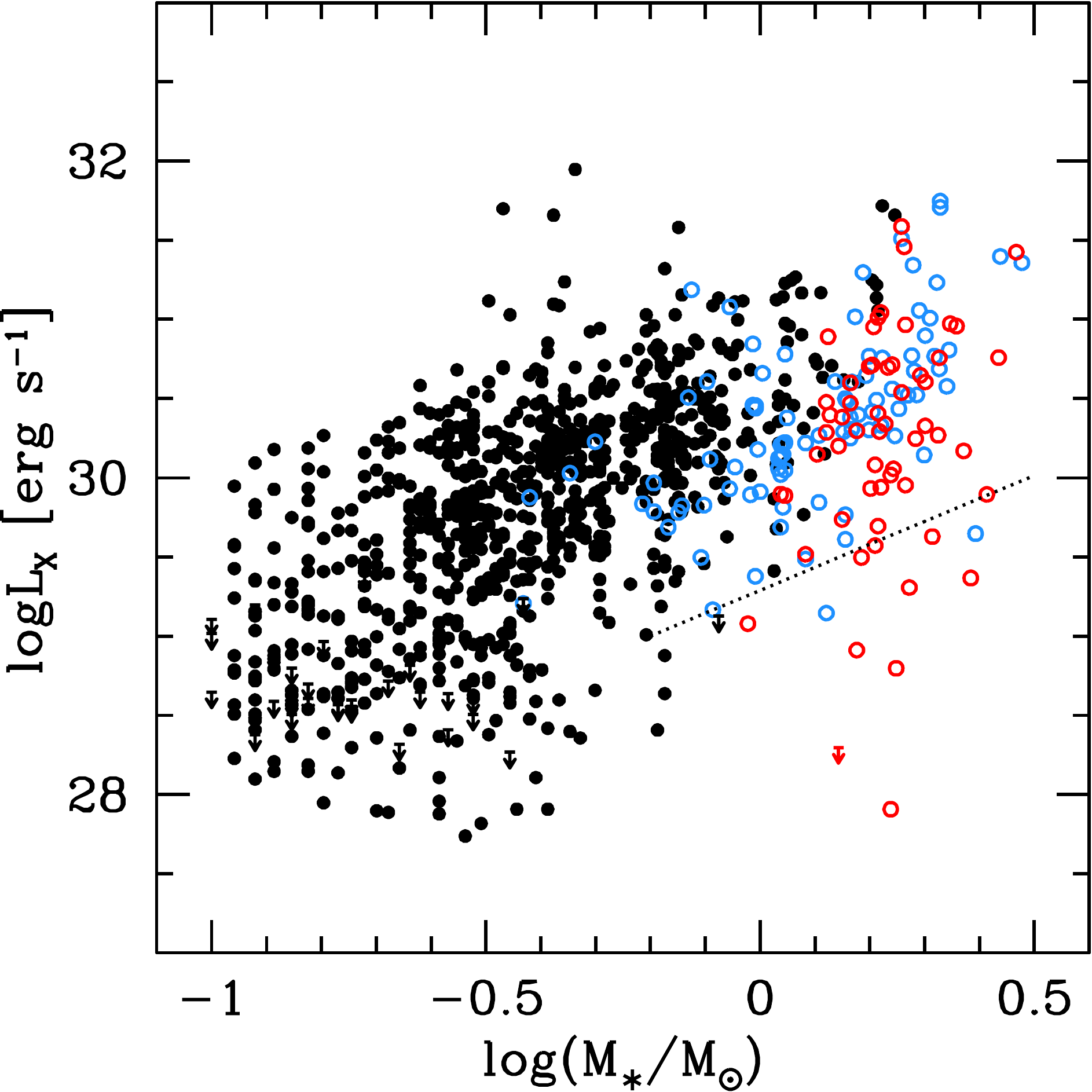}
   \caption{$\log{L_{\rm X}}$ vs $\log{M_\ast}$ for all stars in the five star forming regions.  Fully convective stars are shown as black symbols and partially convective stars as blue/red symbols.  Black and blue symbols combined are stars on Hayashi tracks; those shown as red symbols have evolved onto Henyey tracks.  Stars with $L_{\rm X}$ upper limits (all from IC~348) are shown as downward pointing arrows. The dotted line is defined in section \ref{dashedline}.}
   \label{logLX_logMstar}
\end{figure}

The correlation between $L_{\rm X}$ and $M_\ast$ is plotted in Fig.~\ref{logLX_logMstar}, with linear regression fits from the expectation maximisation algorithm of ASURV presented in Table~\ref{table_logLX_logMstar}.  Probabilities for there not being a correlation from generalized Kendall's $\tau$ tests are also listed, which take account of the $\log{L_{\rm X}}$ upper limits.  Considering the entire sample we find $L_{\rm X}\propto M_{\ast}^{1.45\pm0.06}$, although as with the $L_{\rm X}-L_\ast$ correlation (section \ref{xraycompare}) this hides the differing behaviour with the change in stellar internal structure.  Fully convective stars show a steeper increase in $L_{\rm X}$ with $M_\ast$ with an exponent of 1.84$\pm$0.08. Partially convective stars show a weaker increase with an exponent of 1.32$\pm$0.28, albeit with a large error.  The correlation is marginal at best for Henyey track stars.  It is clear visually from Fig. \ref{logLX_logMstar} that PMS stars with radiative cores are less luminous in X-rays than fully convective stars in the same mass range, with $L_{\rm X}$ having decayed substantially for several Henyey track stars.          

For a more direct comparison to the literature, we also list in Table~\ref{table_logLX_logMstar} the correlation found for $0.1~\le~M_\ast/M_\odot~\le~2$, and for ONC stars in the same mass range. 
The latter offers the most direct comparison with the COUP, the most comprehensive X-ray survey of a young star forming region. \citet{pre05b} find that $L_{\rm X}\propto M_\ast^{1.44\pm0.10}$ with stellar masses estimated from the \citet{sie00} models.  With the same models we find a larger exponent of 1.61$\pm$0.10, although this is consistent with the \citet{pre05b} result within the error.     

\begin{figure*}
   \centering
      \includegraphics[width=0.3\textwidth]{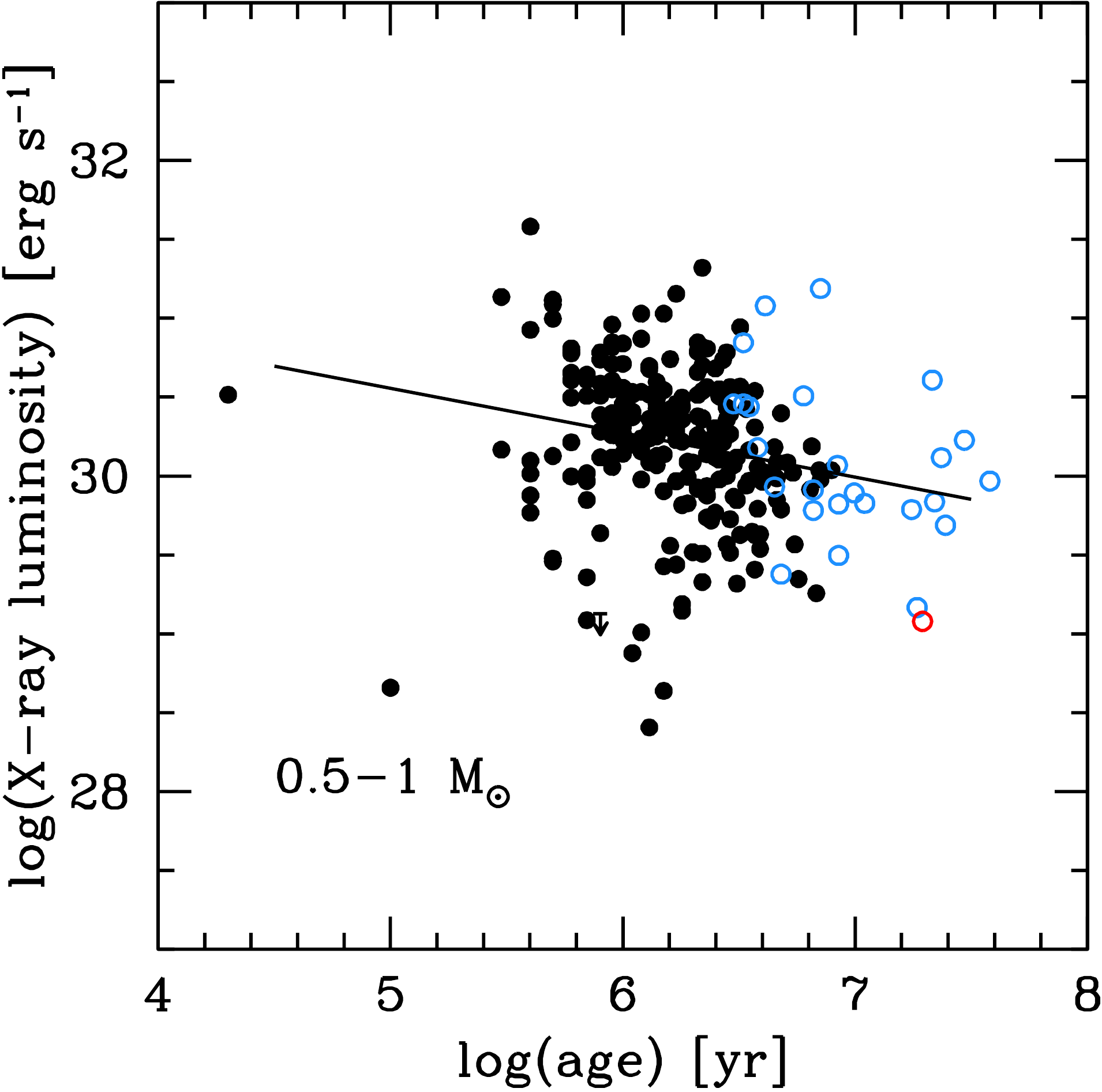} 
      \includegraphics[width=0.3\textwidth]{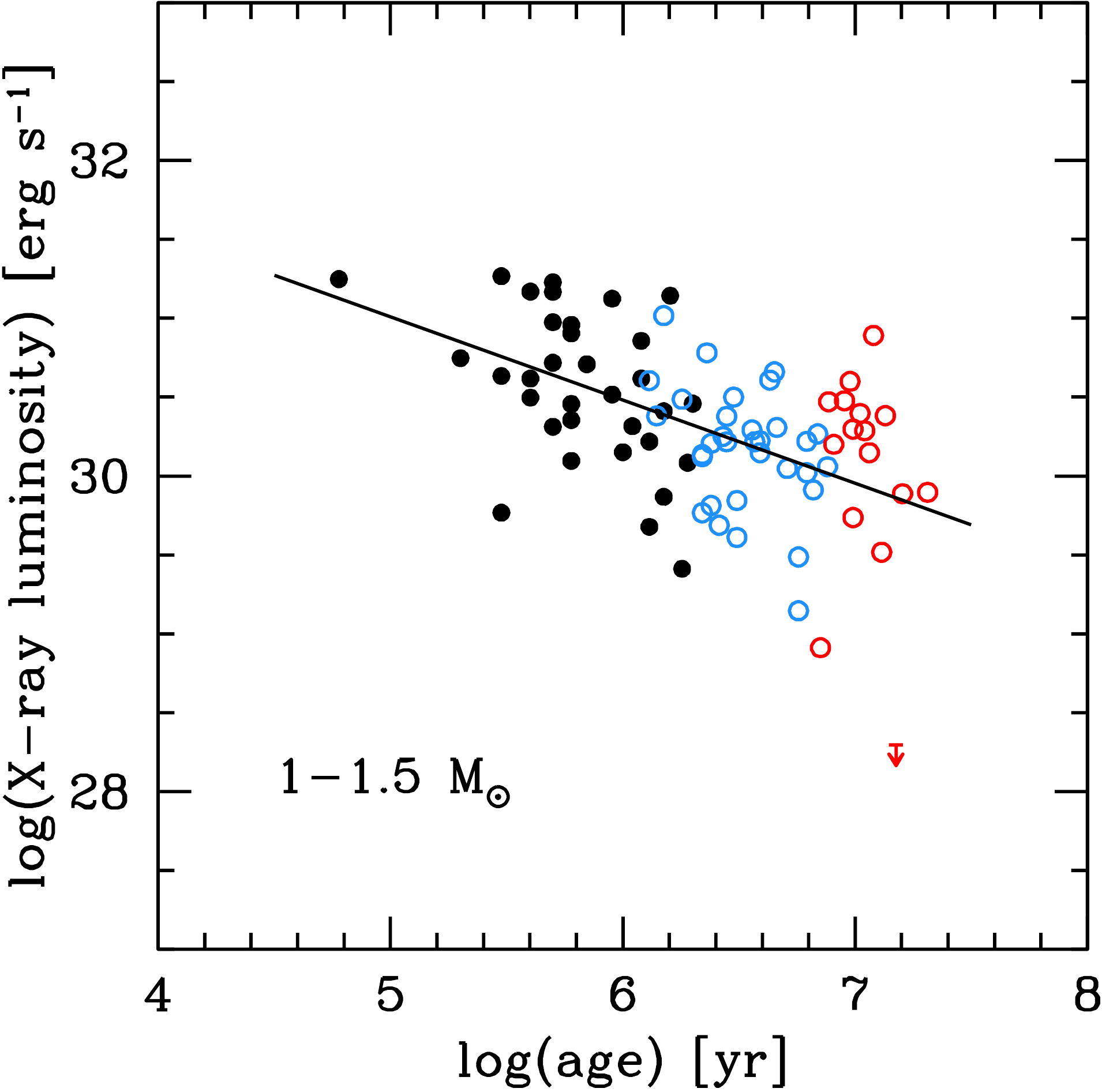} \\
           \includegraphics[width=0.3\textwidth]{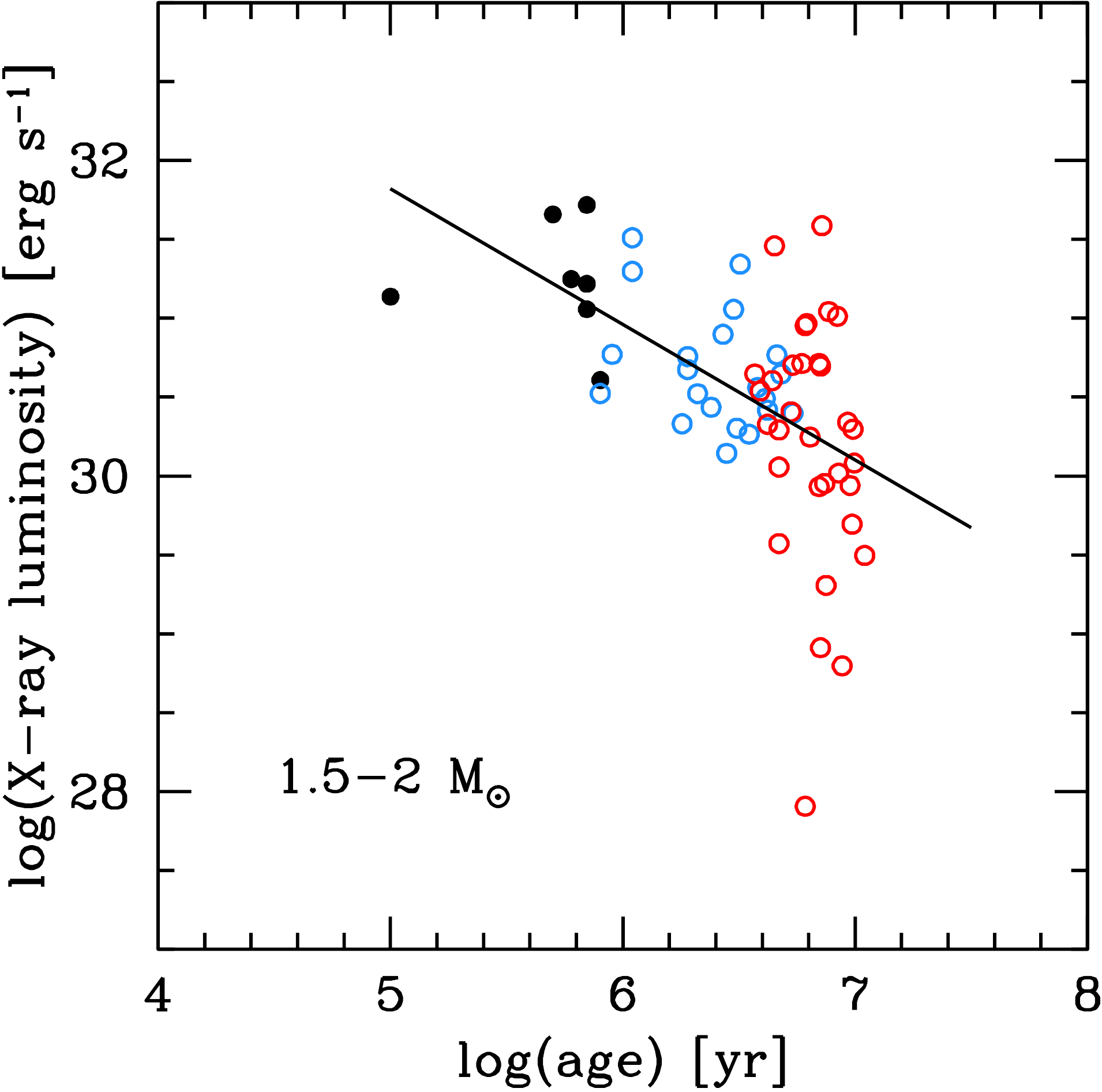}  
      \includegraphics[width=0.3\textwidth]{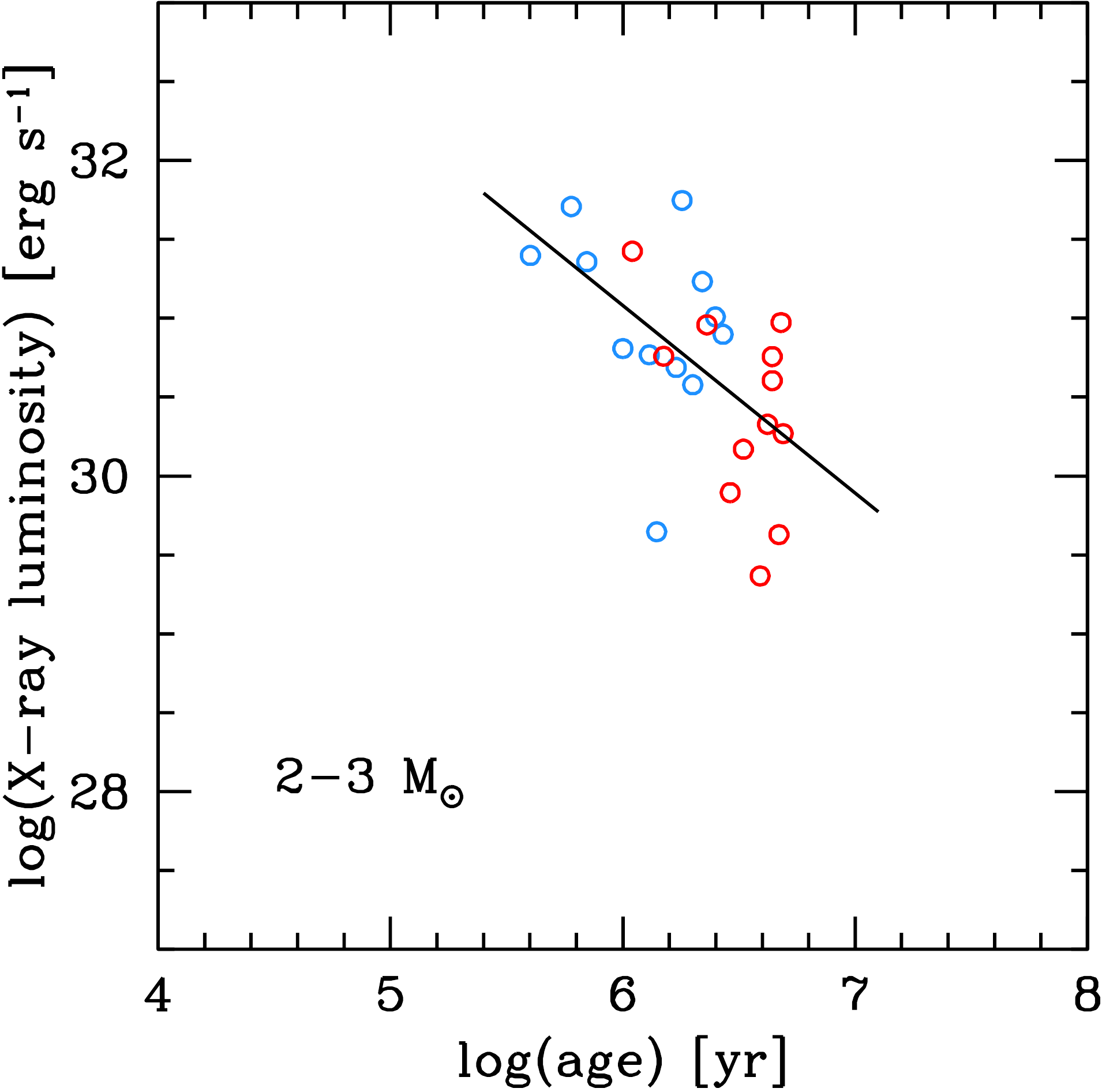}  
   \caption{The decay of X-ray luminosity with stellar age for the indicated mass bins.  Black points are fully convective PMS stars and open points partially convective PMS stars.  Blue/black points are stars on Hayashi tracks, red points are stars on Henyey tracks. The linear
    regression fits from the EM algorithm in ASURV are shown in Table \ref{table_logLX_logage}.}
   \label{plot_logLX_logage}
\end{figure*}


\subsection{$L_{\rm X}$ versus $t$}\label{appendix_age}
\citet{pre05a} report a decay in X-ray luminosity with age in mass stratified samples of PMS stars in the ONC.  Dividing the 
sample into mass bins negates any relationship potentially introduced by the PMS lifetime being a strong function of stellar mass.
Linear regression fits to $\log{L_{\rm X}}$ vs $\log({\rm age})$ for our sample are given in Table \ref{table_logLX_logage}.  The first four
mass bins (0.1-0.2, 0.2-0.4, 0.4-1.0, and 1.0-2.0$\,{\rm M}_\odot$) are chosen to allow a direct comparison between our results and those of \citet{pre05a}.  As our sample extends to 3$\,{\rm M}_\odot$, results are also given for 2.0-3.0$\,{\rm M}_\odot$.  The anti-correlations 
are similar to those reported by \citet{pre05a}, namely a decay of X-ray luminosity with age roughly of the form $L_{\rm X}\propto t^{-1/3}$,
although here, with our larger sample of stars the probability of (anti-)correlations being present from a generalised Kendall's 
$\tau$ test are greater.  It is also notable that in the 1.0-2.0$\,{\rm M}_\odot$ mass bin the decay is steeper: here $L_{\rm X}\propto t^{-0.55\pm0.09}$, 
P(0)$<$5e-5, compared to $L_{\rm X}\propto t^{-0.50\pm0.21}$, P(0)$=$0.053 in \citet{pre05a} with a smaller sample. The decay is steeper still if the mass range is extended 
up to 3$\,{\rm M}_\odot$, $L_{\rm X}\propto t^{-0.61\pm0.09}$.        

As our sample includes almost 1000 PMS stars, we have further divided stars into mass bins of 0.5-1.0, 1.0-1.5, 1.5-2.0, and 
2.0-3.0$\,{\rm M}_\odot$.\footnote{2.5-3.0$\,{\rm M}_\odot$ is not considered as there are only 5 stars in this mass range.} Plots 
of $\log{L_{\rm X}}$ vs $\log t$ shown in Fig. \ref{plot_logLX_logage}. 
As is apparent from Table \ref{table_logLX_logage}, the exponent steadily decreases from about -1/3, to -1/2, to -9/10, to -6/5 across 
these four mass bins.  There is a faster decay in X-ray luminosity with age for stars of higher mass.  The last two columns of Table 
\ref{table_logLX_logage} are the fraction of partially convective stars $f_{\rm rad. core}$, and the fraction of Henyey track stars 
$f_{\rm Hen}$, in each mass bin e.g. $f_{\rm Hen}=0$ (=1) would indicate that all stars in that mass bin are on Hayashi (Henyey) tracks.  This faster decay in $L_{\rm X}$ with age is driven by the greater proportion of partially convective stars in the higher mass bins; and is related to the decay in $L_{\rm X}$ as stars develop substantial radiative cores (section \ref{xraycompare}) and our finding that the longer a star has spent with a radiative core the weaker its X-ray emission becomes (section \ref{sectionevolve}).     

\begin{table}
  \caption{Linear regression fits to $\log{L_{\rm X}}$ vs $\log({\rm age})$ [i.e. $L_{\rm X}\propto t^a$] for stars in different stellar mass bins. $N_{\rm stars}$ is the number of stars in each mass bin with the number in parenthesis the number of $L_{\rm X}$ upper limits that contribute to $N_{\rm stars}$ in that bin.  $f_{\rm rad. core}$ and $f_{\rm Hen}$ are the fraction of stars in a given mass bin that are partially convective and on Henyey tracks respectively.}
  \begin{tabular}{cccccc}
  \hline
   $M_\ast/{\rm M}_\odot$ & a & prob. & $N_{\rm stars}$ & $f_{\rm rad. core}$ & $f_{\rm Hen}$ \\
  \hline
   0.1-0.2 & -0.30$\pm$0.09 & 9e-4 & 142 (20) & 0.00 & 0.00 \\
   0.2-0.4 & -0.29$\pm$0.08 & 1e-4 & 330 (12) & 0.01 & 0.00 \\
   0.4-1.0 & -0.29$\pm$0.07 & $<$5e-5 & 376 (1) & 0.07 & 0.00 \\
   1.0-2.0 & -0.55$\pm$0.09 & $<$5e-5 & 139 (1) & 0.71 &  0.32 \\
   1.0-3.0 & -0.61$\pm$0.09 & $<$5e-5 & 160 (1) & 0.75 &  0.35 \\
  \hline
  0.5-1.0 & -0.28$\pm$0.08 & $<$5e-5 & 257 (1) & 0.10 & 0.00 \\
  1.0-1.5 & -0.53$\pm$0.10 & $<$5e-5 & 81 (1) & 0.59 &  0.19 \\
  1.5-2.0 & -0.86$\pm$0.19 & $<$5e-5 & 60 (0) & 0.88 &  0.53 \\
  2.0-3.0 & -1.19$\pm$0.35 & 0.0059 & 24 (0) & 1.00 &  0.50 \\
  \hline
\end{tabular}
\label{table_logLX_logage}
\end{table}


\section{A comparison of the correlations using different stellar evolution models}\label{appendix_comparison}
In this Appendix we demonstrate that our results are independent of the adopted PMS mass tracks
and isochrones.  Starting with our final sample of 984 PMS stars (including the 34 $L_{\rm X}$ upper limits), see section \ref{siessstuff}, we recalculated the stellar masses, ages, and radiative core masses, using the PMS evolutionary models of \citet{jun07}. Co-author Y.-C. Kim kindly supplied us with more detailed grids than published as well as stellar internal structure information.  All 984 stars fell within the grid of \citet{jun07}. We repeated the process using the models of \citet{tog11}, for which E. Tognelli generously sent us radiative core masses.  However, 0.2$\,{\rm M}_\odot$ is the lowest available mass track of the \citet{tog11} models and 1$\,{\rm Myr}$ the youngest isochrone (the mass tracks themselves do extend to younger ages), so only 759 stars, including 13 $L_{\rm X}$ upper limits, could be used.  The other 225 stars were either of too low mass, too young, or both, to be considered.      

The different models yield different mass and age estimates for individual stars [see \citet{tog11} for a detailed comparison of several PMS evolutionary models].  Some PMS stars have fully convective interiors in one model and a partially convective interior in another. Likewise, some are on a Hayashi track in one model and on a Henyey track in another.  These are stars which lie close to the fully convective limit in the H-R diagram (represented by the solid blue line in Figure \ref{hrd}), or close to the point of transition from the Hayashi to the Henyey track.  Despite the inherent differences between the models, a detailed examination of Table \ref{table_comp} reveals that our results remain the same regardless of the model choice. Fully convective stars have lower, mean, fractional X-ray luminosities than partially convective PMS stars, with the difference being larger when comparing Hayashi to Henyey track stars and when the ONC is excluded (see section \ref{xraycompare}).  There is an almost linear correlation between $L_{\rm X}$ and $L_\ast$ for fully convective and Hayashi track stars, with the exponent of the correlation reducing if we consider partially convective stars.  There is no correlation between $L_{\rm X}$ and $L_\ast$ for Henyey track stars (see section \ref{xraycompare}).  This is likely driven by the decay of $L_{\rm X}$ with time since radiative core development (see section \ref{sectionevolve}).     

The correlation between $L_{\rm X}$ and age, $t$, does initially appear to be model dependent.  For example, in the highest mass bin, 2-3$\,{\rm M}_\odot$, there is no correlation when using the \citet{jun07} or \citet{tog11} models.   There are only 15 and 16 stars in this mass range in the two models, respectively, compared to 24 stars when using the \citet{sie00} models. We suspect that strong correlations would be recovered for all models if a greater sample size could be used.  The different models yield different mass estimates for individual stars. If we consider alternative mass bins to those listed in Table \ref{table_comp} strong correlations are still recovered.  As one example, for mass bins of 0.5-0.7, 0.7-1.0, 1.0-1.5, and 1.5-3.0$\,{\rm M}_\odot$ the exponents of the $L_{\rm X}$ versus $t$ correlation using the models of \citet{tog11} are -0.47$\pm$0.10, -0.63$\pm$0.09, -0.77$\pm$0.16, and -1.00$\pm$0.22 respectively, with $P(0)$<5e-5 for the lowest three mass bins and 1e-4 for the highest mass bin.  Our conclusion that the decrease in $L_{\rm X}$ with age steepens for higher mass bins (see Appendix \ref{appendix_age}), which contain a greater proportion of partially convective stars compared to lower mass bins, remains valid regardless of the chosen PMS evolutionary model.

\begin{table*}
  \caption{A comparison of our main results using three different PMS evolutionary models.}
  \begin{tabular}{ccccc}
  \hline
   & & \citet{sie00} & \citet{jun07} & \citet{tog11} \\
  \hline
  N$_{\rm stars}$ & total                                                      &  984 (34)  & 984 (34)  & 759 (13) \\
  	                   & fully convective                                     &  836 (33)  & 884 (33)  & 620 (12) \\
   		           & radiative core                                       &  148 (1)    & 100 (1)    & 139 (1) \\
                            & Hayashi track                                       &  927 (33)  & 949 (33)  & 720 (12) \\
                            & Henyey track                                        &  57 (1)      & 35 (1)      & 39 (1) \\
                            & Hayashi track with radiative core         &  91 (0)      & 65 (0)      & 100 (0) \\    
  \hline
  $\langle \log(L_{\rm X}/L_\ast)\rangle$ 	& fully convective all regions  		&  -3.44 	&  -3.43 	& -3.35 \\
    							     	& radiative core all regions             	&  -3.57 	&  -3.72 	& -3.56 \\
       						             	& Hayashi track all regions         	&  -3.43 	&  -3.44 	& -3.35 \\
         							& Henyey track all regions       		&  -3.95	&  -4.10 	& -4.14 \\
   								& fully convective ONC excluded      	&  -3.30 	&  -3.31 	& -3.27 \\
						       		& radiative core ONC excluded      	&  -3.68 	&  -3.79 	& -3.71 \\
       								& Hayashi track ONC excluded       	&  -3.32 	&  -3.33 	& -3.30 \\
         							& Henyey track ONC excluded       	&  -3.96 	&  -4.12 	& -4.16 \\
  \hline
                                                 		&              						&   a, P(0)       				& a, P(0) 					& a, P(0) \\
  $L_{\rm X}\propto L_\ast^a$ 		& all stars 						& 0.81$\pm$0.04, $<$5e-5 	& 0.84$\pm$0.04, $<$5e-5 	& 0.67$\pm$0.04, $<$5e-5 \\
   							& fully convective				& 0.93$\pm$0.04, $<$5e-5 	& 0.97$\pm$0.04, $<$5e-5 	& 0.89$\pm$0.05, $<$5e-5 \\
   							& radiative core				     	& 0.33$\pm$0.09, $<$5e-5     	& 0.28$\pm$0.13, 0.0022 		& 0.34$\pm$0.08, $<$5e-5 \\
   							& Hayashi track				     	& 0.92$\pm$0.04, $<$5e-5 	& 0.94$\pm$0.04, $<$5e-5	& 0.83$\pm$0.04, $<$5e-5 \\
   							& Henyey track				     	& 0.13$\pm$0.23, 0.518 		& -0.11$\pm$0.27, 0.5894 	& 0.11$\pm$0.27, 0.8750 \\
  							& Hayashi track with radiative core 	& 0.61$\pm$0.08, $<$5e-5 	& 0.67$\pm$0.15, $<$5e-5 	& 0.60$\pm$0.08, $<$5e-5 \\
  \hline
    $L_{\rm X}\propto t_{\rm since}^a$ 	& radiative core     				& -0.42$\pm$0.09, $<$5e-5 	& -0.45$\pm$0.11, 1e-4 		& -0.40$\pm$0.09, $<$5e-5 \\
     								& Hayashi track with radiative core 	& -0.39$\pm$0.10,  1e-4  		& -0.45$\pm$0.13, 0.0012 	& -0.39$\pm$0.08, $<$5e-5 \\
     								& Henyey track      				& -1.10$\pm$0.36, 0.009  		& -1.34$\pm$0.53, 0.0045 	& -1.02$\pm$0.55, 0.0136 \\ 
  \hline
  $L_{\rm X}\propto t^a$ 	& 0.5-1.0$\,{\rm M}_\odot$ 	& -0.28$\pm$0.08, $<$5e-5 	& -0.40$\pm$0.06, $<$5e-5 	& -0.51$\pm$0.07, $<$5e-5 \\
   					& 1.0-1.5$\,{\rm M}_\odot$	& -0.53$\pm$0.10, $<$5e-5 	& -0.76$\pm$0.11, $<$5e-5  	& -0.77$\pm$0.16, $<$5e-5 \\
   					& 1.5-2.0$\,{\rm M}_\odot$ 	& -0.86$\pm$0.19, $<$5e-5 	& -0.78$\pm$0.25, 7e-4  		& -1.09$\pm$0.26, 1e-4 \\
   					& 2.0-3.0$\,{\rm M}_\odot$	& -1.19$\pm$0.35, 0.0059 	&-0.91$\pm$0.31, 0.0833  	& -0.77$\pm$0.45, 0.4177 \\
  \hline
\end{tabular}
\label{table_comp}
\end{table*}


\bsp	
\label{lastpage}
\end{document}